\documentclass[twocolumn]{aastex631}
\usepackage{amsmath,amssymb}
\usepackage{upgreek}
\usepackage[T1]{fontenc}

\newcommand{\water}{H$_2$O}
\newcommand{\sotwo}{SO$_2$}
\newcommand{\cotwo}{CO$_2$}
\newcommand{\methane}{CH$_4$}
\newcommand{\ammonia}{NH$_3$}

\begin{document}

\title{On Linking Planet Formation Models, Protoplanetary Disk Properties, and Mature Gas Giant Exoplanet Atmospheres}

\author[0000-0002-9464-8101]{Adina~D.~Feinstein}
\altaffiliation{Authors A. D. Feinstein, R. A. Booth, and J. B. Bergner contributed equally to the preparation of this manuscript.}
\affiliation{Department of Physics and Astronomy, Michigan State University, East Lansing, MI 48824, USA}
\affiliation{NHFP Sagan Fellow}

\author[0000-0002-0364-937X]{Richard~A.~Booth}
\altaffiliation{Authors A. D. Feinstein, R. A. Booth, and J. B. Bergner contributed equally to the preparation of this manuscript.}
\affiliation{School of Physics and Astronomy, University of Leeds, Leeds LS2 9JT, UK}

\author[0000-0002-8716-0482]{Jennifer~B.~Bergner}
\altaffiliation{Authors A. D. Feinstein, R. A. Booth, and J. B. Bergner contributed equally to the preparation of this manuscript.}
\affiliation{Department of Chemistry, University of California, Berkeley, Berkeley, CA 94720}

\author[0000-0003-3667-8633]{Joshua~D.~Lothringer}
\affiliation{Space Telescope Science Institute, Baltimore, MD, USA}

\author[0000-0003-0593-1560]{Elisabeth~C.~Matthews}
\affiliation{Max-Planck-Institut für Astronomie, Königstuhl 17, D-69117 Heidelberg, Germany}

\author[0000-0003-0156-4564]{Luis~Welbanks}
\altaffiliation{51 Pegasi b Fellow}
\affiliation{School of Earth \& Space Exploration, Arizona State University, Tempe, AZ, 85257, USA}

\author[0000-0002-0747-8862]{Yamila~Miguel}
\affiliation{SRON Netherlands Institute for Space Research , Niels Bohrweg 4, 2333 CA Leiden, the Netherlands}
\affiliation{Leiden Observatory, Leiden University, Niels Bohrweg 2, 2333CA Leiden, The Netherlands}

\author[0000-0002-8868-7649]{Bertram~Bitsch}
\affiliation{Department of Physics, University College Cork, Cork, Ireland}

\author[0000-0002-8247-6453]{Linn E.\ J.\ Eriksson}
\affiliation{Institute for Advanced Computational Sciences, Stony Brook University, Stony Brook, NY, 11794-5250, USA}
\affiliation{Department of Astrophysics, American Museum of Natural History, 200 Central Park West, New York, NY 10024, USA}

\author[0000-0002-4207-6615]{James Kirk}
\affiliation{Department of Physics, Imperial College London, Prince Consort Road, SW7 2AZ, London, UK}

\author[0000-0002-8573-805X]{Stefan~Pelletier}
\affiliation{Observatoire astronomique de l’Université de Genève, 51 chemin Pegasi 1290 Versoix, Switzerland}

\author[0000-0002-8873-6826]{Anna~B.T.~Penzlin}
\affiliation{Ludwig-Maximilians-Universit{\"a}t M{\"u}nchen, Universit{\"a}ts-Sternwarte, Scheinerstr.~1, 81679 M{\"u}nchen, Germany}
\affiliation{Department of Physics, Imperial College London, Prince Consort Road, SW7 2AZ, London, UK}

\author[0000-0002-4487-5533]{Anjali A.\ A.\ Piette}
\affiliation{School of Physics and Astronomy, University of Birmingham, Edgbaston, Birmingham B15 2TT, UK}

\author[0000-0002-2875-917X]{Caroline~Piaulet-Ghorayeb}
\affiliation{Department of Astronomy \& Astrophysics, University of Chicago, Chicago, IL 60637, USA}

\author[0000-0002-6429-9457]{Kamber~R.~Schwarz}
\affiliation{Max-Planck-Institut für Astronomie, Königstuhl 17, D-69117 Heidelberg, Germany}

\author[0000-0002-1923-7740]{Diego Turrini}
\affiliation{Turin Astrophysical Observatory, National Institute of Astrophysics (INAF), via Osservatorio 20, 10025, Pino Torinese (TO), Italy}

\author[0000-0002-9147-7925]{Lorena Acu\~{n}a-Aguirre}
\affiliation{Max-Planck-Institut f\"ur Astronomie, K\"onigstuhl 17, D-69117 Heidelberg, Germany}

\author[0000-0003-0973-8426]{Eva-Maria Ahrer}
\affiliation{Max-Planck-Institut f\"ur Astronomie, K\"onigstuhl 17, D-69117 Heidelberg, Germany}

\author[0000-0002-8399-472X]{Madyson~G.~Barber}
\altaffiliation{NSF Graduate Research Fellow}
\affiliation{Department of Physics and Astronomy, The University of North Carolina at Chapel Hill, Chapel Hill, NC 27599, USA}

\author[0000-0002-2072-6541]{Jonathan Brande}
\affiliation{Department of Physics and Astronomy, University of Kansas, Lawrence, KS, USA}

\author[0000-0001-6703-0798]{Aritra Chakrabarty}
\altaffiliation{NASA Postdoctoral Program (NPP) Fellow}
\affiliation{NASA Ames Research Center, Moffett Field, CA 94035, USA}

\author{Ian J.\ M.\ Crossfield}
\affiliation{Department of Physics and Astronomy, University of Kansas, Lawrence, KS, USA}
\affiliation{Max-Planck-Institut f\"ur Astronomie, K\"onigstuhl 17, D-69117 Heidelberg, Germany}

\author[0000-0002-2919-7500]{Gabriel-Dominique~Marleau}
\affiliation{Max-Planck-Institut f\"ur Astronomie, K\"onigstuhl 17, D-69117 Heidelberg, Germany}
\affiliation{Universit\"at Duisburg--Essen, Lotharstra\ss{}e 1, 47057 Duisburg, Germany}

\author[0009-0002-5134-3911]{Helong Huang}
\affiliation{Department of Astronomy, Tsinghua University, Haidian DS 100084 Beijing, China}

\author[0000-0002-5893-6165]{Anders Johansen}
\affiliation{Globe Institute, University of Copenhagen and Lund Observatory, Department of Physics, Lund University}

\author[0000-0003-0514-1147]{Laura Kreidberg}
\affiliation{Max-Planck-Institut f\"ur Astronomie, K\"onigstuhl 17, D-69117 Heidelberg, Germany}

\author[0000-0002-4881-3620]{John H.\ Livingston}
\affiliation{Astrobiology Center, 2-21-1 Osawa, Mitaka, Tokyo, 181-8588, Japan}
\affiliation{National Astronomical Observatory of Japan, 2-21-1 Osawa, Mitaka, Tokyo, 181-8588, Japan}
\affiliation{Department of Astronomical Science, The Graduate University for Advanced Studies, 2-21-1 Osawa, Mitaka, Tokyo, 181-8588, Japan}

\author[0000-0002-4671-2957]{Rafael Luque}
\affiliation{Department of Astronomy \& Astrophysics, University of Chicago, Chicago, IL 60637, USA}
\affiliation{NHFP Sagan Fellow}

\author[0000-0002-3327-1072]{Maria~Oreshenko}
\affiliation{Institute for Particle Physics and Astrophysics, ETH Z\"urich, Wolfgang-Pauli-Strasse 27, 8093 Zurich, Switzerland}

\author[0000-0003-1096-7656]{Elenia~Pacetti}
\affiliation{Institute for Space Astrophysics and Planetology, National Institute of Astrophysics (INAF), Via Fosso del Cavaliere 100, I-00133, Rome, Italy}

\author[0000-0002-8545-6175]{Giulia Perotti}
\affiliation{Max-Planck-Institut f\"ur Astronomie, K\"onigstuhl 17, D-69117 Heidelberg, Germany}
\affiliation{Niels Bohr Institute, University of Copenhagen, NBB BA2, Jagtvej 155A, 2200 Copenhagen, Denmark}

\author[0009-0009-8749-9513]{Jesse Polman}
\affiliation{Division of Space Research and Planetary Sciences, Physics Institute, University of Bern, Gesellschaftsstrasse 6, 3012 Bern, Switzerland}

\author[0000-0001-7216-4846]{Bibiana~Prinoth}
\affiliation{Lund Observatory, Division of Astrophysics, Department of Physics, Lund University, Box 118, 221 00 Lund, Sweden}
\affiliation{European Southern Observatory, Alonso de Córdova 3107, Vitacura, Región Metropolitana, Chile}

\author[0000-0002-3913-7114]{Dmitry~A. Semenov}
\affiliation{Institut f\"ur Theoretische Astrophysik, Zentrum f\"ur Astronomie der Universit\"at Heidelberg, Albert-Ueberle-Str. 2, 69120 Heidelberg, Germany}
\affiliation{Max-Planck-Institut f\"ur Astronomie, K\"onigstuhl 17, D-69117 Heidelberg, Germany}

\author[0000-0002-3771-8054]{Jacob~B.~Simon}
\affiliation{Department of Physics and Astronomy, Iowa State University, Ames, IA, 50010, USA}

\author[0009-0008-2801-5040]{Johanna Teske}
\affiliation{Earth and Planets Laboratory, Carnegie Institution for Science, 5241 Broad Branch Road, NW, Washington, DC 20015, USA}
\affiliation{The Observatories of the Carnegie Institution for Science, 813 Santa Barbara St., Pasadena, CA 91101, USA}

\author[0000-0001-8818-1544]{Niall Whiteford}
\affiliation{Department of Astrophysics, American Museum of Natural History, Central Park West at 79th Street, New York, NY 10034, USA}

\begin{abstract}

Measuring a single elemental ratio (e.g., carbon-to-oxygen) provides insufficient information for understanding the formation mechanisms and evolution that affect our observations of gas  giant planet atmospheres. Although the fields of planet formation, protoplanetary disks, and exoplanets are well established and interconnected, our understanding of how to self-consistently and accurately link the theoretical and observational aspects of these fields together is lacking. To foster interdisciplinary conversations, the Max-Planck Institut für Astronomie (MPIA) hosted a week-long workshop called, ``Challenge Accepted: Linking Planet Formation with Present-Day Atmospheres.''  Here, we summarize the latest theories and results in planet formation modeling, protoplanetary disk observations, and atmospheric observations of gas giant atmospheres to address one of the challenges of hosting interdisciplinary conferences: ensuring everyone is aware of the state-of-the-art results and technical language from each discipline represented. Additionally, we highlight key discussions held at the workshop. Our main conclusion is that it is unclear what the ideal observable is to make this link between formation scenarios and exoplanet atmospheres, whether it be multiple elemental abundance ratios, measuring refractory budgets, or something else. Based on discussions held throughout the workshop, we provide several key takeaways of what the workshop attendees feel need the most improvement and exploration within each discipline.

\end{abstract}

\keywords{Planet Formation (1241) -- Protoplanetary Disks (1300) -- Exoplanet Atmospheres (487) -- Exoplanet Formation (492) -- Exoplanet Atmospheric Composition (2021)}

\section{Motivation}\label{sec:motivation}

Modern exoplanet science aims to not only characterize planets around other stars, but to explain why and how the observed diversity of planets arises. Early attempts to link giant planet atmospheric compositions to a formation mechanism began in our own solar system, following the Galileo probe measurements of Jupiter's atmosphere \citep[e.g.][]{Owen1999}. For example, \cite{Lodders2004} found that when the elemental abundances in Jupiter's atmosphere are normalized by sulfur, there may be a factor of 2 depleted in oxygen. Given these abundances, \cite{Lodders2004} proposed a new formation explanation that suggests Jupiter may have formed near the carbonaceous condensation/evaporation line, rather than the water ice line. As it subsequently became possible to characterize the atmospheres of planets beyond our solar system, the idea emerged that measurements of elemental ratios in a planet's atmosphere could provide a powerful constraint on its formation origins \citep{Madhusudhan2011}.  Around this time, \citet{Oberg2011} introduced the idea that volatile snowlines can dramatically influence the C/O ratio of gas vs.~solids at different radial positions within a protoplanetary disk.  This simple and elegant picture suggests that a giant planet's atmospheric C/O ratio can be directly mapped to its formation location within the parent disk \citep{molliere2022}. This idea found particular resonance with the community, and has been a major driver of exoplanet science over the past decade.  

\begin{figure*}[!htb]
    \centering
    \includegraphics[width=\textwidth]{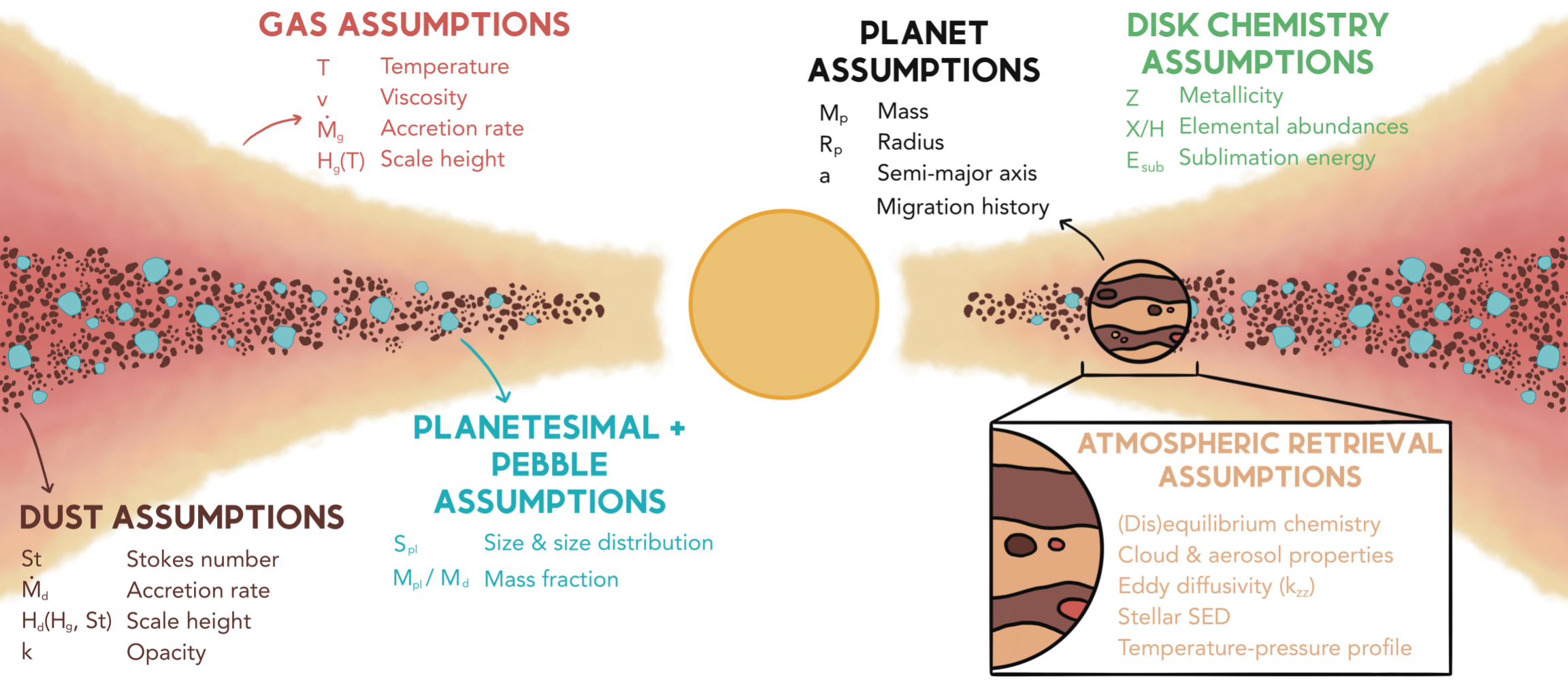}
    \caption{
    A summary schematic of assumptions that are made in planet formation modeling, protoplanetary disk modeling, and exoplanet atmospheric modeling. We note that several of the presented variables can be directly measured (e.g., planet mass, radius, and semi-major axis), but not necessarily in all systems. 
    }\label{fig:assumptions}
\end{figure*}

The processes involved in planet formation and evolution are, however, complex, interdependent, and dynamic (Figure~\ref{fig:assumptions}).  It is \textit{widely acknowledged} that a single elemental ratio is insufficient for linking giant planet atmospheres to their formation origins. Understanding how planets form remains an important goal for the field, but achieving it will require collaborative efforts across the communities of protoplanetary disk physics and chemistry, planet formation, atmospheric evolution, and exoplanet characterization.  In the coming years, maintaining open communication between these fields will be especially important as operational facilities (e.g., the Atacama Large Millimeter/sub-millimeter Array and JWST) continue to deliver and near-future space- and ground-based facilities (e.g., the Extremely Large Telescope, Square Kilometre Array, and ESA's PLATO and ARIEL missions) will deliver an unprecedented volume and quality of constraints on exoplanet atmospheres and the physics and chemistry of protoplanetary disks.  

The aim of this work, building off of the ``Challenge Accepted: Linking Planet Formation with Present-Day Atmospheres''\footnote{\url{https://the-great-link.github.io/}} workshop hosted by the Max-Planck-Institut f\"ur Astronomie, is to provide an accessible overview of the current progress and limitations in making the connection between gas giant planet ($R_p \geq 4 R_\oplus$) atmosphere observations and their formation. This article is intended to reflect the discussions and perspectives of workshop participants, and not to provide a comprehensive review of the fields covered.

This work is organized as follows. In Section \ref{sec:planet_formation} we outline leading theoretical frameworks of planet formation.  In Section \ref{sec:disks} we describe the current understanding of protoplanetary disk physics and chemistry which may influence planetary atmosphere compositions. In Section \ref{sec:planet_observations}, we discuss existing constraints on elemental abundances in giant planet atmospheres. Lastly, in Section \ref{sec:link}, we comment on potential promising avenues for making the link from planet formation to mature atmospheres.

\section{State of Planet Formation Models}
\label{sec:planet_formation}
There are two main theories of giant planet formation. One is the `bottom-up' core accretion theory, where the planet's rocky core forms first (by accreting small (mm- to cm-sized) pebbles, larger ($\sim$km-sized) planetesimals, or a combination of the two) before accreting a gaseous envelope from the protoplanetary disk. The second theory is `top-down', where the planet forms directly via the fragmentation of a gravitationally unstable disk. We briefly review the essentials of these ideas.

\subsection{``Bottom-up'': Core Accretion}

Planet formation via core accretion is thought to proceed as follows \citep[e.g.][]{Pollack1996}. First, the planet's core  grows by the accretion of solids. The high temperature during core growth initially prevents significant amounts of gas accretion. However, as core growth slows, the planets can start to accrete gas. Initially, gas accretion is slow and limited by the ability of the planet to cool \citep{Lee2015}, but the planet enters the rapid runaway gas accretion phase once the gas mass becomes comparable to the mass of solids accreted \citep{Mizuno1980,Stevenson1982,Bodenheimer2000,Rafikov2006}. At this stage, the planet quickly grows to become a gas giant. 

Largely, the picture is the same whether the solids are primarily accreted in the form of dust/pebbles ($\lesssim$~cm particles) or planetesimals ($\gtrsim$~km-size solid bodies), although details such as the efficiency,  speed of growth, and preferred range of locations where gas giants form differ between the two scenarios. Both scenarios can reproduce some key features of the exoplanet demographics \citep[e.g.][]{emsenhuber2021,Drazkowska2023}, but no model yet explains the full population \citep[see also discussion in][]{Burn2025}. Here, we discuss the key issues for linking planet formation to the giant exoplanet atmospheres and refer the reader to reviews by \citet{Goldreich2004}, \citet{Johansen2017}, and \citet{Mordasini2024} for further details.

Some key differences between planetesimal and pebble accretion are when and how much solid material is accreted. The composition of these components can also differ between the scenarios. In planetesimal accretion, the initial phase of core growth slows once the core has accreted all of the planetesimals within its `feeding zone' \citep{Lissauer1987}. However, planetesimals continue to be accreted because the size of the feeding zone increases as the planet grows via gas accretion \citep[e.g.][]{Pollack1996}. Furthermore, planet migration and the diffusion of planetesimals into the feeding zone continue to supply fresh planetesimals \citep{Tanaka1999,Turrini2021}. As a result, the gas giants can continue to accrete planetesimals throughout their growth. 

Conversely, the accretion of pebbles terminates once the planet reaches the `pebble isolation mass' -- the mass at which the planet starts to open a gap in the protoplanetary disk \citep{Lambrechts2014,Bitsch2018, Ataiee2018}. Above the pebble isolation mass, pebbles are trapped in a pressure maximum outside the planet's orbit. They are no longer accreted, resulting in little enrichment of solids once run-away gas accretion has begun (except perhaps in the innermost disk regions where the pebbles are small, and the turbulence is strong, \citealt{Morbidelli2023}). Another difference between the two mechanisms is in the composition of the material added to the forming planet. Because of their small size, pebbles quickly reach thermal equilibrium with the surrounding gas, meaning that the material accreted by the growing planet always has the elemental composition characteristic of solids in the surrounding disk environment. The bulk composition of planetesimals records instead that of their formation region, meaning that they can enrich the accreting planets in elements not otherwise readily accessible \citep{Turrini2015,seligman2022_comets,Sainsbury-Martinez2024,Polychroni2025}.
The two scenarios also likely prefer different formation locations for the planets; however, conclusions about where planets can and cannot form by a given mechanism are sensitive to a number of the assumptions made \citep[e.g.][]{Goldreich2004,Brugger2020,emsenhuber2021,Drazkowska2023}.

In principle, on-going accretion of planetesimals versus halted accretion of pebbles could result in an atmospheric ``metallicity'' (elements heavier than H and He) difference. However, internal mixing may be important (although the efficiency is poorly constrained), and moreover atmospheric metallicities are likely different from bulk metallicities \citep[e.g][]{Vazan2015,Polman2024,2024ApJ...977..227K}. Fortunately, the thermal evolution of the planet is also affected by internal mixing, suggesting that these processes may be possible to constrain observationally. Similarly, measurements of a planet's Love number via its orbital evolution may help to constrain the internal structure \citep[e.g.,][]{Baumeister2020,Bernabo2024}.  

There are, however, other factors that complicate the interpretation of a planet's composition. For example, the composition of planetesimals accreted may not reflect the local disc conditions since planetesimals may largely retain their original composition when moving to warmer regions \citep[e.g.][]{Turrini2015,Shibata2020,Turrini2021,Sainsbury-Martinez2024,Polychroni2025}. It should also be noted that the actual differences between planetesimal and pebble accretion end-member scenarios may be smaller than currently predicted because often-neglected secondary processes may help enrich the envelopes of giant planets formed by pebble accretion. Similarly, planetesimals and pebbles likely co-exist, both contributing to planet growth to some extent. Examples of neglected processes include the formation of planetesimals out of the pebbles trapped in the pressure maxima \citep{carrera2021, carrera2022, Eriksson2022}, the creation of dust and pebbles by planetesimal collisions \citep{Turrini2019,Turrini2023,Bernabo2022}, pebbles fragmenting to smaller sizes that can be accreted along with gas \citep{Szulagyi2022, Stammler2023, Petrovic2024} and the possibility that even large pebbles may be accreted onto giant planets in the inner disc \citep{Morbidelli2023}. Similarly, the recycling of atmospheric material back into the disk may further alter the composition of planets, allowing a planet to become depleted in volatiles, such as H$_2$O, even if the planet forms exterior of the sublimation line \citep{Johansen2021, Wang2023, Muller2024}.

In addition to growth processes, migration via interaction with the protoplanetary disk, planet-planet scattering and resonant interactions, and the Kozai-Lidov mechanism are important as they all imply that the final orbital architecture of the system may be different from how it was when the planets accreted their gas. Interactions between different planets in the same system are often neglected in studies aiming to predict exoplanet compositions. However, the study of the Solar System clearly highlights their importance in shaping the characteristics of planetary bodies, from the cometary impacts on its most massive planet Jupiter \citep{Zahnle2003,Turrini2015,Hueso2018} and their atmospheric effects \citep{Taylor2004,Cavalie2013,Turrini2015,Flagg2016} to the collisional contamination of its smallest solid bodies recorded by asteroids \citep{McCord2012,Turrini2014,Turrini2016,Tatsumi2021,DellaGiustina2021} and primitive meteorites  \citep{Nittler2019,Kebukawa2020}. N-body processes were even more intense at the time of the solar nebula, when they have been argued to have played a key role in delivering water and organics to the inner Solar System \citep{Turrini2011,Turrini2018b,Turrini2014b,Raymond2017,pirani2019}.

Exoplanetary studies have recently begun including the effects by simulating the N-body dynamics \citep[e.g.][]{Turrini2015,emsenhuber2021,seligman2022_comets,Sainsbury-Martinez2024,Polychroni2025}. Furthermore, N-body interactions can be important for reproducing exoplanet demographics \citep{Limbach2015,izidoro2017,Zinzi2017,Laskar2017,Turrini2020,emsenhuber2021, Izidoro2022}, but the degree to which N-body dynamics affect exoplanet compositions is currently unknown and strongly dependent on the specific architecture of the planetary system (e.g. sequential giant planet formation seems to have only minimal effects on the composition of inner planets, see \citealt{Eberlein2024}). At the simplest level, the very existence of hot Jupiters highlights the need to account for some orbital evolution, as they unlikely to form in situ \citep[e.g.][]{dawson2018}.

The qualitative similarity between pebble and planetesimal accretion models means that studies predicting planet bulk compositions come to broadly similar conclusions if the same disk compositions are used and if the planets compared have at the same metallicity \citep[c.f.][]{madhusudhan2014,madhu2017,Hobbs2022}, though details can vary somewhat \citep[c.f.][]{Bitsch2022,Turrini2021}. However, it has long been recognised that large pebbles drift inward through the protoplanetary disk faster than the gas, carrying volatiles with them in the form of ices \citep{Cuzzi2004,Oberg2016,Booth2017}. Since the inner parts of disks are warmer, these volatiles evaporate once the pebbles cross ice lines. As a result, the disk becomes enhanced in volatiles inside their respective ice lines. Including this effect in planet formation models results in the formation of a population of planets that acquire super-solar C/H, N/H and O/H abundances by accreting volatile-rich gas while having lower refractory elements (including sulfur) because the refractories remain solid \citep{Booth2017,Schneider2021b,Danti2023}. Conversely, high volatile abundances due to the accretion of pebbles/planetesimals should be associated with higher refractory abundances. Thus, the volatile/refractory ratio has been suggested as a way to differentiate between planet formation via pebble and planetesimal accretion \citep{Schneider2021b, Turrini2021,Pacetti2022,Chachan2023,crossfield23}. However, caution should be taken not to over-interpret volatile/refractory ratios because these ratios more directly constrain the efficiency of pebble drift than whether the solids accreted were predominantly pebbles or planetesimals. Further, these end-member models that include only pebbles or planetesimals are likely unrealistic, and combined pebble-planetesimal models are needed to address this.
Such models are starting to be developed \citep[e.g.][]{Danti2023}. One caveat is that refractory species do eventually evaporate, both sufficiently close to the start within the protoplanetary disc \citep[e.g.][]{Estrada2016,Pacetti2022,Danti2023} and also within planetary envelopes \citep[e.g.][]{Brouwers2018, Steinmeyer2023}. Evaporation of refractories within the discs could mask the signature from the accretion of solids, if giant planets can accrete sufficient gas under such conditions. Further, sulfur sublimation produces H$_2$S at around 750~K \citep{Steinmeyer2023}, and thus may mix more readily into the atmosphere or back into the disc than the more refractory oxygen-bearing silicates, potentially affecting the S/O ratio. Care will be needed to distinguish these factors affecting the atmospheric volatile/refractory ratios.

\subsection{``Top-down'': Gravitational Instability}
For a comprehensive review of gravitational instability (GI), see \citet{Kratter2016}.

Protoplanetary disks are `gravitationally unstable' when their masses are high enough to overcome stabilizing forces due to pressure and rotation. The famous Toomre Q parameter, $$Q=\frac{c_s \Omega}{\uppi G \Sigma} \lesssim 1,$$ \citep{safronov1960, toomre1964} determines whether disks are gravitationally unstable. Here, the sound speed is given by $c_s$ and the Keplerian angular velocity, $\Omega$, which act against the local disk mass represented by the disk surface density $\Sigma$.

While $Q < 1$ is necessary for GI, the formation of planets or brown dwarfs depends on how quickly the disk can cool and thereby reduce the local pressure in the gas. If cooling is sufficiently slow, heating due to spiral shocks generated by GI will balance cooling, and the disk will enter a quasi-steady state \citep{gammie2001,rice2003}. These works also showed that the disk fragments when the cooling time is short, $t_{\rm c} \Omega \equiv \beta \lesssim 3$. The precise condition that delineates whether fragmentation occurs has been the subject of much debate \citep{gammie2001,rice2005,meru2011,young2015,booth2019gi} with $\beta$ up to 20 being found to produce fragmentation in some studies. Additionally, whether a given disc fragments may be subject to some stochasticity \citep{paardekooper2012,young2016}. Still, the consensus is that a critical $\beta \approx 3$ is reasonable and that the precise value is unimportant because $\beta$ decreases rapidly with increasing distance from the star. 

We can estimate where GI may typically form planets by assuming a typical disk temperature structure, $T(R) = 150 (R/{\rm au})^{-0.5}\,{\rm K}$, which together with $Q=1$ determines $\Sigma(R)$. Assuming that the disk cools as a blackbody at $T(R)$ allows us to determine $\beta (R) \approx 5 (R/50\,{\rm au})^{-1.75}$. The planet masses can be estimated from $M_{\rm p} \approx \Sigma \lambda^2$, where $\lambda = 2\uppi H$ ($H=c_s/\Omega$) is the typical length scale of the instability, resulting in  $M_{\rm p}\approx 10 M_{\rm J} (R/50\,{\rm au})^{0.75}$. We see that GI produces super-Jupiters or brown dwarfs beyond 50--100~au from their host star. While our treatment here is very simplified, the results are borne out by more detailed calculations \citep{rafikov2005gi,clarke2009}. The fate of these fragments is somewhat sensitive to how they evolve because rapid migration through the disk can lead to tidal stripping, causing the fragments to lose mass or be destroyed \citep[e.g.][]{nayakshin2010}. However, population synthesis models by \citet{forgan2013} generally found that tidal stripping is insignificant and that the planets tend to either remain giants or are completely destroyed.

\begin{figure}[h!]
\includegraphics[width=\columnwidth]{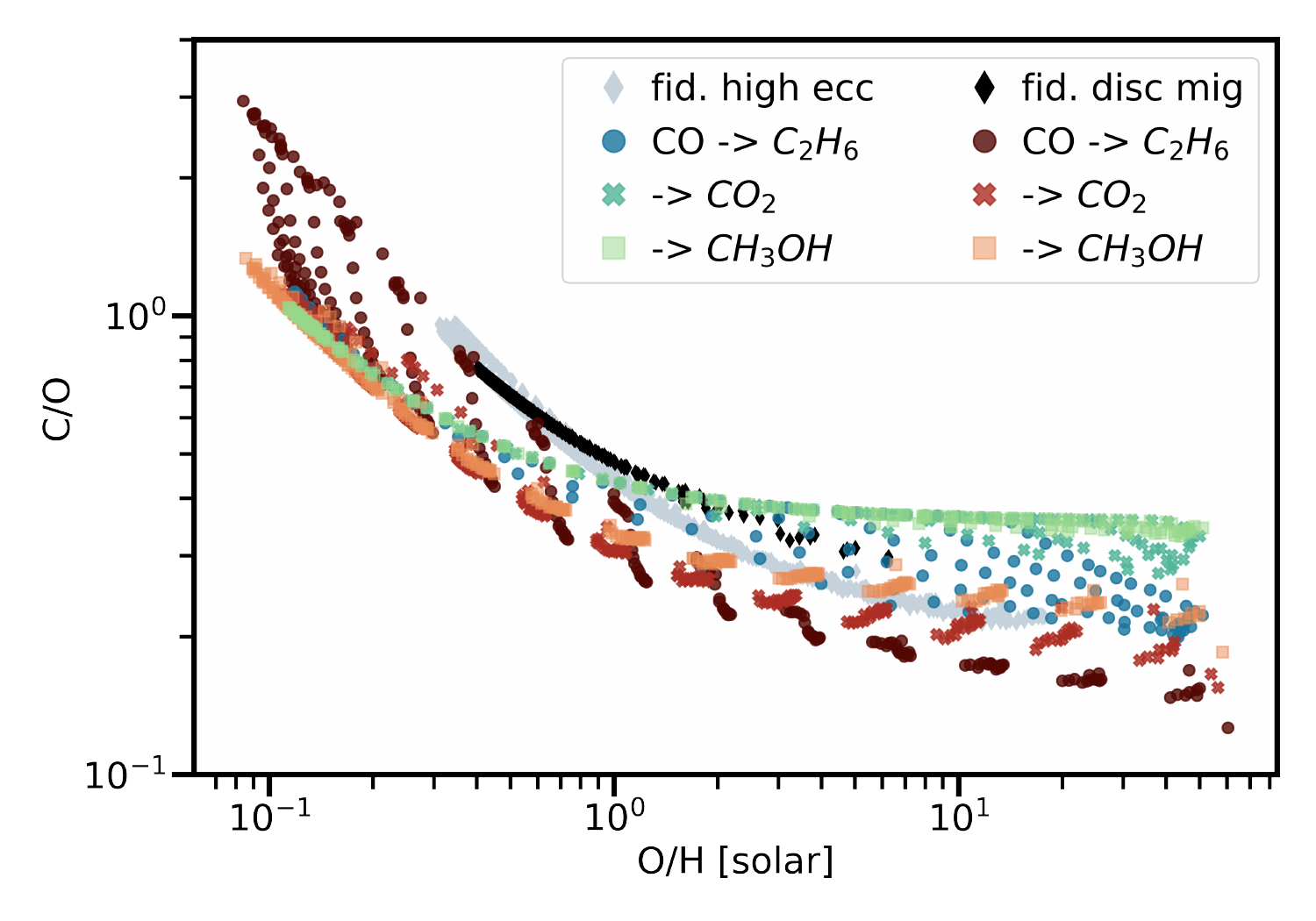}
\caption{The impact of CO depletion on the composition of hot Jupiters using the formation model of \citet{Penzlin2024}. Here, 90\% of the CO in the fiducial model has been replaced with either C$_2$H$_6$, CO$_2$, or CH$_3$OH. In all of the different disk models, the composition of hot Jupiters that underwent high-eccentricity migration differs from those that did not. However, different assumptions about CO depletion in the disk produce larger changes in the C/O ratio than the differences in formation history.}
\label{fig:co_comp}
\end{figure}

The atmospheric composition of planets formed by GI is not well understood. It is often assumed that planets formed by GI initially acquire a composition similar to the bulk composition of the disk, thus having a similar composition to their host stars. This is, however, not consistent with the archetypal system of wide orbit giant planets, HR~8799 \citep[e.g.][]{Bergin2024}. Some studies have considered that GI planets may then have their envelopes enriched by planetesimal accretion \citep[e.g.][]{Boley2011}. However, the formation process is considerably more complex, and several factors may lead to changes in their composition. For example, the spirals and fragments lead to large temperature perturbations in the discs, perturbing ice lines and even producing local ice lines surrounding the fragments \citep{ilee2017,molyarova2021}. Any preferential accretion of pebbles during fragmentation would thus lead to an increase in the planet's metallicity and likely a decrease in the C/O ratio \citep{nayakshin2015}. Previous work has shown that GI planets and spirals can effectively concentrate dust \citep[e.g.][]{rice2004,boley2010,booth2016,baehr2019}, but existing works have focussed on dust grain sizes that are larger than those observed ($>10\,{\rm cm}$ when observed sizes are $\sim{\rm mm}$; see section \ref{sec:disc_dust}) and no work has yet demonstrated how this impacts the planet's composition. Since these processes are essentially the same as those that modify the composition of planets formed by core accretion, it may be challenging to disentangle the populations based on composition alone (see, e.g., \citealt{madhusudhan2014}, where the GI and core accretion planets overlap in composition). More work is needed to understand the bulk and atmospheric compositions of planets formed by GI before conclusive predictions can be made about whether composition can distinguish planets formed via GI from those formed by core accretion.

\subsection{Testable Predictions}

It is difficult to make robust, testable predictions of exoplanet atmosphere compositions that can be uniquely identified to a particular formation scenario due to the aforementioned uncertainties in the disk composition, formation processes, and interior mixing. This is illustrated in Figure~\ref{fig:co_comp}, which shows that differences in the disk composition can introduce changes in the planet's C/O ratio that exceed the differences produced due to different formation histories. For this example, we used the pebble-planetesimal accretion models of \citet{Penzlin2024}, which assume steady-state disk compositions set by the transport of gas and ices. Other formation models also show a similar influence of changing the disk composition \citep[e.g.][]{Schneider2021b}.

Nevertheless, atmosphere compositions likely do contain imprints of the planets' evolutionary history; in most scenarios the disc composition \emph{does} vary throughout the the disk (Figure~\ref{fig:co_comp}). These differences are however not large, and distinguishing between different scenarios will be challenging given the precision achievable with observations (see the discussion in Sections \ref{sec:lowres} \& \ref{sec:direct_imaging}) given the degeneracies. Similarly, variations in composition due to stellar properties or planetary system architecture must be accounted for when interpreting the data. 

\begin{figure}[h!]
\includegraphics[width=\columnwidth]{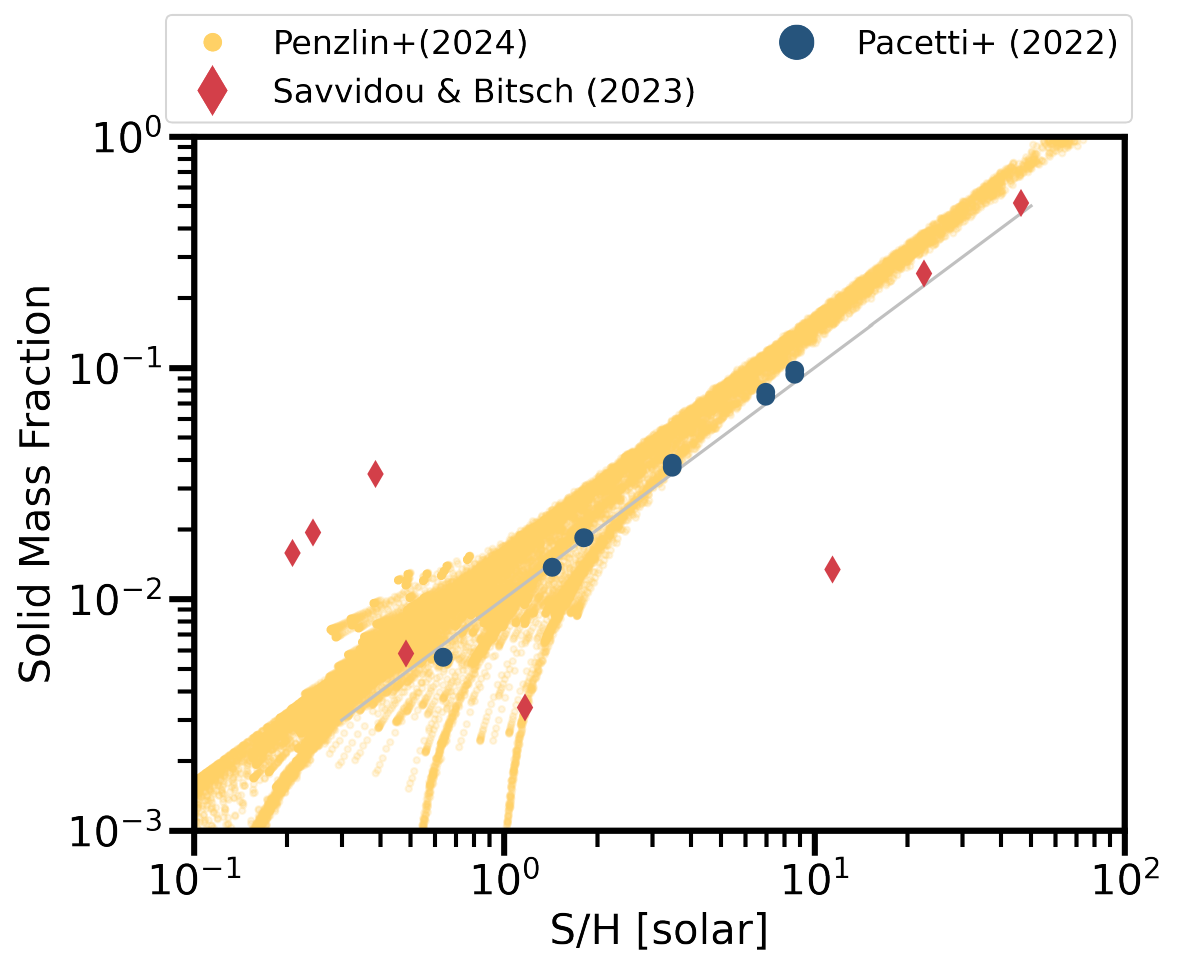}
\caption{
Correlation between the sulfur abundance and the amount of solids accreted for three different formation models: \citealt{Penzlin2024}, \citealt{Savvidou2023}, \& \citealt{Pacetti2022}. Sulfur is a good tracer of the amount of solids accreted due to its lower volatility than carbon and oxygen \citep[as revealed by both meteoritic data and disk observations,][]{Kama2019,Turrini2021}. Hot Jupiters formed close to their parent star in the pebble accretion model of \citealt{Savvidou2023} do not follow the correlation due to the enrichment of gas-phase sulfur in the disk via the release of H$_2$S from pebbles that migrate to the inner disk.
}
\label{fig:sulfur}
\end{figure}

Further, combining \emph{relatively} easily constrained elemental ratios such as C/O with other constraints undoubtedly helps produce more meaningful constraints on the planets formation history. For example, measuring the abundances of other species in addition to carbon and oxygen will provide additional constraints that can help disentangle the planets' formation history. It may also be possible to use any refractory species (e.g. sulfur, silicon, sodium, iron) to trace the amount of solid material accreted by the planets (Figure~\ref{fig:sulfur}) because they are almost entirely accreted in solid form throughout most of the disc, while nitrogen largely traces the gas accreted due to its high volatility \citep[e.g.][]{Turrini2021, Lothringer21, Pacetti2022, Chachan2023,crossfield23}. However, some care will be needed when interpreting sulfur abundances because refractory sulfur can sublimate into the volatile H$_2$S at lower temperatures than silicates evaporate \citep[e.g.][]{Steinmeyer2023}, which might in principle modify the planet's S/O ratio.

Bulk metallicities determined from the internal structure models may further help distinguish formation scenarios (see Section~\ref{sec:interiors}), e.g., because pebble accretion models predict that the accretion of solids ends before most gas is accreted. At the same time, planetesimals can continue to be accreted. As a result, these scenarios may predict different bulk/atmospheric metallicity ratios.

Similarly, combining abundance constraints with other information known about a planet's history will help understand how planets form. One example of this is to take planets with different migration histories and see how their compositions differ -- to this end, \citet{kirk2024} suggested using orbital (mis)-alignment with the star's spin axis as a proxy for planets that may have either migrated through the disc or undergone high-eccentricity migration. If these sub-populations of planets are found to have different compositions then it will provide strong evidence that planet composition and formation history are indeed linked. Another example is studying planets with different ages to understand how potential tracers of planet formation may or may not evolve with time \citep{barat24a, barat24b, feinstein24, thao24}.

\section{State of Protoplanetary Disks}\label{sec:disks}

Observations of protoplanetary disks directly constrain the composition and distribution of planet-forming gas, ice, and dust.  In principle this can be used to inform predictions of the end composition of planets, which form through different mechanisms.  The past decade has seen tremendous advances in disk observations thanks to modern telescope facilities like the Atacama Large Millimeter/submillimeter Array (ALMA) and JWST.  Here, we outline the current understanding of the volatile and dusty building blocks from which planets assemble.

\subsection{Disk dust properties}
\label{sec:disc_dust}

Extensive continuum surveys of protoplanetary disks at mm-wavelengths conducted with ALMA provide constraints on the dust mass in disks. As this topic has recently been reviewed extensively for PPVII by \citet{Manara2023}, we present only the critical details here. A key result of these surveys is that the typical Class II protoplanetary disk with an age of $1-10\,{\rm Myr}$ contains only a few Earth masses of dust, far below what is required at the onset of planet formation \citep[e.g.][]{Ansdell2016,Ansdell2017,Barenfeld2016,Pascucci2016,Cieza2019,Williams2019}. This is usually taken as evidence that planet formation must be well underway by $1-3\,{\rm Myr}$ \citep[e.g.][]{Manara2018, Cridland2022, Savvidou2024}, but selection biases and uncertainties in the disc mass may provide a resolution to the mass budget problem if planet formation is very efficient \citep[e.g.,][]{Mulders2021,LiuY2024}. There are also hints that the disk mass evolution may not be monotonic \citep{Testi2014,Testi2022}, pointing at the possibility of second generation dust due to collisional disruption of planetesimals \citep{Turrini2012,Turrini2019,Bernabo2022} or late infall from the molecular cloud \citep{Kuffmeier2023,Winter2024b}. An increasing number of observations show discs accreting from their parent molecular cloud \citep{Huang2021,Huang2023,Pineda2023,Gupta2024,Cacciapuoti2024,Valdivia-Mena2024}; however, many of these are in regions that are young and still forming new stars.

However, most dust masses are derived by scaling the fluxes at a single wavelength under the assumption that the disks are optically thin and have similar temperatures. This may significantly underestimate the mass of all but the largest discs since multi-wavelength observations show that many of the smaller disks may indeed be optically thick \citep[e.g.][]{Tychoniec2020,Tazzari2021,Chung2024}, meaning the dust masses may be underestimated.

The size of dust grains in protoplanetary disks can be constrained using multi-wavelength continuum observation via the disk spectral indices \citep[i.e., how the flux varies with wavelength; see][for a review]{Testi2014}. This is best done at high spatial resolution and using a wide range of wavelengths to break degeneracies between temperature, grain size, and optical depth. There have been a number of such detailed studies \citep{Carrasco-Gonzalez2019,Macias2021,Sierra2021,Guidi2022}, with results falling into two categories. One solution is that disks typically contain large grains with sizes in the mm to cm size regime, consistent with what was derived from surveys of disk-integrated spectral indices \citep[e.g.][]{Ricci2010, Tazzari2021}. The other solution is that grains may be closer to $100\,\micron$ in size \citep[e.g.][]{Guidi2022}, which is more consistent with sub-mm polarization studies conducted with ALMA \citep{Kataoka2016,Dent2019,Mori2021}. Part of this discrepancy may be attributed to uncertainties in the opacity of dust grains arising because their structure and composition are poorly known.

\subsubsection{Empirical constraints on dust transport}

It is well understood that large dust grains should settle to the midplanes of disks and drift towards their host stars \citep{Weidenschilling1977,Dubrulle1995,Takeuchi2002}. Dust settling is well confirmed empirically, through both SED modeling \citep{DAlessio2006} and by comparison between direct measurements of the disk thickness with ALMA and at shorter wavelengths \citep{Pinte2016,Villenave2022,Pizzati2023,Duchene2024}; however, the evidence in support of large-scale radial drift is much weaker.

The most direct way to probe radial drift would be via observations that demonstrate the dust disk has shrunk either in size or mass compared to the gas. Precise measurements of the dust-to-gas ratio are difficult to obtain due to the difficulties in measuring both the dust and gas masses. Robust measurements of disk sizes are easier to obtain. Comparison between dust disk sizes measured in the continuum with ALMA to gas disk sizes probed via CO lines routinely show that the dust disks are a factor $\sim 3$ smaller \citep{Sanchis2021}. However, large-scale radial drift is \emph{not} needed to explain these differences, which can be explained by CO line emission remaining optically thick to lower column densities \citep{Trapman2019,Toci2021}. There are a few exceptions where the dust disk is notably smaller \citep{Facchini2019,Trapman2020}, which might point to strong radial drift occurring in 
a subset of the disk population. 

Further evidence against large-scale radial drift in most discs comes from disk spectral indices. Dust growth and radial drift models predict that large grains are rapidly lost onto the star, resulting in low fluxes and high spectral indices that are incompatible with observations \citep{Birnstiel2010}. Trapping of dust in pressure maxima has been invoked to resolve these issues and is successful, at least qualitatively \citep{Pinilla2012,Zormpas2022,Delussu2024}. Indeed, the observed substructures \citep[e.g.][]{Andrews2018} require some radial drift and dust trapping to explain their existence \citep{Lee2024}. However, the overall efficiency of radial drift and its importance for disk compositions and planet formation is less clear. Despite these difficulties, models including radial drift have can reproduce the evolution of some disc properties, such as the dust mass budget \citep{Sellek2020,Appelgren2023}. In the next few years, we will have another way to constrain the efficiency of drift by comparing disk evolution models that include volatile evolution \citep{Booth2017,Booth2019,Schneider2021} to the compositions of the outer and inner regions of protoplanetary disks measured with ALMA and JWST. Indeed, JWST observations are already finding a range of compositions \citep[e.g.][]{Tabone2023,Banzatti2023,Grant2023}, the interpretation of which is not yet settled \citep[e.g.][]{Arulanantham2025}.

\subsection{Disk volatiles}

T
The disk gas and ice compositions are influenced not only by freeze-out of major volatile molecules \citep{Oberg2011}, but also by chemistry and transport.  Realizing the goal of inferring a planet's formation history from its atmospheric composition therefore requires observational benchmarks of the elemental inventories carried by planet-forming gas and ice.

\subsubsection{Gas masses \label{subsubsec:gasmass}}
Most of the disk mass is gas, and most of the disk gas is H$_2$.  Disk gas masses and gas-to-dust ratios are very challenging to measure, as outlined in detail in \citet{Miotello2023}.  Briefly, indirect tracers must be used to estimate disk gas masses since H$_2$ is not emissive throughout most of the disk.  One approach is through spectroscopic observations of HD, CO, and other trace chemicals, which requires detailed understanding of the underlying chemical and physical conditions \citep{Bergin2013, Williams2014, Trapman2022}.  Disk mass is also commonly estimated from the dust continuum, with uncertainties introduced by the assumption of optically thin emission and the adopted gas-to-dust ratio and dust opacity \citep{Williams2011}.  More recently, disk mass estimates have been obtained through signatures of self-gravity in the gas kinematics \citep[e.g.][]{Andrews2024}.  For TW Hya, the `best-studied' disk, different approaches lead to mass estimates spanning two orders of magnitude \citep{Miotello2023}.  This is a major limitation for the current understanding of planet formation physics.  

\subsubsection{Carbon and oxygen: C/H and O/H}
\begin{figure}[h!]
\includegraphics[width=\columnwidth]{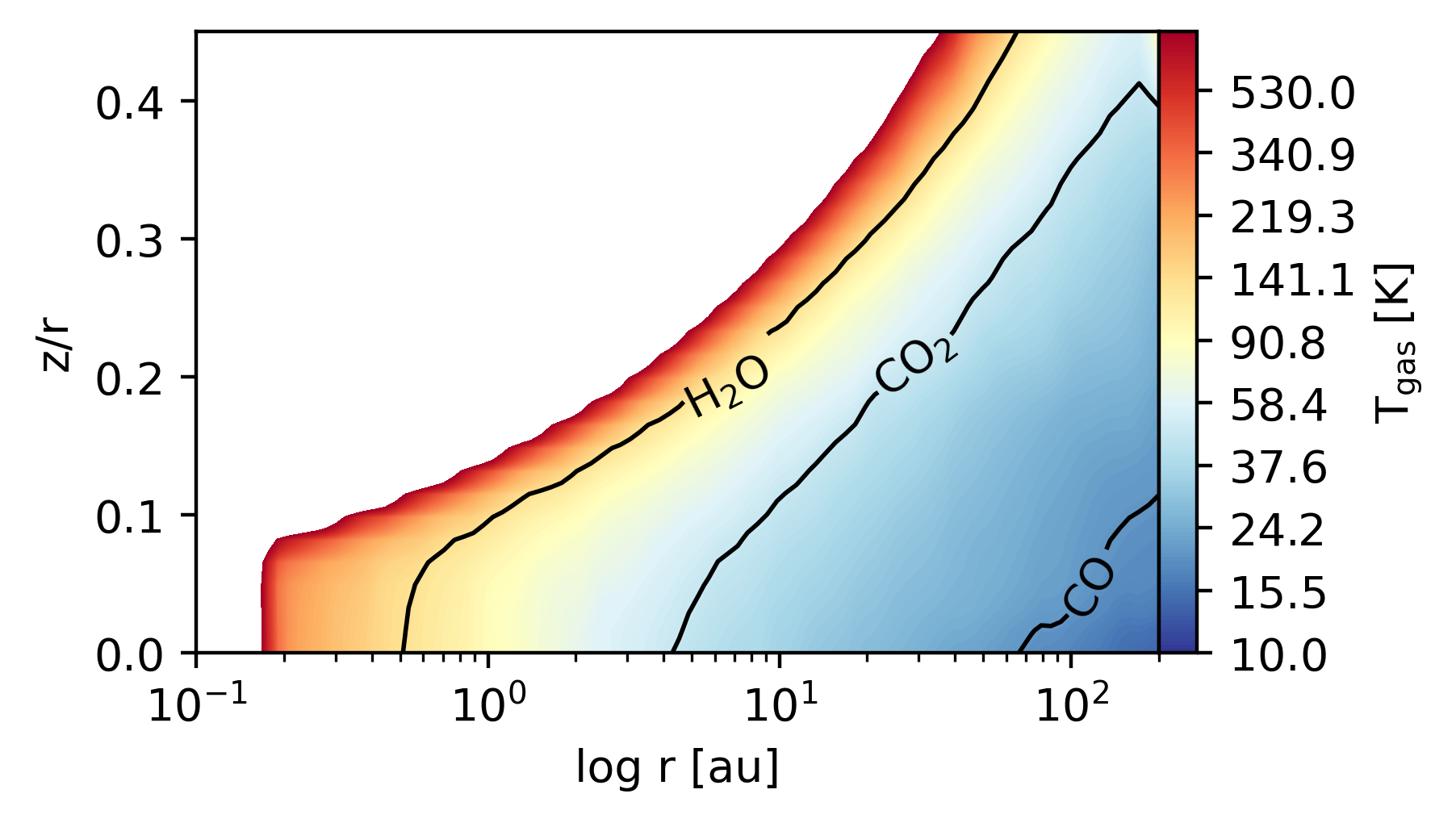}
\caption{Illustration of the temperature structure and the location of snow surfaces in an example protoplanetary disk, taken from the 0.03 M$_\odot$ model of \citet{Schwarz2018}.  Each species is primarily gaseous interior to its snow surface, and frozen (ice) beyond its snow surface.  The radial location of the snow surface at the midplane ($z/r=0$) is termed the snowline.}
\label{fig:disk_snowlines}
\end{figure}

Prior to the protoplanetary disk stage, several tens of percent of the cosmic oxygen budget, and about half of the cosmic carbon budget, are thought to be stored in refractory carriers \citep[e.g.][]{Bardyn2017,Oberg2021}.  Here, we use `refractory' to refer to compounds including silicates and macromolecular organics that have sublimation temperatures greater than a few hundred Kelvin, and are solid throughout most dense astrophysical environments.  The remainder is partitioned among volatile molecules, with CO and CO$_2$ being the main carbon carriers, and H$_2$O, CO, and CO$_2$ being the main oxygen carriers.  These molecules can be either gases or ices depending on the local temperature and density.  Figure \ref{fig:disk_snowlines} illustrates the snow surface locations for H$_2$O, CO$_2$, and CO within a protoplanetary disk model, marking the transition where they are available primarily within gaseous versus solid planetary building blocks.

\begin{figure*}[!hbt]
    \centering
    \includegraphics[width=\textwidth]{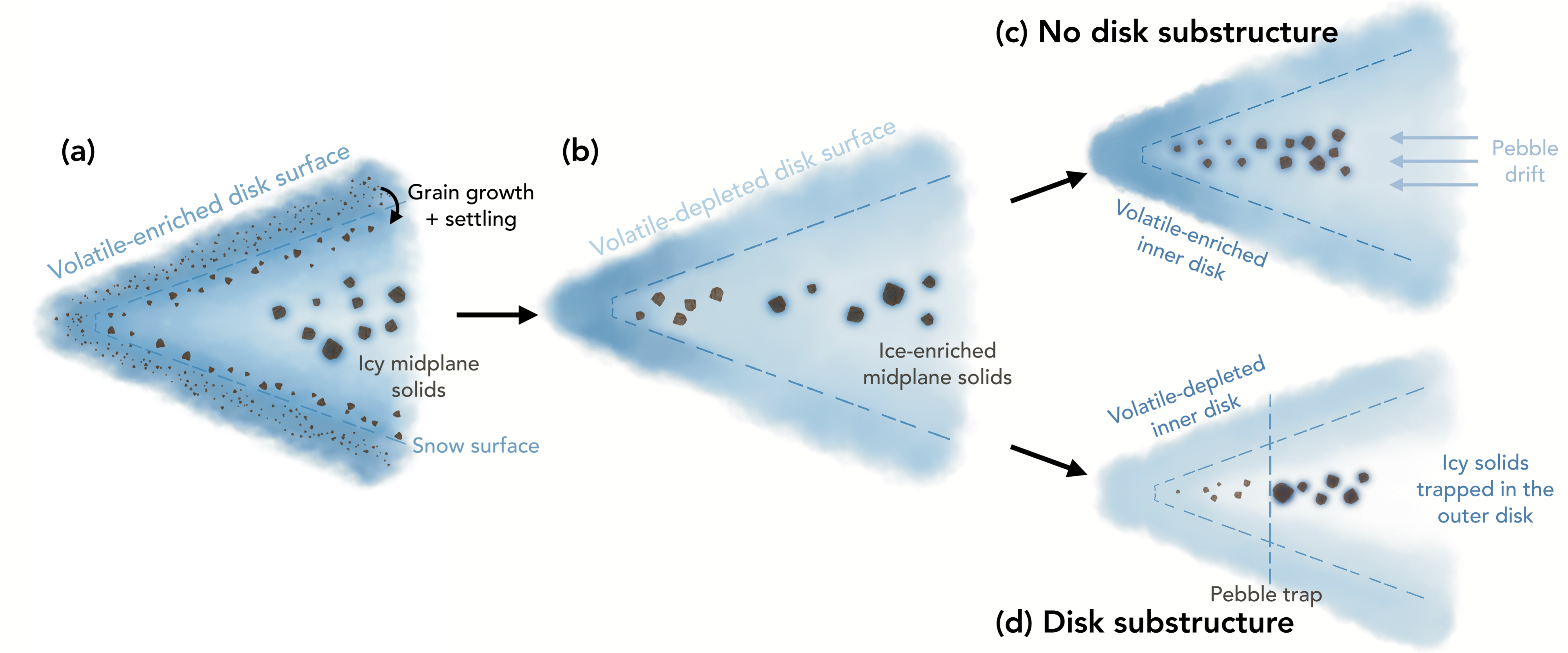}
    \caption{
    Cartoon of volatile redistribution in disks due to dust evolution and dynamics.  Blue shading represents the volatile enrichment of the gas.  Growth and settling of icy grains is expected to deplete volatiles from the surface layers of the outer disk (a,b).  Subsequent radial drift within the midplane carries the ice-enriched solids inwards towards the star.  If drift is unimpeded, sublimation of icy grains enhances the inner disk with volatiles (c).  Alternatively, pressure traps due to disk substructures may halt the inward drift of icy pebbles (d).
    }\label{fig:disk_structure}
\end{figure*}

The gas in the cool, outer regions of the disk (which makes up most of the total mass of the disk) appears to have low absolute abundances of oxygen and carbon, or low O/H and C/H, traced by low abundances of the molecular carriers CO and H$_2$O \citep[e.g.][]{Cleeves2018, Bosman2021}.  A leading explanation is that the spatial distribution of volatiles within the disk is reshaped by the dynamics and evolution of the dust population \citep[Figure \ref{fig:disk_structure};][]{Bergin2016, Krijt2020, vanClepper2022}.  In broad strokes, the gravitational settling of ice-coated grains to the disk midplane removes abundant ice species (including H$_2$O, CO$_2$, and CO) from the gas phase within the elevated surface layers of the disk.  To explain the observed low CO and H$_2$O abundances, this ice reservoir must be prevented from re-entering the gas.  This could be achieved by chemical conversion of simple ices to less-volatile molecules, sequestration of ice within growing planetesimals, drift of icy pebbles to the inner disk, or some combination thereof \citep[e.g][]{Reboussin2015, Du2015}.

The large reservoir of disk ices implied by this framework should, in principle, be detectable.  Early results from JWST indeed indicate a H$_2$O-enriched disk ice composition \citep{Bergner2024}, though a better demographic picture should emerge the coming years.  Note that due to the complex radiative transfer affecting ice observations in disks, ice abundance retrievals require detailed modeling \citep{Sturm2023}.  

Another avenue to shed light on the role of volatile transport in disks is characterization of the gas content in the inner disk \citep[e.g][]{Banzatti2020,Henning2024}.  As noted in Section \ref{sec:planet_formation}, icy pebbles around millimeter/centimeter in size should rapidly drift inwards towards the star, and enrich the gas with volatile molecules as they pass interior to snowline locations \citep[e.g.][]{Piso2015,Booth2019}.  There is currently evidence that, in at least some systems, the gas CO abundance is indeed enhanced interior to the CO snowline, particularly for hotter disks around higher-mass stars \citep{Zhang2019}.  It is also expected that enrichment of the inner disk with icy volatiles should only be observed in disks with unhindered radial drift, and not in disks that contain pressure bumps which can halt radial drift.  There are indeed hints that this is the case in some disks \citep{Banzatti2020, Banzatti2023}, though others do not follow this trend \citep{Perotti2023, Pinilla2024,Jang2024, Gasman2025}, and factors such as stellar multiplicity, disk age, gap depth, and continuum optical depth may complicate the picture \citep{Banzatti2024, Houge2025, Gasman2025}.  The statistics should improve considerably in the coming years thanks to JWST.

\subsubsection{Carbon and oxygen: C/O}

While both C and O appear to be influenced by this chemical-dynamical evolution, they may not be affected to the same extent.  Constraints on the C/O ratio come from observations of gas emission lines combined with source-specific disk chemical models.  The cool outer-disk gas with low C/H and O/H is also characterized by a high C/O ratio ($>$1), traced in particular by the bright emission of gas-phase hydrocarbon molecules \citep{Bergin2016, Miotello2019, Bergin2024b} and a high CS/SO ratio \citep{Semenov2018, LeGal2021}.  Models show that the gas C/O ratio may be increased up to a ratio of $\sim$unity by photochemical processing of CO, which ultimately leads to O-rich ices and C-rich gas \citep{vanClepper2022}.  First results from JWST observations of disk ices point to low ice-phase C/O ratios \citep{Bergner2024}, consistent with the idea that ices act as an oxygen sink.  Moreover, observations of disks with active ice sublimation point to an O-rich ice reservoir \citep{BoothA2021, Tobin2023}.  Still, disk observations imply an even higher gas C/O ratio of 1--2 \citep[e.g.][]{Bosman2021}, which requires an additional source of gas-phase carbon in addition to sequestration of O in ices.  The UV ablation of refractory carbon grains in the disk surface layers was proposed as the additional carbon source \citep{Bosman2021b}.  

In the hot inner disk, the volatile content seems to be interrelated with the radial transport of icy pebbles and therefore dependent on the large-scale structure of the disk \citep{Kalyaan2021,Banzatti2023,Mah2024}, as also noted above.  Drift of icy pebbles is expected to enhance the oxygen content of the inner disk gas.  Moreover, the detection of gas-phase H$_2$O in the inner region of the transition disks SY Cha and PDS 70 implies that water may be transported inwards via small dust and gas in addition to pebbles \citep{Perotti2023, Schwarz2024}.  Another factor that may strongly affect the inner-disk C/O is the `soot' line: the sublimation front of refractory carbonaceous material, expected around $\sim$500 K, interior to which a carbon-dominated gas chemistry should prevail \citep{Cridland2019, vantHoff2020,Li2021, Turrini2021, Pacetti2022, Bergin2023}.  The detection of an extremely carbon-rich gas chemistry in the inner disks around several very low mass M stars \citep{Tabone2023, Kanwar2024, Arabhavi2024} could be a signature of this, though it is not universal \citep{Xie2023, Perotti2025}.  Alternatively, a carbon-rich inner-disk gas may be explained by pebble drift and sublimation: due to the short viscous timescales interior to the snowlines around faint M-dwarf stars, the initial oxygen-rich material can be accreted and replaced with carbon-rich vapor from the outer disk within $\sim$2 Myr \citep{Krijt2020,Mah2023}.  Note that the effect of late infall on inner-disk volatile content has yet to be rigorously simulated, and could complicate the picture \citep{Pineda2023}.

Snowlines of major volatile molecules should, to some degree, influence the radial variation in disk gas vs.~ice C/O ratios \citep{Oberg2011}, though at present there are few observational constraints on this.  There are numerous factors that will affect how important snowlines are as transition points in the C/O ratio.  Disk thermal evolution and dynamics can smear the location of snowlines in time and space \citep[e.g.][]{Dodson2009, Piso2015}, and pebble drift almost certainly complicates the radial distribution of volatiles \citep{Booth2019}.  Moreover, chemistry in the disk may convert major C and O carriers (H$_2$O, CO$_2$, CO) into other molecules with different volatility \citep{Yu2016, Eistrup2018, Schwarz2018, Schwarz2019}.  This would reduce the importance of the `major' snowlines, and lead to a larger number of more gradual transitions in the gas vs.~ice C/O.  Along similar lines, sequestration of O and C in more refractory phases like hydrated silicates and refractory organics will reduce the importance of ice snowlines \citep{Turrini2021, Pacetti2022}.  The composition and microphysics of disk ices can also influence the nature of snowlines.  Notably, the binding energies of volatile molecules depend on the surface they are adsorbed to; as an example, lab-measured CO binding energies span a factor of $\sim$2 depending on whether CO is bound to other CO molecules, water, graphite, or silicates \citep{Minissale2022}.  Additionally, volatile molecules can be `entrapped' within an ice mixture and remain in the solid phase interior to their nominal snowline location \citep[e.g.][]{Collings2003, Simon2023}.  Indeed, early results from JWST reveal that CO ice is likely entrapped in H$_2$O and CO$_2$ ice, which will reduce the importance of the CO snowline in setting the gas and ice C/O ratios \citep{Bergner2024}.

In summary, current observational constraints point to a cool outer disk characterized by high-C/O gas and low-C/O ice, and a hot inner disk in which the C/O ratio is influenced by radial transport of O-rich volatiles and perhaps in some cases the sublimation of refractory carbon. While snowlines are undoubtedly a major factor influencing disk C/O ratios, there are important nuances that preclude a simple step-function radial variation.

\subsubsection{Nitrogen}
The nitrogen budget in protoplanetary disks is not as well constrained as C and O, in major part because the main N carrier (N$_2$) is spectroscopically inactive.  Inference about the N budget must therefore be made using trace N carriers such as HCN.  Current evidence suggests that N is not depleted from cool disk gas in the same way as C and O \citep{Cleeves2018,Leemker2023}.  The higher volatility of N$_2$ compared to CO, coupled with its chemical inertness, means that it is not efficiently sequestered in icy solids \citep{Furuya2022}.  While at present we may assume that transport is not significantly affecting the N budget, dedicated work will be needed if N is to be used as a reliable tracer for exoplanet atmospheres.  The curious enhancement of N in Jupiter's atmosphere suggests that it can provide an important lever for constraining planet formation \citep{Oberg2019, Bosman2019, Cridland2020}.  

In hot inner-disk gas, the trace N carrier HCN is commonly detected \citep{Pascucci2009}.  However, we are not aware of attempts to extract a gas-phase N abundance for the inner disk.  

Based on interstellar ice observations and early results from JWST disk ice observations, volatile ices (NH$_3$, OCN$^-$) can account for only a few percent of the cosmic N budget \citep{McClure2023, Sturm2023}.  There could be a large component of ice-phase N$_2$, but as a homonuclear diatomic it is not detectable.  Solar system comets also appear under-abundant in ice-phase nitrogen, though the recent discovery of ammonium salts suggests that these semi-refractory carriers may make up a substantial part of the total N budget \citep{Altwegg2020, Poch2020}.  If so, then the ammonium salt sublimation fronts, interior to the water snow line \citep{Kruczkiewicz2021}, may be an important transition point for disk gas vs.~solid N/H and N/O ratios.  The detection of ammoniated minerals on other minor bodies like Ceres \citep{DeSanctis2015} indicates that ammonia or ammonium must have been relatively abundant in the ice phase in parts of the solar system.

\subsubsection{Sulfur}
While numerous S-bearing molecules have been detected in cool disk gas, they represent only a small fraction ($\lesssim$10\%) of the total S budget \citep{LeGal2021, Keyte2024}.  Similarly, volatile ices can likely account for only a few percent of the sulfur budget \citep{McClure2023, Sturm2023}.  The remainder of sulfur is expected to be in a refractory form, with a volatility consistent with minerals like FeS \citep{Kama2019}.  This is consistent with solar system constraints: volatile forms of sulfur are inferred to account for a small fraction ($\sim$10\%) of the solar sulfur budget \citep{Lodders2010,Palme2014,Turrini2021}, whereas refractory sulfur within primitive CI meteorites has a near-solar abundance \citep{Lodders2009}. Taken together, it appears that the bulk of S in circumstellar disks is linked to rocks and should remain in solid phase until the inner-most disk regions where temperatures exceed 700 K \citep{Fegley2010}.  This makes it a good tracer of the accretion of solid material in general and rocky material in particular  \citep{Turrini2021}.  

\subsubsection{Isotope fractionation}\label{sec:isotopes:disk}
Molecular carriers of the rare stable isotopes D, $^{13}$C, $^{17}$O, $^{18}$O, and $^{15}$N have been detected in protoplanetary disks \citep[e.g.][]{Nomura2022}.  In the case of D, $^{18}$O, and $^{15}$N, there are clear signatures of fractionation: or, deviation of the isotope ratio within a molecular reservoir compared to the cosmic isotope ratio.  D/H fractionation is driven by low temperatures \citep{Millar1989}, while $^{18}$O/$^{16}$O and $^{15}$N/$^{14}$N fractionation appear to be driven by isotope-selective photodissociation \citep{Heays2014}.  While the origin of disk fractionation signatures remains a topic of active study, it is clear that there is a wide range of molecular D/H, $^{18}$O/$^{16}$O, and $^{15}$N/$^{14}$N ratios detected during the star-forming sequence, across protoplanetary disks, and within solar system bodies.  This variation implies that these ratios could provide an independent constraint on the source of gas and solids accreted onto exoplanets.  Meanwhile, $^{12}$C/$^{13}$C ratios are remarkably consistent across interstellar contexts and in diverse solar system bodies \citep{Nomura2022}.  On the other hand, early attempts to constrain $^{12}$C/$^{13}$C in disk gas hint at the presence of two distinct carbon isotopic reservoirs \citep{Bergin2024}.  The utility of this ratio for tracing the origin of planet-forming material should become clearer as more measurements are made.

\section{State of Gas Giant Exoplanet Atmospheres}
\label{sec:planet_observations}

Our current understanding of planet formation, and potential and migration mechanisms, does not yet fully explain the presence of gas giant exoplanets ($R_p > 4 R_\oplus$). While a system's architecture and/or atmospheric composition could hold clues to how and where planets originally formed within the protoplanetary disk, making this connection is challenging. There are several lines of evidence which can be used to link current day exoplanet atmospheres to their natal formation conditions. However, the observables are dependent on the orbital parameters of the planet itself. For example, the detection of molecular species such as \water\ and \cotwo\ is possible in transiting (hot Jupiters; $a \lesssim 0.1$~AU) and directly imaged ($a \geq 5$~AU), while the detection of refractory elements is limited to ultra-hot planets ($T_{eq} > 2000$~K).

Here, we present a broad overview of current atmospheric observations and trends for gas giant exoplanets from low-resolution observations, high-resolution observations and direct imaging.

\subsection{Atmospheric Modeling and Inferences}\label{sec:atm_modeling}

Before discussing the results of observations of exoplanet atmospheres, it is important to define the framework in which these observations are interpreted. The theory and modeling of exoplanet atmospheres had been thoroughly explored. One of the first mentions of exoplanetary atmospheres is in \citet{burrows93} and \citet{seager98}, where the authors explore interactions between the radiation from a given host star and a planetary atmosphere. This laid the foundation for using radiative transfer models to both predict and interpret observations of exoplanet atmospheres ranging from small rocky worlds to gas giants.

Broadly, an atmospheric model produces a synthetic exoplanet spectrum, regardless of the geometry of the planet (e.g., transmission for transiting exoplanets, emission, or reflected-light). Atmospheric models consist of the computation of radiative transfer on a parcel of gas representing a portion of a planetary atmosphere. These models can vary in complexity by using either one-dimensional or multi-dimensional radiative transfer, or assuming hydrostatic equilibrium or not, as examples. The complexity of the model depends on the purpose of the model. Furthermore, the chemical composition and thermal properties of the parcel of gas being modeled will depend based on the specific application. For example, atmospheric models may assume that the gas is in a state of radiative-convective-thermochemical equilibrium for a given chemical composition. It is important to understand the assumptions used in any given atmospheric model. Beyond the aforementioned assumptions, models may vary in their considerations of: the presence of absorbing aerosols, reflecting and scattering light, interactions with a planetary surface, variations in the stellar flux interacting with the planetary atmosphere (e.g., starspots), winds, and magnetic fields. These modeling efforts are also referred to as \textit{forward models}.

The nomenclature of \textit{forward models} surfaces from a desire to distinguish the above modeling effort from inverse techniques aimed to infer the parameters of a model of interest from a given dataset. These inverse methods have been well established in astronomy and other science fields. Due to  similarities between the inverse problems in planetary sciences, the field of exoplanetary sciences adopted the term \textit{atmospheric retrievals} to indicate the use of inverse models to infer model parameters and thus  the atmospheric properties of interest of a planet from observations \citep[for a more thorough review, see][]{madhusudhan19}. At their core, atmospheric retrievals couple an observable (e.g., a planet spectrum) with an atmospheric model (e.g., a forward-model) through a parameter estimation framework (e.g., Markov Chain Monte Carlo or Nested Sampling) to derive estimates of the model parameters \citep[Figure~1 in][]{Welbanks2021}. When a Bayesian framework is implemented, a retrieval considers priors for the parameters in the model, a likelihood function, and outputs posterior probability distributions for each model parameter. The use of Nested Sampling has simplified the calculation of the marginal likelihood (e.g., model evidence) which has in turn been widely used as a means to estimate goodness of fit, model comparisons, and quantification of model preference \citep[for a more thorough review, see][]{welbanks23}.

The atmospheric retrieval requires the comparison of a dataset to a family of model evaluations to determine the regions of parameter space that ``best'' explain the dataset; the best-fit is derived by a statistical metric. There are two predominant methods to produce the family of models used within the atmospheric retrieval framework. The first method is by generating the model evaluation with every single iteration within the parameter estimation framework. The second method is by exploring a set of pre-computed model, or \textit{grid of models}, evaluations. The first efforts to perform exoplanet atmospheric retrievals \citep[e.g.,][]{madhusudhan09} relied on a large grid of models to cover a sufficiently large parameter space that may be compatible with observations. This was often sufficient, but the interpretation of the observations was inherently limited by how the grid of models was computed. However, the advent of parametric treatments has enabled rapid computation of atmospheric models, allowing for ``on the fly'' model generation alongside the parameter estimation. Such parametric treatments include, but are not limited to (i) the vertical temperature structure of the planetary atmosphere, while still capturing the basic features of temperature structures, (ii) relaxations to the assumption of chemical equilibrium, and (iii) different physical and chemical atmospheric processes \citep[e.g. clouds,][]{lecavelier08b, lecavelier08a, line16, heng16}. Nonetheless, computing atmospheric models simultaneously with the parameter estimation is computationally expensive, and the necessity for more complex, physically-motivated, models has resulted in a rise in grid-based approaches within the era of JWST. These grid-based efforts are sometimes referred to as \textit{gridtrievals}, despite their existence dating to the beginning of the \textit{retrieval} effort. 

Multiple considerations are important to keep in mind when using atmospheric models to infer the atmospheric properties from a set of observations. First, as with any data-model technique, model assumptions may significantly impact the resulting planetary inferences \citep{welbanks19}. Among the assumptions explored in the literature are the use of one-dimensional models \citep{feng16, caldas19, Welbanks2021, nixon21}, the use of different vertical temperature structure parameterizations \citep[e.g.,][]{blecic17}, and uncertainties in system properties (e.g., mass -- \citealt{dewit13, batalha19}; or stellar flux -- \citealt{pinhas18, rackham18}). Second, many of the claimed detections and planet properties in the literature rely on model comparisons based on a given goodness of fit metric. 

While this is a consideration not unique to exoplanet atmospheres, it is important to note that a model preference may not mean the detection of a gas or atmospheric property. This is because the model being preferred may be inappropriate (e.g., un-physical, too flexible, or too constraining) or the goodness of fit metric is  misleading (e.g., assuming a Gaussian Likelihood for the data-model comparison in the presence of correlated noise in the observations); see \cite{welbanks25} for further discussion. Any atmospheric detection or inference is stronger when interpreted within the larger context of the data and model limitations \citep{welbanks23}. Third, our model inferences are affected by the limitations in observables. The combination of wavelength coverage, instrument resolution, and instrument sensitivities inhibit our ability to fully characterize the atmospheric properties of a given planet \citep[][Welbanks et al. in prep.]{carter24}.  

Finally, for the purposes of linking atmospheres to planet formation, the output of an atmospheric retrieval is a model parameter estimate and \textit{not} a direct signature of a given planet formation pathway. This point can be self-evident, but it is important to remember that a retrieval is not informed by any formation mechanism or disk chemistry; that interpretation depends on the practitioner.

\begin{figure}[!ht]
    \centering
    \includegraphics[width=\linewidth]{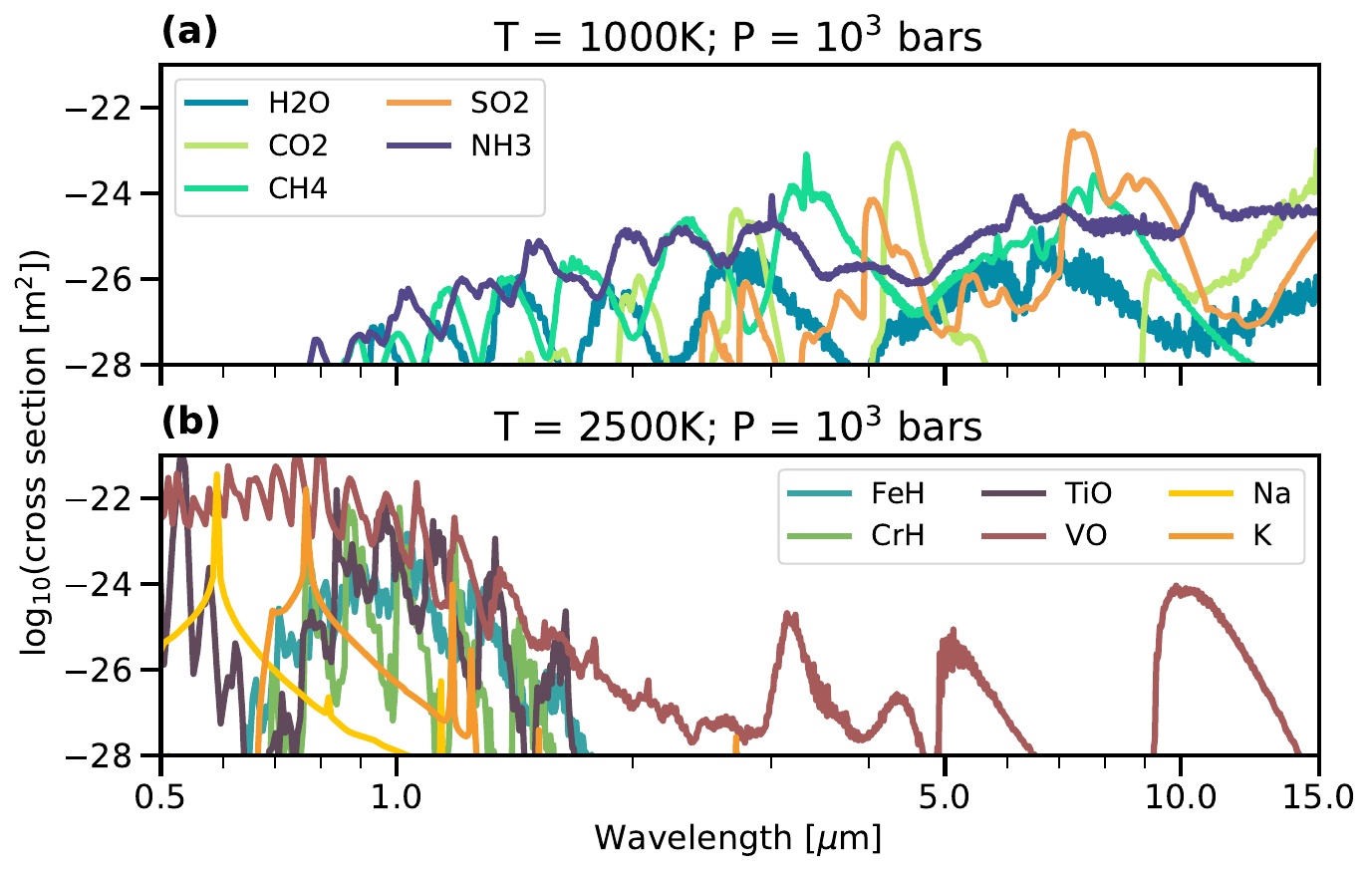}\\
    \caption{The cross-sections for molecular species which could exist in a (a) hot Jupiter at 1000\,K and (b) ultra hot Jupiter at 2500\,K atmosphere. The wavelengths plotted here are within the range of JWST's NIRISS, NIRCam, NIRSpec, and MIRI/LRS. With access to the broader wavelength coverage of JWST, we are now able to potentially detect multiple absorption features from the same species and multiple species, simultaneously. Cross sections computed following the methods and sources of opacity listed in \cite{Welbanks2021}. We note that several versions of this plot already exist in the literature \citep[e.g.,][]{madhusudhan19}.}
    \label{fig:spaghetti}
\end{figure}

\subsection{Observations of Giant Transiting Exoplanets}\label{sec:lowres}

In an effort to constrain exoplanet formation mechanisms from atmospheric properties, dozens of observations of transiting exoplanets have been obtained with the \textit{Hubble} Space Telescope (HST), \textit{Spitzer} Space Telescope, ground-based facilities, and more recently with JWST. Unlike exoplanets discovered via radial velocities, microlensing, or astrometry, transiting exoplanets have a favorable geometric orientation which allows for a detailed characterization of their atmospheres. Each of the aforementioned observatories brings a unique wavelength regime to the picture. For example, HST's Wide Field Camera 3 (WFC3) and Space Telescope Imagine Spectrograph (STIS) instruments probed the presence of ionized metals and \water, while \textit{Spitzer}'s 3.6 and 4.5~\micron\ channels probe \methane\ and \cotwo. The successful launch and deployment of JWST has revolutionized our understanding of gas giant exoplanet atmospheres. The broad continuous wavelength coverage of JWST allows for the direct observation of C- and O-bearing molecules, such as \water\ ($1 - 3$~\micron), \cotwo\ ($2.8, 4.2$~\micron), CO ($2.4, 4.8$~\micron), and \methane\ ($3.2$~\micron), along with many other molecular species (Figure~\ref{fig:spaghetti}), even out to the mi-infrared with MIRI/MRS \citep{deming24}.

\subsubsection{Molecular and Refractory Species in Transiting Giants}

\begin{figure}[!ht]
    \centering
    \includegraphics[width=\linewidth]{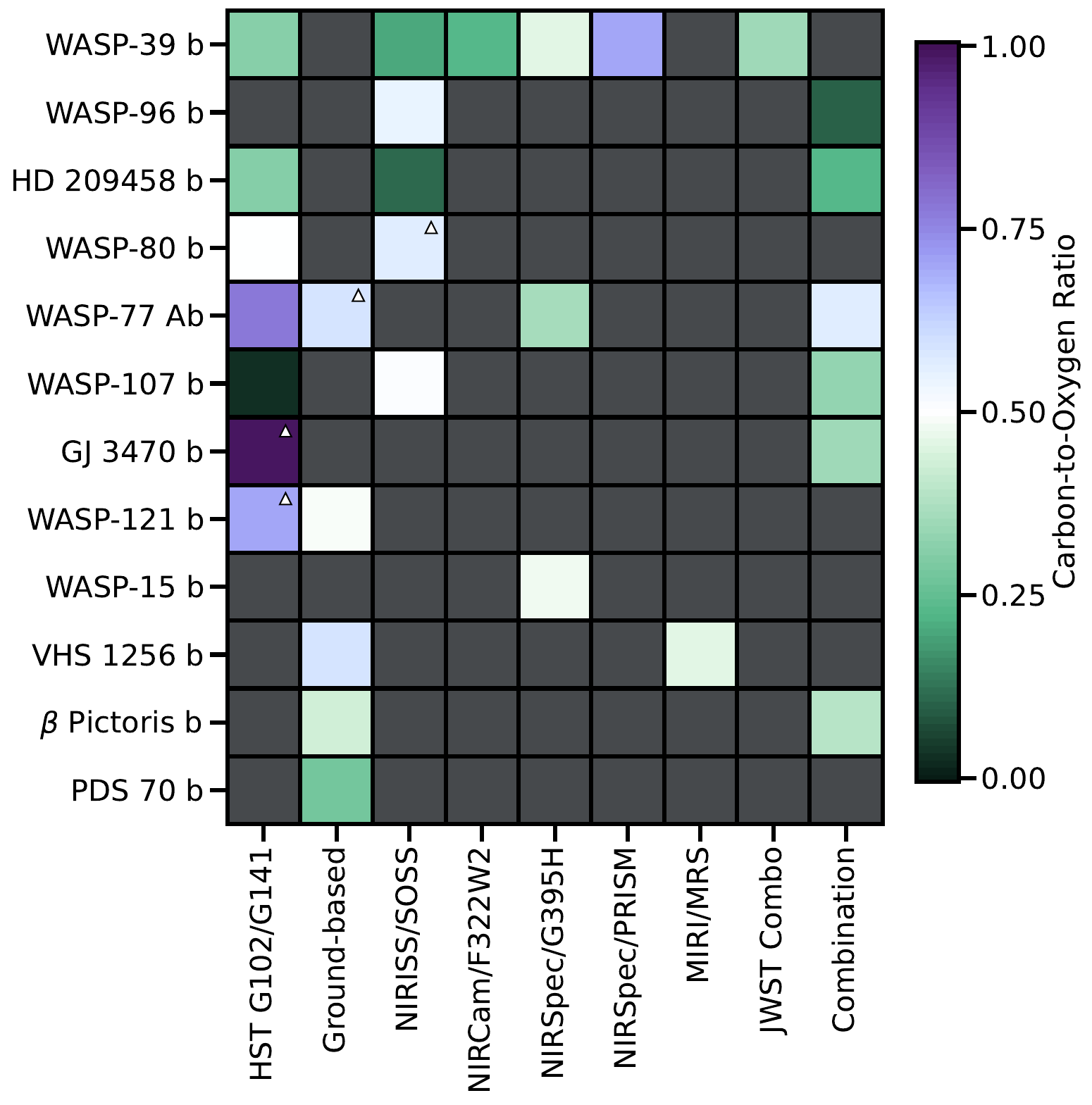}
    \caption{A comparison of the measured carbon-to-oxygen (C/O) ratio for planets, as observed with different instruments/wavelength coverages (see Table~\ref{tab:exoplanet_c2o} for details). Panels marked with $\triangle$ highlight upper limits on the C/O. Grayed out squares do not have C/O measurements for those planets with those instruments. While the majority of measurements of C/O for each of these planets are consistent within $2\sigma$, the difference in wavelength coverage, and thus which molecular species are detected, change the measured C/O for the same planet. }
    \label{fig:exoplanet_c2o}
\end{figure}

A cornerstone example of what JWST can deliver was provided by the Transiting Exoplanet Community Early Release Science program \citep{stevenson16, bean18}, which completed a comprehensive exploration of WASP-39\,b ($R_p = 14.93^{+0.38}_{-0.35} R_\oplus$; $M_p = 89.9 \pm 13.0 M_\oplus$) with five instrument modes. WASP-39\,b was observed using NIRISS/SOSS ($0.6-2.8 \mu$m), NIRCam/F322W2 ($2.4 - 4.0 \mu$m), NIRSpec/G395H ($2.8 - 5.2 \mu$m), NIRSpec/PRISM ($0.6 - 5 \mu$m), and MIRI/LRS ($5-12\mu$m). Cumulatively, the observations revealed five \water\ absorption features \citep{ahrer23, feinstein23}, \cotwo, CO \citep{alderson23, jtec23}, Na \citep{rustamkulov23}, K \citep{feinstein23, rustamkulov23}, and \sotwo\ \citep{alderson23, rustamkulov23, powell24}.  The metallicity and C/O were derived independently across four datasets (all excluding MIRI/LRS) and agree that the C/O is between $0.2 - 0.7 \times$ solar C/O \citep[top panel of Figure~\ref{fig:exoplanet_c2o};][]{feinstein23, rustamkulov23}; however, this is quite a range of solutions that provides no clear evidence of a single formation pathway. A recent joint analysis of the complete $0.6 - 5$~\micron\ spectrum (again, excluding MIRI/LRS) for WASP-39\,b reconfirmed a sub-solar C/O$= 0.35\times$ solar \citep{carter24} using grid models informed by atmospheric retrievals of the full data set (Welbanks et al. in prep); this is consistent with the previous analyses. \citet{constantinou23} noted that considering only the $3-5$\,\micron\ data alone could affect abundance estimates by up to $\sim 1$~dex, highlighting the necessity of broad wavelength coverage for transmission spectra. We note that the host star WASP-39 has a metallicity and C/O consistent with solar \citep{polanski22}, thus the derived abundances here are reflective of the formation history of the system, rather than reliant on assumptions about stellar abundances.

While WASP-39\,b was observed with a combination of different instruments, these datasets highlight that the derived atmospheric metallicity and C/O will vary greatly depending on the wavelength coverage and resolution of each dataset, and that determining the ``true'' atmospheric abundances for a given planet will be challenging. This has been demonstrated for other planets observed with multiple JWST instruments \citep[e.g., WASP-107\,b;][]{sing24, welbanks24}. Unsurprisingly, there can be significant differences between the measured C/O with HST and JWST. We summarize reported values of C/O in the literature as a function of planet and observing mode in Figure~\ref{fig:exoplanet_c2o}. How to determine the ``true'' atmospheric C/O will continue to be an open question within exoplanet atmospheres, as we explore how e.g., (i) wavelength coverage (ii) spectral resolution (iii) fitting for potential offsets between instruments affect the resulting observed transmission spectrum for a given planet. 


Furthermore, it is through population studies that we can begin to understand how short period gas giants form and evolve. \cite{sing16} completed a homogeneous survey of the atmospheres of ten transiting gas giants with $R_p = 0.96 - 1.89 R_J$ and $T_\textrm{eq} = 960 - 2510$~K with HST and \textit{Spitzer} with a broad wavelength coverage of $0.3 - 5$\,\micron. The spectra revealed a diversity of absorption features due to \ion{Na}{1}, \ion{K}{1}, and \water. Additionally, these observations revealed a strong optical scattering slope in two of the ten targets. The strength of the \water\ feature at 1.4~\micron\ ranges from very pronounced to absent. The absence of \water\ could be due to the lack of \water\ (i.e., the planet accreted its atmosphere in a \water-depleted protoplanetary disk) but is more likely due to the \water\ absorption bands being obscured by clouds \citep{knutson14, kreidberg14}. The relationship between cloudy and clear atmospheres appears to be nonlinear with $T_\textrm{eq}$ \citep{fu17, gao20}, unlike what is seen for low-mass brown dwarfs \citep{burgasser02, cushing05}. The limited wavelength coverage and resolution of HST + \textit{Spitzer} did not yield any conclusive results about the measured C/O and metallicity in this sample.

Prior to JWST, the observed C/O for transiting exoplanets had a wide spread from C/O = $0.3 - 1.6$ \citep{changeat22, hoch23}. However, now with the broader wavelength coverage of JWST, which allows for the direct detection of both C- and O-bearing molecules, this range in values is condensed to C/O = $< 0.75\times$ solar across all published literature (Figure~\ref{fig:exoplanet_c2o}). With this sample of seven transiting exoplanets, there may be a trend arising that gas giant atmospheres are consistent with C/O = sub-solar to solar values ($< 0.57$); this was also pointed out in \citep{fu25}, which analyzed only NIRSpec/G395H observations of eight gas-giant planets. However, there are several caveats to this observation. First, in order to interpret these results with respect to the formation of the individual system, we must compare the C/O to the \textit{stellar} C/O, not solar C/O. Second, the C/O discussed above are measured with different instruments (i.e., wavelength coverage) and have been interpreted using different atmospheric models. Thus, it is important to note that without observing and interpreting exoplanet atmospheric observations in a homogeneous manner, it may be hard to draw generalized conclusions about how and where these planets may have formed.

\subsubsection{Direct Detection of Sulfur- and Silicon-bearing Molecules}

Beyond molecular abundances, the refractory species offer another perspective on the composition of the giant planet. One of the direct observations of refractories on hot Jupiters is the presence of clouds. As cloud particles introduces scattering features at near-IR wavelengths, it often weakens the strength of molecular lines. This allows the inference of clouds on many giant planets, for example, WASP-127 b \citep{boehm25}, HAT-P-18 b \citep{fournier24} and WASP-39 b \citep{feinstein23}. With the resolving power of JWST MIRI, silicate clouds have been directly identified through the absorption features at 8-10 microns. The planets with silicate features include WASP-107 b \citep{dyrek24}, WASP-17 b \citep{Grant2023} and HD 189733 b \citep{inglis24}. The more and more cases of planets confirmed hosting silicate clouds are hopefully offering indications about the formation location of the planets and the relative strength of gas accretion and solid accretion.

The JWST observations of WASP-39\,b revealed the presence of \sotwo\ \citep{alderson23, rustamkulov23}, which was subsequently confirmed with MIRI/LRS \citep{powell24} at $> 3\sigma$. \cite{tsai23} explained the presence of \sotwo\ through photochemical processing from the parent star i.e., that it is produced by the oxidation of sulfur after H$_2$S, the primary S-bearing species, is destroyed via energetic stellar photons. As such, the presence of \sotwo\ may be more sensitive to the atmospheric metallicity. Besides WASP-39\,b, \sotwo\ has only been detected in GJ 3470\,b \citep{beatty24} and WASP-107\,b \citep{sing24, welbanks24}. \sotwo\ was not been detected in WASP-80\,b \citep{bell23}, HD~209457\,b \citep{xue24}, or WASP-77\,Ab \citep{august23}; to date, no other giant exoplanets have published JWST data at the relevant wavelengths for \sotwo. The ability to measure sulfur in warm and hot Jupiter atmospheres is a promising new avenue for probing the amounts of solid material accreted during formation, although the relatively complex chemistry of sulfur-bearing species \citep{hobbs21, polman23, tsai23, deGruijter25} may mean that more refractory species such as silicon or iron still act as better tracers of the refractory content if fully in the gas phase \citep{Turrini2021,Pacetti2022,Chachan2023}.

\subsubsection{Refractory Species in Ultra-Hot Jupiters with High Resolution Spectroscopy}

In addition to the low spectral resolution approaches described above, high spectral resolution techniques have been developed with great success (e.g. \citealt{snellen10,birkby:2018}; see also the review by \citealt{Snellen2025}). Also referred to as high-dispersion spectroscopy, this technique effectively treats the planet-star system as a spectroscopic binary, whereby the planet's light is resolved spectrally, rather than spatially or temporally, through the Doppler shift of the planetary spectrum due to the planet's orbit relative to the stellar and telluric lines. An important benefit to this technique is that the atmospheres of non-transiting planets can also be observed as has been demonstrated by observations of, e.g., 51 Peg b \citep{rodler12, Birkby17}. However, these observations require stable, high-resolution instrumentation, which restricts their use to large ground-based observatories.

Detections are most often made by cross-correlating the observations with a Doppler-shifted template or model spectrum. Peaks in the cross-correlation function indicate areas in velocity space where the planet's spectrum matches the template or model, indicating a detection. Turning this cross-correlation function into a likelihood value to compare to atmosphere models and measure elemental abundances has been less than straightforward, in part because the observations are continuum-normalized during the reduction process. However, techniques have been developed to successfully obtain elemental abundances and/or abundance ratios from high-resolution observations \citep{Brogi19,Gibson20}.

\begin{figure*}[t]
    \centering
    \includegraphics[width=1\linewidth]{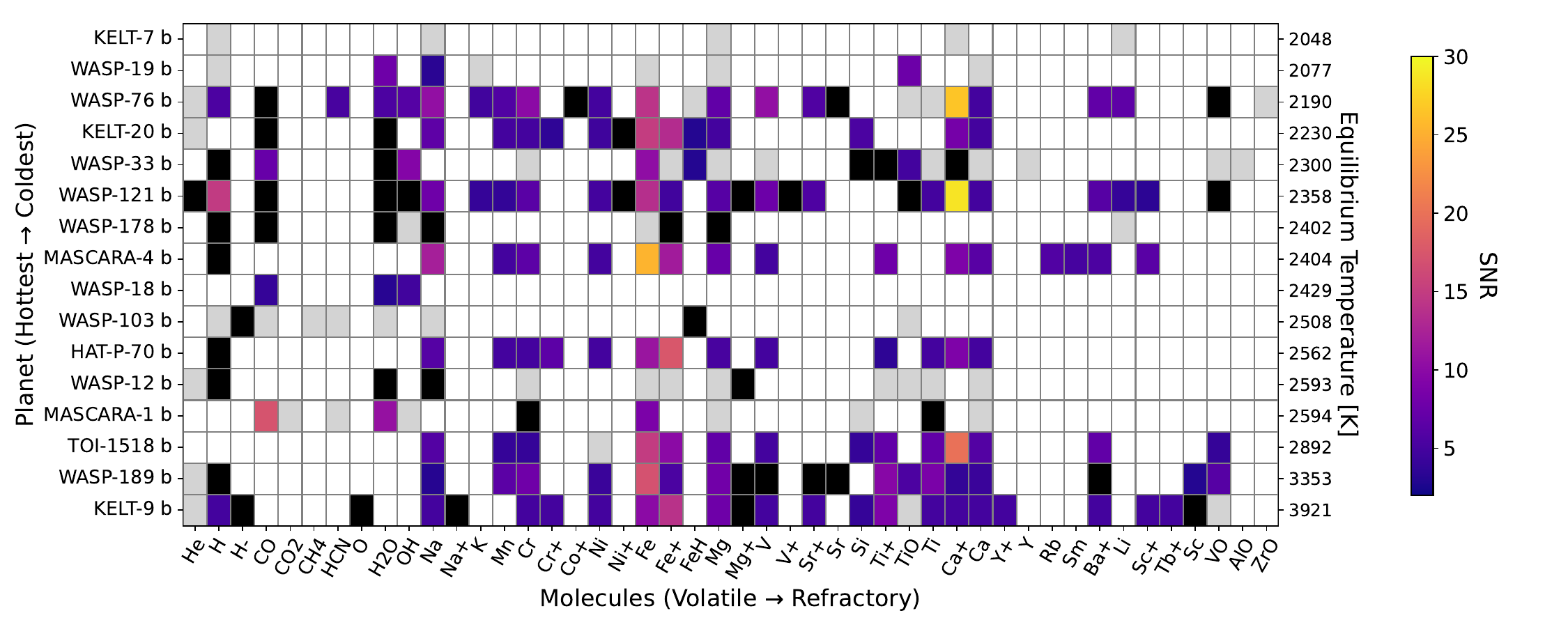}
    \caption{Heatmap of atomic and molecular detections in ultra-hot Jupiters, in order of volatile to refractory from left to right. Detection significances are color-coded with black indicating a detection only (i.e., no available significance) and grey indicating only an upper-limit or non-detection. Detection significances data taken from The IAC Community Database (\url{https://research.iac.es/proyecto/exoatmospheres/table.php}) with detection significances taken from \cite{savel:zenodo}.}
    \label{fig:hires}
\end{figure*}

For example, \cite{line21} used $R\sim$45,000 IR thermal emission observations of WASP-77Ab to measure H$_2$O and CO in the planet's atmosphere, finding a sub-solar metallicity and a solar C/O ratio. These results were the confirmed by an independent analysis \citep{Bazinet24}, follow-up IGRINS observations \citep{smith24}, and JWST NIRSpec data \citep{august23, smith24}. The star was recently measured to have a super-solar metallicity and a sub-solar C/O, meaning that the planet appears to have a more sub-stellar metallicity and super-stellar C/O ratio than previously thought \citep{Reggiani22}. These results combined suggests that the planet's envelope was accreted beyond the H$_2$O ice line through gas-rich accretion \citep{Khorshid23}. Similar analyses have now been performed for a number of other planets \citep[e.g.,][]{Carleo2022,Guilluy2022}. High spectral resolution also enables the measurement of isotopic ratios, which can provide addition clues to planet formation and evolution (see Section~\ref{sec:isotopes:disk} and \ref{sec:isotopes:DI}). \cite{line21} measured a sub-solar $^{12}$C/$^{13}$C ratio, which \cite{coria:2024} confirmed to be sub-stellar.

Simultaneous with the advent of high-resolution atmospheric observations were breakthroughs regarding our theoretical understanding of ultra-hot Jupiters. Loosely defined as planets above  $\approx$2000~K, ultra-hot Jupiters are giant planets that reach high enough temperatures that processes like ionization and molecular dissociation become significant \citep{Parmentier18,Lothringer18,Kitzmann18}. One of the most important effects of the high temperatures is the evaporation of clouds, i.e., the atmosphere is so hot that even the most refractory elements are in their gaseous form. For our purposes here, the evaporation of refractory species is exceedingly important because it opens up the entire reservoir of rocky material with which a planet may have formed to remote sensing with both transmission and emission spectroscopy, which when measured relative to the volatile content may provide insight to the planet formation process \citep[e.g.,][]{Lothringer21,Schneider2021b,Turrini2021,Pacetti2022,Chachan2023}. In one sense, we can measure the composition of an ultra-hot Jupiter in more detail than the composition of the giant planets in our own Solar System.

The refractory species in ultra-hot Jupiters can be observed with low-spectral resolution at UV-optical wavelengths where atoms and ions absorb strongly \citep{Sing19,Lothringer20,Lothringer22} and at IR wavelengths where metal hydrides and oxides absorb \citep{flagg23}. However, high-resolution observations of ultra-hot Jupiters have shown especially bountiful success, where entire inventories of atoms and ions have been measured. For example, in WASP-121b alone, H, He, CO, H$_2$O, OH, Na, K, Mn, Cr, Ni, Ni+, Fe, Fe+, Mg, Mg+, V, V+, Sr+, VO, TiO, Ti, Ca, Ca+, Ba+, Li, and Sc+ \citep{arcangeli18, hoeijmakers:2020,cabot_detection_2020,ben-yami:2020,Borsa21,gibson_detection_2020,Gibson22,Langeveld22,silva_detection_2022,maguire_high-resolution_2023,ouyang_detection_2023, hoeijmakers:2024, Pelletier24,Smith24b} (though the detections of VO and TiO are contested \citep{merritt_non-detection_2020,hoeijmakers:2024}). A plot of atomic and molecular species measured in ultra-hot Jupiters with high-resolution is shown in Figure~\ref{fig:hires}, demonstrating the plethora of measurements taken in recent years.



As mentioned above, turning these detections into abundance measurements through atmosphere retrievals is difficult, however efforts have begun to build compilations of such abundances \citep[e.g.,][]{Gibson22, maguire_high-resolution_2023, Gandhi2023B, Pelletier23}. Exploring population level trends, \cite{Gandhi2023B} measured the refractory abundances of six ultra-hot Jupiters, including Fe, Mg, Ti, V, Ca, Cr, Ni, Na, and K. In that study, Fe abundances were mostly found to be consistent with stellar metallicity, while Ca, Ti, and Na were sub-stellar in abundance, likely due to ionization and/or nightside condensation and rainout (of, e.g., CaTiO$_3$). Comparisons to the volatile content of these planets could provide great insight into the relative amounts of rock and ice that enriched these planets' atmospheres. Such a measurement was attempted with the low-resolution \textit{Hubble Space Telescope} transit spectra of WASP-121b \citep{Lothringer21}, inferring a refractory-rich formation, though subsequent high-resolution follow-up has found evidence for both refractory-rich \citep{Smith24b} and a volatile-rich compositions \citep{Pelletier24}, highlighting the difficulty in obtaining such measurements in exoplanetary atmospheres.

Open questions remain including 1) the degree to which clouds may remain on the nightside, horizontally cold-trapping refractory species, 2) how effects like ionization, molecular dissociation, and atmospheric escape may be affecting the measured elemental abundances, and 3) how representative the refractory elements measured in the atmospheric envelope are of the bulk refractory abundance, especially for planets formed via core-accretion. Another difficulty can be obtaining accurate stellar abundances \citep[e.g.,][]{Reggiani22}, which is necessary for inferring a reasonable primordial elemental abundance, because many ultra-hot Jupiters have relatively early-type host stars. Such stars can often have high rotational velocities, peculiar abundances, or are simply too hot to precisely measure some elements. Even with these open questions, high-spectral resolution observations of giant planets, including ultra-hot Jupiters, hold immense promise for clues to the story of planet formation, migration, and evolution due to the many measurable species in their atmospheres.

\subsubsection{Interior Structures of Gas Giants}
\label{sec:interiors}
The often-missing link between planetary atmospheres and formation scenarios lies in the interiors of planets. Planetary interiors contain more than 90\% of the total mass, which highlights their fundamental importance \citep{2023RemS...15..681M}. Traditionally, it has been assumed that giant planet interiors are fully convective (e.g. \citet{2016A&A...596A.114M, miller11, thorngren16}), allowing atmospheric composition to be directly interpreted in the context of formation pathways. However, recent gravity field measurements from the Juno and Cassini missions have revealed that the interiors of Jupiter and Saturn are not fully convective \citep{2017GeoRL..44.4649W, 2021NatAs...5.1103M, 2022A&A...662A..18M, 2022PSJ.....3..185M}. Furthermore, it is known that the interiors of giant planets may contain deep radiative zones \citep{1994Icar..112..337G, 2025A&A...693A.308S}, which influence the transport of material from the interior to the atmosphere. This implies that atmospheric abundances may not reflect the planet's overall composition, limiting our ability to infer formation processes from atmospheric constituents.
This paradigm shift highlights the need to reassess our models, and places increased emphasis on accurately characterizing planetary interiors to robustly connect atmospheric properties with formation processes. 

In this context, new planetary interior codes have emerged for exoplanets \citep{acuna24}, but also established tools \citep{2020MNRAS.495.2994M, 2022A&A...665L...5K, 2023MNRAS.523.6282B} have been improved to allow the study and exploration of exoplanets with more complex and realistic interiors. In particular, the open-source code GASTLI \citep{acuna24} allows for the modeling of the evolution and interior structure of giant exoplanets and 
is optimized to rapidly calculate mass-radius relations, making it a valuable tool for interpreting observational data. Currently, its applicability is limited to non-inflated exoplanets with equilibrium temperatures below 1000 K. While the model assumes an interior structure composed of a core and a fully convective envelope, the interior is coupled to a grid of self-consistent, cloud-free atmospheric models that ensure a more realistic interaction between the interior and the atmosphere.

Another widely used tool in the field is the well established MESA code \citep{2013ApJS..208....4P}. MESA has been successfully applied to a broad range of planetary science questions. For example, it has been used to investigate how planetary migration affects the thermal evolution of inflated exoplanets \citep{2020MNRAS.495.2994M}, and to analyze the effects of compositional gradients on the internal mixing of gas giants \citep{2024ApJ...977..227K}. The latter study, by \citet{2024ApJ...977..227K}, highlights how initial compositional gradients and entropy conditions can influence the long-term evolution of a planet’s interior. The authors modeled giant exoplanets with internal structures similar to those inferred for Jupiter and Saturn, including compositional gradients in the region just above the core, sometimes referred to as the ``dilute core.'' Their results show that, when such gradients are present, the convective envelope can efficiently mix material during the planet’s early evolution (within $10^7$ years). However, if the planet starts with a low primordial entropy, these composition gradients can persist for billions of years. In such cases, the atmospheric composition may remain decoupled from the bulk interior composition, complicating the interpretation of atmospheric measurements in terms of the planet’s formation history and internal makeup.

Although the internal structures of exoplanets appear more complex than initially expected, upcoming and ongoing observations with JWST provide new opportunities to improve our understanding. In  \citet{2023MNRAS.523.6282B}, the authors develop a retrieval framework for modeling exoplanet interiors, building on the CEPAM code, which has been widely used for solar system giants \citep{1995A&AS..109..109G, 2016A&A...596A.114M}. With this approach, they show that atmospheric properties such as equilibrium temperature and metallicity, which can be measured with JWST, provide invaluable constraints for the interior structure of giant exoplanets. In their analysis, they consider both simplified models, where a planet consists of a core and a homogeneous envelope, and more realistic ones, incorporating inhomogeneous envelopes and marking one of the first efforts to explore this added ``dilute core'' layer in giant exoplanets.  

Another key observational constraint for probing exoplanet interiors is the measurement of the Love number \citep{2025vanDijk}. This parameter quantifies how much a planet deforms under the tidal forces of its host star and are directly related to how mass is distributed within the planet. As such, it offers a unique window into exoplanet internal structures. Love numbers have already been estimated for a handful of exoplanets \citep{Batygin2009_firstkeccentricity, Ragozzine_2009, Buhler_2016, Hardy_2017, love_number_wasp18b, Hellard_2019, Barros_2022, Akinsanmi-2024-2, Bernabo2024}. However, not all of these measurements have provided strong constraints on interior properties.

In \citet{2025vanDijk}, the authors used their interior retrieval code to demonstrate that a precision of at least 40\% is required in the Love number determination to meaningfully constrain planetary core masses. Encouragingly, such precision is within reach using JWST, and several ongoing and approved Cycle 4 programs are designed with this goal in mind. These observations are expected to yield critical data that will significantly improve our understanding of exoplanet interiors. 

Looking ahead, upcoming ESA missions such as PLATO and Ariel are expected to provide valuable insights into stellar ages and the atmospheric compositions of a large sample of giant exoplanets. These data will offer important constraints for studying planetary evolution and internal structure. By expanding the observational baseline, these missions will help advance our understanding of how planetary characteristics relate to their formation histories, bringing us closer to linking observed properties with formation pathways.

\subsubsection{Giant Planets in Time}

A potentially more direct link between atmospheric composition and initial formation conditions is to observe planetary atmospheres as they are forming or right after they formed. This would minimize the impact of processes which cause significant compositional changes such as fractionated atmospheric mass-loss \citep{Louca2023, malsky23}, photochemical processing \citep{tsai23}, cometary bombardment \citep{Turrini2015, rimmer20, piette20, Welbanks2021, seligman2022_comets,Polychroni2025}, atmosphere-interior interactions \citep{2024ApJ...977..227K, 2025A&A...693A.308S}, or the formation of refractory oxides \citep{burrows99, fonte23}; most of these processes are unconstrained and unaccounted for in atmospheric modeling efforts.

One transiting exoplanet has been observed with JWST thus far. \cite{thao24} presented the first JWST transmission spectrum of the young planet HIP~67522\,b, a $\sim 1 R_J$ giant on a 6.96~day orbital period around a 17~Myr G0V star \citep{rizzuto20}. This single NIRSpec/G395H observation the presence of \water\ ($7\sigma$), CO ($3.5\sigma$), and \cotwo\ ($11\sigma$), and a weak preference for H$_2$S and \sotwo\ at $2.1$ and $1.8\sigma$, respectively. Atmospheric retrieval models suggest the atmosphere is consistent with a solar-to-subsolar C/O and an atmospheric metallicity of $3-10\times$~solar. These results are consistent with the picture that gas giants form \textit{beyond} the water snowline. Furthermore, based on the scale height of the observed absorption features \citep{dewit13}, the mass of HIP~67522\,b was constrained to $\sim 14 M_\oplus$, suggesting the planet is more likely a precursor to the sub-Neptune population. Other young planets with similar radii (e.g., V1298 Tau\,b) have also been noted to have masses consistent with sub-Neptunes, rather than true gas giants \citep[$\sim100$s $M_\oplus$;][]{barat24a}.

While observations of young planets may be more insightful, it is only through observations of both the planet \textit{and} its natal disk that we can draw more definitive conclusions about how and where gas giants form. IRAS 04125+2902\,b is a recently discovered transiting gas giant orbiting a 3~Myr $0.7M_\odot$ pre-main sequence star in the Taurus Molecular Cloud \citep{barber24}. This system is highly unique, as the host star still hosts its primordial transition disk. The transits of IRAS 04125+2902\,b were detectable because the disk is in a unique nearly face-on ($i \simeq 30^\circ$) configuration. Future observations of this system will allow for characterization of the planet's atmosphere \textit{and} the disk composition \citep[e.g. JWST GO 8597, Co-PIs Feinstein \& Bergner; ][]{feinstein25}. Should the elemental abundance ratios between the inner or outer region of the disk be consistent with what is measured in the planet, it could provide the strongest evidence of where gas giants originally form. If the disk and planet compositions are inconsistent, this would significantly challenge our ability to link planetary atmospheric compositions to formation scenarios, especially for mature ($>1$~Gyr) systems. 

\subsection{Observations of Directly Imaged Exoplanets}\label{sec:direct_imaging}


A few dozen exoplanets have been directly imaged, namely, their emitted flux has been spatially resolved from their host stars. These planets are typically several Jupiter masses, several tens of AU from their host stars, and orbit stars younger than a few 100Myr due to inherent biases of the high-contrast imaging technique (see \citealt{Currie2023} for a recent review). Based on the mass/period distribution of exoplanets identified in a large VLT/SPHERE imaging survey, \citet{Vigan2021, chomez25} suggested that the directly imaged planet population is consistent with GI formation but that GI rarely forms giant planets. It remains unclear whether these planets are forming similarly to the most massive transiting planets which are then migrating differently, or whether the formation of wide-separation super-Jupiters is fundamentally different than the close-in super-Jupiter planets. With sufficient photometric or spectroscopic coverage their bulk luminosities can be derived, allowing for an analysis of the initial entropy budget of these planets, as well as any time delay between formation of the star and its planet \citep[e.g.][]{Marley2007,Fortney2008,Spiegel2012,Macintosh2015,Zhang2024}.

Directly imaged exoplanets are readily accessible to spectroscopic characterization. By 2021, 175 exoplanets had spectroscopic measurements in the NASA Exoplanet Archive \citep{Currie2023}; directly detected planets made up a substantial fraction of these planets (42/175)\footnote{According to the NASA Exoplanet Archive inclusion criteria, whereby any object with mass $<$30M$_J$ is considered a planet}, and in fact \textit{every} known directly imaged planet had spectroscopic measurements at the point of publication of that work. We refer the reader to the PPVII review by \citet{Currie2023} for a detailed discussion of the state of pre-JWST high-contrast imaging spectroscopy, but summarize a few highlights here.

Ground-based observations with integral field spectrograph (IFS) instruments routinely allow for low-resolution (R$\leq$100) spectroscopy between 1-2.5\micron, leading to detections of broadband \water\ and CO features in a number of planets \citep[e.g.][]{bonnefoy2016,Chilcote2015,Chauvin2017}, as well as a clear \methane\ absorption in the cooler ($\sim$700K) exoplanet 51~Eri~b \citep{Macintosh2015}. In recent years, VLT/GRAVITY interferometry has allowed R$\sim$a few hundred spectroscopy of several resolved exoplanets, including those at very tight spatial separations. Higher resolution observations (R$\sim$several thousand) can disentangle stellar and planetary light based on individual molecular lines, usually by the correlation with molecular templates. \citet{Konopacky2013} detected CO and \water\ in HR8799c with R$\sim$4000 Keck/OSIRIS measurements, and found a super-stellar C/O ratio. A larger Keck/OSIRIS survey of planetary mass companions found that most imaged planets have C/O ratios around solar \citep[e.g.][]{hoch23}; VLT/CRIRES+ observations are expected to further expand this sample in the coming years \citep[see e.g.][for an early CRIRES+ detection of an imaged planet]{Parker2024}. HR8799bcde is perhaps the best-characterized directly-imaged exoplanet system, with four gas giant planets with similar mass providing the opportunity for comparative exoplanetology in the same system. \citet{Nasedkin2024} found enriched metallicities and stellar to superstellar C/O atmospheric ratios for all four HR8799bcde exoplanets in a combined analysis of several ground-based spectra.

Molecules containing atoms other than H, C and O have so far remained largely elusive in directly imaged planets. No sulfur species have yet been identified, though H$_2$S has been seen in a few cool brown dwarfs \citep{Hood2024,Lew2024}. Similarly, no nitrogen species have yet been confidently detected in a directly imaged planet, though \citet{Whiteford2023} claimed to detect NH$_3$ in the atmosphere of 51~Eri~b at a confidence of 2.7$\sigma$. With its exquisite mid-IR sensitivity, JWST should make \ammonia\ detections in several cool exoplanets; this has been demonstrated technically with Spitzer and JWST detections of \ammonia\ in cool brown dwarfs \citep[e.g.][]{Roellig2004,Barrado2023,Beiler2023}. 

\subsubsection{Direct Imaging Spectroscopy with JWST}

JWST spectroscopic observations have the potential to provide high-quality constraints on the molecular and isotopic abundances of exoplanets resolved from their host stars, for sufficiently widely separated exoplanets. Moderate-resolution spectra (R$ \sim$1000-3500) can be measured from 0.8-23\,\micron\ for a subset of imaged exoplanets, by combining observations with NIRSpec and MIRI MRS. To date, there are only a few exoplanets and planetary-mass objects with published JWST spectra \citep[e.g.][]{Miles2023,Worthen2024}. Several other imaged exoplanets have been observed or scheduled for JWST spectroscopy (e.g. GO 1188, 3522, 3647, 4892), and further trends in their C/N/O/S content may emerge. A handful of cold brown dwarfs have also been observed or scheduled (e.g. GO 1189, 1275), and should provide a useful comparison point to benchmark these exoplanet spectra and analysis methods. In particular, measurements of specific molecular and isotope abundance ratios will provide a useful zero-point against which to compare directly imaged exoplanet abundance ratios, and will indicate the intrinsic scatter in abundance ratios between these objects (e.g. due to different local formation environments). For both directly imaged exoplanets and brown dwarfs, we are studying emission spectra -- a ``like for like'' comparison that is perhaps simpler than comparing transmission spectra of close-in giants and emission spectra of field brown dwarfs. Direct Imaging spectroscopy of giant exoplanets therefore provides a "bridge" between these populations of objects.

The best-characterized planetary-mass object to date is VHS1256b, for which there are medium-resolution JWST spectra from 1-20\,\micron \citep{Miles2023}. That work identified \water, \methane, CO, \cotwo, Na, and K in the atmosphere, and found evidence for disequilibrium chemistry and direct evidence of clouds in the form of a silicate absorption feature. \citet{Petrus2024} compared these spectra to forward-model grids (e.g. ATMO, \citealt{Tremblin2015}, and Exo-REM, \citealt{Charnay2018}) and found $\sim$solar C/O ratios of $0.46 \pm 0.16$ and $0.38 \pm 0.20$ using the ATMO and Exo-REM model grids respectively. That work also highlighted that these estimates can vary substantially when using different atmospheric model grids and different individual spectral windows (see their Figure 4), making a careful and multi-wavelength analysis key to deriving accurate constraints \citep{whiteford2024}.

There are also published JWST observations of $\beta$-Pictoris b, a $\sim$9M$_J$ exoplanet orbiting a young ($\sim$23Myr; \citealt{mamajek14}) A-star. MIRI/MRS observations of this target revealed water vapor, but are only sensitive to $\sim$7\,\micron\ since the bright debris disk \citep{Smith1984} obscures the planet signal at longer wavelengths \citep{Worthen2024}. That work found a slightly sub-stellar C/O ratio of $0.39^{+0.10}_{-0.06}$ or $0.36^{+0.13}_{-0.05}$ (depending on the atmosphere model used, ATMO or Exo-REM) when fitting the JWST data alongside spectra from GRAVITY \citep{GRAVITYCollaboration2020} and GPI \citep{Chilcote2017}, and photometry from Magellan, Gemini NICI and VLT/NACO \citep{Males2014, Bonnefoy2013}. As the sample of exoplanets and brown dwarfs with high-quality JWST spectra grows, it will be possible to start drawing trends among this population.

\subsubsection{Isotopes in Directly Imaged Exoplanet Atmospheres}\label{sec:isotopes:DI}

Isotope ratios may also provide hints about how a planet formed \citep[e.g][]{Molliere2019,Morley2019}. These isotopic species are beginning to be detected in directly imaged exoplanets. \citet{Molliere2019} suggested that deuterium and $^{13}$C molecules should be detectable from the ground with high-resolution spectroscopy, and a few detections of $^{13}$C in exoplanet atmospheres have been made from the ground with VLT/SINFONI data (in TYC 8998-760-1 b, \citealt{Zhang2021}) and with VLT/CRIRES+ data (in YSES 1b, \citealt{Zhang2024}).
JWST spectra, meanwhile, have already led to the identification of C, N, and O isotopes in cold brown dwarfs. \citet{Barrado2023} found $^{15}$NH$_3$ in the atmosphere of the free-floating brown dwarf WISE J1828, and measured a $^{14}$N/$^{15}$N ratio of $670^{+390}_{-211}$, tentatively indicating depletion in $^{15}$N relative to the ISM, but consistent with Jupiter and Saturn \citep{Fletcher2014}. \citet{Gandhi2023} measured $^{13}$C, $^{17}$O and $^{18}$O in the spectrum of the planetary-mass VHS1256b \citep{Miles2023}, and found all abundance ratios to be enhanced in the minor isotope (i.e., that containing $^{13}$C, $^{17}$O or $^{18}$O) relative to the ISM and the solar system. D/H ratios should also be measurable for some cool brown dwarfs with JWST, via the detection of CH$_3$D \citep{Morley2019}. Indeed, very recently the isotopes CH$_3$D \citep{Rowland2024} and $^{15}$NH$_3$ \citep{kuehnle2025} have both been detected in the coldest brown dwarf (WISE-0855, \citealt{Luhman2014}). With similar quality spectra, these isotopic species would also be detectable in cool exoplanets, and we can expect such detections in the coming years, providing a new lever with which to study exoplanet formation.


\section{Where to go from here}
\label{sec:link}

Here, we have provided a broad overview of the state-of-the-art theories and results for planet formation, protoplanetary disk chemistry, and exoplanet atmospheres. We conclude below with several ideas discussed at the MPIA workshop for what next steps could take place to advance all three subfields highlighted throughout this work. This list is meant to represent the viewpoints of workshop attendees and is not a complete list of directions we need to go in order to advance our understanding of how to link exoplanet atmospheres to their formation conditions.

\subsection{Ways Forward for Planet Formation}

\begin{itemize}
    \item {\bf Understanding the properties of discs in detail.} The core properties of discs, such as their masses, composition, the size of pebbles and the physics of planetesimal formation, are still debated. These are key inputs to formation models and likely vary with time, position in the disc and stellar properties. Uncertainties in these properties can drive uncertainties comparable to the differences between formation models.  Chemical surveys with JWST and ALMA are starting to provide the data. Soon, the SKA may further help to constrain cm-sized grains in the inner regions of discs (which are optically thick at ALMA wavelengths).
    
    \item {\bf Understanding how heavy elements are distributed inside of the planets.} The bulk properties of planets are relatively easy to predict, but it is critical to know how these are linked to the atmospheric properties. The extent to which the cores can be eroded to pollute the envelopes, how these envelopes mix and the subsequent impact on the atmospheric metallicity and composition needs to be understood. A resolution to the radius inflation problem \citep[e.g.][]{Thorngren2024} would help by improving constraints on the bulk metallicity from the planet's mass, radius and temperature.
    
    \item {\bf Population-level statistics of exoplanet compositions may be essential}:  solutions for a single system will be degenerate \citep[e.g.][]{molliere2022}, but these degeneracies might be broken at a population level \citep[e.g.][]{kirk2024}.
    
    \item {\bf Constraining a wider range of elemental species in planet atmospheres will help.} More refractory species (e.g. sulfur, silicon, iron, sodium) are better tracers of the amount of solids accreted. Meanwhile, volatiles (e.g. nitrogen) better trace the gas. If these quantities can be better estimated, constraints on the formation will likely improve \citep[e.g.][]{Turrini2021,Pacetti2022,Chachan2023,crossfield23}.
    
    \item {\bf Understanding how atmospheric sulfur abundances may be biased.}. It is unclear as to what extent the atmospheric sulfur abundance is biased by the preferential dredging of ultra-volatile H$_2$S released from FeS in the core \citep{Steinmeyer2023}. In this case sulfur may trace core-dredging rather than direct enrichment of the envelope pebble/planetesimal accretion.
    
    \item {\bf Understanding the physics of formation in detail.} Some key questions are: how does the formation process modify the abundances of gas and solids? How do the properties of solids vary through the disk and time? How much do N-body interactions influence the composition of growing giant planets \citep[e.g.][]{Turrini2015,Turrini2021,seligman2022_comets,Sainsbury-Martinez2024,Polychroni2025}? How much does the secular evolution of planetary systems alter or overwrite the compositional signatures of planet formation? Models for these processes are likely oversimplified and can impact the planet's final composition.
    
    \item \textbf{Understanding the life cycle of solids during the disk lifetime and its compositional implications.} How do solids partition between small dust, pebbles and planetesimals? How quickly does this partition set (i.e. how much dust mass is converted into planetesimals during Class0/I, e.g. \citealt{Cridland2022})? How much planetesimal mass is reverted to dust and pebbles by collisions during the planet formation process and on which timescales \citep{Turrini2019,Turrini2023,Bernabo2022}? What role does continued infall from the parent molecular cloud play?
\end{itemize}

\subsection{Ways Forward for Disk Chemistry}

\begin{itemize}
    \item \textbf{Measuring elemental abundances beyond C \& O.} We need better constraints on other observable elements. In particular, dedicated work on the N and S budget could be particularly valuable, especially when linking to gas giant atmospheres. The main forms of N and S-bearing species are difficult or impossible to detect (e.g.~N$_2$, solid S carriers). With this in mind, we need to better understand the uncertainties when translating between trace observable species and bulk reservoirs.
    
    \item \textbf{Connecting volatile transport and disk structure.} Transport appears to profoundly influence the distribution of volatiles across a disk, and this is likely sensitive to the structure and evolution of a given disk. Significant progress is anticipated in coming years as ALMA \& JWST measure outer \& inner disk gas compositions towards more statistical samples \citep[e.g.][]{Gasman2025}, complemented by constraints on ice compositions towards edge-on disks.
    
    \item \textbf{Measuring more isotopic ratios.} Expanding the number of D/H, $^{12}$C/$^{13}$C, $^{15}$N/$^{14}$N, and $^{18}$O/$^{16}$O measurements is a high priority for interpreting emerging constraints on rare isotopes in exoplanet atmospheres. Key areas include understanding how isotope ratios vary across the disk and whether there are separate molecular reservoirs of isotopic enrichment and depletion.
    
    \item \textbf{Measuring properties that can be used for comparisons to exoplanet atmospheres.} The C/O ratios measured in outer-disk gas are systematically higher than those inferred towards wide-separation exoplanet atmospheres \citep{Bergin2024b}, though the sample is currently limited and biased.  Programs with larger and more representative disk samples (e.g.~DECO ALMA Large Program) are needed before formation scenarios can be inferred.
    
    \item \textbf{Expanding the wavelength coverage to the far-IR.} The lack of a far-IR telescope facility is a major limitation for disk chemistry: this wavelength range is needed to quantify disk gas masses and water budgets, as well as improve constraints on the nitrogen budget \citep[e.g.][]{Kamp2021,Bergner2022,Miotello2023,Schwarz2024}.
    
    \item \textbf{Constraining disk refractory compositions.} The budget and composition of refractory building blocks is poorly understood.  Avenues for further characterization include IR signatures of silicates and minerals detectable by JWST \citep{Jang2024}; polluted Herbig stars \citep{Kama2019}; and protostellar shocks \citep{Ginsburg2019,Bergner2022b}.
    
    \item \textbf{Understanding the environmental context of planet formation.} Protoplanetary disks form and evolve within complex larger-scale star-forming environments.  The effect of the larger-scale environment on planet formation remains poorly understood, which may be especially important for the young embedded disk stage, systems with late-stage infall, and highly irradiated regions.  It is also important to better understand the influence of the star (e.g. flares and accretion-driven outbursts) on planet-forming materials and disk chemistry. 
\end{itemize}

\subsection{Ways Forward for Planetary Atmospheres}

\begin{itemize}
    \item \textbf{Measuring multiple elemental abundance ratios.} Measuring the C/O ratio was a necessary starting point towards understanding how atmospheres may hold clues to their initial formation conditions. However, the increased IR wavelength coverage of JWST now allows us to detect new non C- or O-bearing molecules (e.g. \sotwo). This, in combination with ground-based facilities used to measure refractory abundances (e.g., Fe, Mg, Na) provides crucial context for understanding the characteristics of exoplanet atmospheres. Measuring multiple elemental abundance ratios will ensure a more wholistic, and less biased, insight into where and how a given planet may have formed.
    
    \item \textbf{Measuring stellar abundances and elemental ratios.} Measuring stellar C/O to appropriately compare the planet's atmosphere to its own system, and also investigating the dependence on other stellar elemental ratios.
    
    \item \textbf{Understanding how elemental abundance ratios evolve.} How much are elemental abundance ratios expected to evolve over time? Coupling planet evolution tracks (e.g. mass-loss) and atmospheric chemistry (also atmosphere-interior interactions). What timescales to these processes operate on?
    
    \item \textbf{Detecting young planets in protoplanetary disks.} Direct imaging exoplanets are inherently biased towards younger populations, when the planets are still self-luminous. However, only \textit{two} planets have been detected thus far in a system where the protoplanetary disk has not yet dissipated: PDS70~b \citep{keppler18} and IRAS~04125+2902~b \citep{barber24}. The detection of more such systems are crucial, as it would grant us the opportunity to directly put constraints on the timescale of formation, characteristics of the formation environment, as well as to the migration history. 
    
    \item \textbf{Building homogeneous data sets of atmospheric observations.} How do we define a ``truth'' for atmospheric abundances? What wavelength coverage and resolution will get us there? Are we able to combine non-simultaneous transmission observations?
    
    \item \textbf{Comparing planets within and without the same system.} Intra- and extra-system comparisons of exoplanet atmospheres: this may allow us to break degeneracies in planet formation processes and test differences in mechanisms.
    
    \item \textbf{Exploring conditional properties of gas giant planets.} How do gas giant compositions change in different system configurations? Are atmospheric compositions different when there is an inner terrestrial companion, an outer giant companion, or are compositions insensitive to other planets? Do compositions change as a function of their obliquity \citep[e.g.,][]{kirk2024, Penzlin2024}?

    \item \textbf{Understanding the uncertainty of atmosphere modeling.} It is important to investigate How the uncertainties of our models affect the inferred planet parameters, including the role of clouds and atmospheric variability. By disentangling the uncertainty introduced by the models may help us understand how reliable our interpretation about the observational data is.

    \item \textbf{Combining interior structure modeling with atmospheric characterization}. By coupling interior and atmospheric models, we may provide a more physical understanding of the bulk metal enrichment in exoplanets. 

    \item \textbf{Obtaining more accurate opacities and equations of state.} JWST and upcoming missions have and will provide unprecedented precision when studying both disks and exoplanet atmospheres. All the while, the model uncertainties are starting to be the limiting factor in atmospheric and interior models. We require mode accurate opacities (used in both atmospheric and interior models) and equations of state (essential for computing the density and thermal evolution in interior models. Specifically, we require laboratory-measured opacities at high pressures and temperatures, which are the most relevant for planetary interior conditions.
    
\end{itemize}

We are in an era where our picture and understanding of planet formation, protoplanetary disks, and exoplanet atmospheres is rapidly evolving. Therefore, we recognize that what is outlined here may not be as broadly applicable within the next few years.  If a key astrophysical question is understanding how planets form, then it is \textit{essential} members of each subfield continue to communicate results with each other. We hope this highlights the importance of continuing to host interdisciplinary workshops, such that the community continues to have these important conversations. 

\vspace{4mm}
\textbf{Contributions}

ADF, RAB, and JBB contributed equally to this organization, writing, and editing of this work. ADF wrote Section 1, 4, 4.2, and 5.3, and created Figures 1-7 and Table 1 in the manuscript. RAB wrote Section 1, 2, and 5.1, and contributed to Figures~\ref{fig:co_comp} and \ref{fig:sulfur}. JBB wrote Section 1, 3, and 5.2, and contributed to Figure~\ref{fig:disk_structure}. JDL wrote Section 4.2.2 and created Figure~\ref{fig:hires}. ECM wrote Section 4.3. LW wrote Section 4.1 and provided the cross-sections for Figure~\ref{fig:spaghetti}. YM wrote Section 4.2.4. BB, AAAP, CPG, KRS, and DT provided a ''red team" review of the manuscript to ensure all topics covered at the MPIA workshop were sufficiently accurately represented. KRS created Figure~\ref{fig:disk_snowlines}. LAA, EMA, MGB, JB, AC, IJMC, GDM, HH, AJ, LK, JHL, RL, MO, EP, GP, JP, BP, DAS, JBS, JT, and NW provided detailed feedback on the manuscript.

\vspace{4mm}
\textbf{Acknowledgments}

We acknowledge the Max-Planck-Institut f\"ur Astronomie in Heidelberg, which funded and hosted this meeting. We thank Ian Crossfield and Laura Kreidberg for initiating this conference. We thank the members of the Scientific and Local Organizing Committees for their roles in ensuring a successful conference. Members of the Scientific Organizing Committee include:  Bertram Bitsch, Ian Crossfield, Adina Feinstein, Sasha Hinkley, Laura Kreidberg, Paul Mollière, Christoph Mordasini, Anjali Piette, and Niall Whiteford. Members of the Local Organizing Committee include: Lorena Acu\~{n}a-Aguirre, Eva-Maria Ahrer, Duncan Crhsitie, Carola Jordan, Xiang Luo, and Elisabeth Matthews.

ADF acknowledges funding from NASA through the NASA Hubble Fellowship grant HST-HF2-51530.001-A awarded by STScI. JK acknowledges financial support from Imperial College London through an Imperial College Research Fellowship grant. RAB and AP acknowledge support from the Royal Society through a University Research Fellowship and Enhanced Expenses Award. CPG acknowledges support from the NSERC Vanier scholarship, and the Trottier Family Foundation. CPG also acknowledges support from the E. Margaret Burbidge Prize Postdoctoral Fellowship from the Brinson Foundation. KS and DS acknowledge support from the European Research Council under the Horizon 2020 Framework Program via the ERC Advanced Grant Origins 83 24 28. SP acknowledges the financial support of the SNSF under grant 51NF40\_205606. GP gratefully acknowledges support from the Max Planck Society and from the Carlsberg Foundation, grant CF23-0481. YM acknowledges funding from the European Research Council (ERC) under the European Union’s Horizon 2020 research and innovation programme (grant agreement no. 101088557, N-GINE). D.T and E.P. acknowledge support from the European Research Council under the Horizon 2020 Framework Programme ERC Synergy ``ECOGAL'' Project GA-855130 and from the ASI–INAF grants no. 2016-23-H.0 plus addendum 2016-23-H.2-2021 and no. 2021-5-HH.0 plus addenda 2021-5-HH.1-2022 and 2021-5-HH.2-2024. MGB acknowledges support from the NSF Graduate Research Fellowship (DGE-2040435). AC acknowledges support from the NASA Postdoctoral Program at NASA Ames Research Center, administered by Oak Ridge Associated Universities under contract with NASA. RL acknowledges support by NASA through the NASA Hubble Fellowship grant HST-HF2-51559.001-A awarded by STScI.

\begin{longrotatetable}
\begin{deluxetable}{p{2cm} p{5.5cm} p{1.5cm} p{3cm} p{2cm} p{2cm} p{3.5cm}}
\tabletypesize{\footnotesize}
\tablecaption{Measured C/O in transiting and direct imaging exoplanets from novel JWST observations. \label{tab:exoplanet_c2o}}
\tablehead{
\colhead{Planet} & \colhead{Instrument} &  \colhead{$\lambda$ Range [$\mu$m]}  & \colhead{Species Detected} & \colhead{C/O} & \colhead{Method} & \colhead{Ref.} }
\startdata
WASP-39\,b & HST/WFC3 G102/G141 & $0.8 - 1.7$ & Na, K, \water\                     & $0.31_{-0.05}^{+0.08}$  & Retrieval & \cite{wakeford18} \\
           & JWST NIRCam/F322W2 & $2.5 - 4.0$ & \water, \cotwo\                    & 0.23                    &  Grid     & \cite{ahrer23}\\
           & JWST NIRSpec/G395H & $2.8 - 5.2$ & \water, \cotwo, \sotwo\            & 0.46                    &  Grid     & \cite{alderson23}\\
           & JWST NIRISS/SOSS   & $0.6 - 2.8$ & K, \water\                         & 0.20                    &  Grid     & \cite{feinstein23}\\
           & JWST NIRSpec/PRISM & $0.6 - 5.2$ & Na, K, \water, CO, \cotwo, \sotwo\  & $\sim 0.7$ &  Grid &\cite{rustamkulov23}\\
           & JWST NIRISS/SOSS, NIRCam/F322W2, NIRSpec/G395H/PRISM & $0.6 - 5.2$ & & 0.35 &  Grid &\cite{carter24}\\
\hline
WASP-96\,b & VLT, HST/WFC3, Spitzer/IRAC & $0.4 - 4.5$ & Na, K, \water\ & $0.10_{-0.07}^{+0.19} $ &  Retrieval & \cite{nikolov22} \\
           & JWST NIRISS/SOSS            & $0.6 - 2.8$ & \water. Detected at $\sim 2 \sigma$: Na, K, CO, \cotwo, H$_2$S & 0.55  & Grid & \cite{radica23} \\
           & JWST NIRISS/SOSS            & $0.6 - 2.8$ & \water. Detected at $\sim 2 \sigma$: Na, K, CO, \cotwo\        & $0.50_{-0.29}^{+0.24}$ & Retrieval &\cite{taylor23} \\
\hline
HD~209458\,b & HST WFC3/G141                     & $1.1 - 1.7$ & \water\         & $0.18 - 0.43$          &  Retrieval & \cite{changeat22} \\
             & JWST NIRCam/F444W                 & $2.3 - 5.1$ & \water, \cotwo\ & $0.11_{-0.06}^{+0.02}$ &  Retrieval & \cite{xue24} \\
             & HST WFC3/G141, JWST NIRCam/F444W  & $1.1 - 5.1$ & \water, \cotwo\ & $0.23_{-0.15}^{+0.12}$ &  Retrieval & \cite{xue24} \\
\hline
WASP-80\,b & HST STIS/G430L/G750L, WFC3/G141  & $0.4 - 1.7$ & \water\   & $0.50_{-0.17}^{+0.14}$ &  Retrieval & \cite{wong22}\\
           & JWST NIRCam/F322W2               & $2.5 - 4.0$ & \methane\ & $< 0.57$               &  Retrieval & \cite{bell23}\\
\hline
WASP-77\,Ab & Gemini-S/IGRINS         & $1.43 - 2.42$ & \water, CO & $0.59 \pm 0.08$        &  Retrieval & \cite{line21}\\
            & HST WFC3/G141           & $1.1-1.7$     & \water\    & $<0.78$                &  Grid      & \cite{mansfield22}\\
            & JWST NIRSpec/G395H      & $2.8 - 5.2$   & \water, CO & $0.36_{-0.09}^{+0.10}$ &  Retrieval &  \cite{august23}\\
            & NIRSpec/G395H, Gemini-S/IGRINS & $1.5 - 5.2$   & \water, CO & $0.57 \pm 0.06$        &  Retrieval &  \cite{smith24}\\
\hline
WASP-107\,b & HST WFC3/G141 & $1.1 - 1.7$ & \water\ & $0.03_{-0.01}^{+0.09}$ &  Retrieval & \cite{kreidberg18}\\
            & HST WFC3/G141, JWST NIRCam/F322W2/F444W, MIRI/LRS & $0.8 - 12.0$ & \water, \cotwo, \methane, CO, \sotwo, \ammonia\ & $0.33_{-0.05}^{+0.06}$ &  Retrieval & \cite{welbanks24}\\
            & JWST NIRSpec/G395H & $2.8 - 5.2$ & \water, \cotwo, \methane, CO, \sotwo\ & $0.51_{-0.21}^{+0.27}$ &  Retrieval & \cite{sing24}\\
\hline
 GJ 3470\,b & HST WFC3/G141                           & $1.1 - 1.7$ & ---   & $< 1$ & Retrieval & \cite{ehrenreich14}\\
            & HST STIS/G750L, WFC3/G141, Spitzer/IRAC & $1.1-4.5$ & \water\ & Not reported &   Free retrieval & \cite{benneke19}\\
            & HST WFC3/G141, NIRCam F322W2/F444W      & $1.1-5.0$ & \water, \cotwo, CO, \methane, \sotwo\ & $0.35 \pm 0.1$ &  Retrieval & \cite{beatty24}\\
\hline
WASP-121\,b & Gemini-S/IGRINS & $1.43 - 2.42$ & \water, CO & $0.70_{-0.10}^{+0.07}$ & Retrieval & \cite{Smith24b}\\
& HST/WFC3 G102 & $1.08 - 1.1$ & \water & $0.49_{-0.37}^{+0.65}$ & Retrieval & \cite{mikalevans19}\\
\hline
WASP-15\,b & NIRSpec/G395H & $2.8 - 5.2$ & \water, \cotwo, CO, \sotwo\ & $0.48_{-0.16}^{+0.11}$ & Retrieval & \cite{kirk2025} \\
\hline
VHS 1256\,b  & Keck/OSIRIS & & & $0.590_{-0.354}^{+0.280}$ & & \cite{hoch22} \\
& JWST MIRI/MRS & $1-12$ & \water, \methane, CO, \cotwo, Na, K & $0.46 \pm 0.16$ &  Forward model & \cite{Miles2023, Petrus2024}\\
\hline 
$\beta$ Pictoris\,b  & VLTI/GRAVITY & $2-2.5$ & \water\ & $0.43_{-0.06}^{+0.10}$ &  method & \cite{GRAVITYCollaboration2020}\\
  & Gemini/GPI, VLTI/GRAVITY, JWST MIRI/MRS, Magellan/VisAO, VLT/NACO, Gemini/NICI & $1 - 7.5$ & \water\ & $0.39_{-0.06}^{+0.10}$ &  method & \cite{Worthen2024}\\
\hline 
PDS\,70\,b  & Keck/KPIC & $2.29 - 2.49$ & \water, CO & $0.28_{-0.12}^{+0.20}$ &  Retrieval & \cite{hsu24}\\
\enddata
\end{deluxetable}
\end{longrotatetable}

\bibliography{main}{}

\begin{thebibliography}{}
\expandafter\ifx\csname natexlab\endcsname\relax\def\natexlab#1{#1}\fi
\providecommand{\url}[1]{\href{#1}{#1}}
\providecommand{\dodoi}[1]{doi:~\href{http://doi.org/#1}{\nolinkurl{#1}}}
\providecommand{\doeprint}[1]{\href{http://ascl.net/#1}{\nolinkurl{http://ascl.net/#1}}}
\providecommand{\doarXiv}[1]{\href{https://arxiv.org/abs/#1}{\nolinkurl{https://arxiv.org/abs/#1}}}

\bibitem[{{Acu{\~n}a} {et~al.}(2024){Acu{\~n}a}, {Kreidberg}, {Zhai}, \& {Molli{\`e}re}}]{acuna24}
{Acu{\~n}a}, L., {Kreidberg}, L., {Zhai}, M., \& {Molli{\`e}re}, P. 2024, \aap, 688, A60, \dodoi{10.1051/0004-6361/202450559}

\bibitem[{{Ahrer} {et~al.}(2023){Ahrer}, {Stevenson}, {Mansfield}, {Moran}, {Brande}, {Morello}, {Murray}, {Nikolov}, {Petit dit de la Roche}, {Schlawin}, {Wheatley}, {Zieba}, {Batalha}, {Damiano}, {Goyal}, {Lendl}, {Lothringer}, {Mukherjee}, {Ohno}, {Batalha}, {Battley}, {Bean}, {Beatty}, {Benneke}, {Berta-Thompson}, {Carter}, {Cubillos}, {Daylan}, {Espinoza}, {Gao}, {Gibson}, {Gill}, {Harrington}, {Hu}, {Kreidberg}, {Lewis}, {Line}, {L{\'o}pez-Morales}, {Parmentier}, {Powell}, {Sing}, {Tsai}, {Wakeford}, {Welbanks}, {Alam}, {Alderson}, {Allen}, {Anderson}, {Barstow}, {Bayliss}, {Bell}, {Blecic}, {Bryant}, {Burleigh}, {Carone}, {Casewell}, {Changeat}, {Chubb}, {Crossfield}, {Crouzet}, {Decin}, {D{\'e}sert}, {Feinstein}, {Flagg}, {Fortney}, {Gizis}, {Heng}, {Iro}, {Kempton}, {Kendrew}, {Kirk}, {Knutson}, {Komacek}, {Lagage}, {Leconte}, {Lustig-Yaeger}, {MacDonald}, {Mancini}, {May}, {Mayne}, {Miguel}, {Mikal-Evans}, {Molaverdikhani}, {Palle}, {Piaulet}, {Rackham}, {Redfield}, {Rogers}, {Roy}, {Rustamkulov},
  {Shkolnik}, {Sotzen}, {Taylor}, {Tremblin}, {Tucker}, {Turner}, {de Val-Borro}, {Venot}, \& {Zhang}}]{ahrer23}
{Ahrer}, E.-M., {Stevenson}, K.~B., {Mansfield}, M., {et~al.} 2023, \nat, 614, 653, \dodoi{10.1038/s41586-022-05590-4}

\bibitem[{{Akinsanmi} {et~al.}(2024){Akinsanmi}, {Barros}, {Lendl}, {Carone}, {Cubillos}, {Bekkelien}, {Fortier}, {Flor{\'e}n}, {Collier Cameron}, {Bou{\'e}}, {Bruno}, {Demory}, {Brandeker}, {Sousa}, {Wilson}, {Deline}, {Bonfanti}, {Scandariato}, {Hooton}, {Correia}, {Demangeon}, {Smith}, {Singh}, {Alibert}, {Alonso}, {Asquier}, {B{\'a}rczy}, {Barrado Navascues}, {Baumjohann}, {Beck}, {Beck}, {Benz}, {Billot}, {Bonfils}, {Borsato}, {Broeg}, {Buder}, {Charnoz}, {Csizmadia}, {Davies}, {Deleuil}, {Delrez}, {Ehrenreich}, {Erikson}, {Farinato}, {Fossati}, {Fridlund}, {Gandolfi}, {Gillon}, {G{\"u}del}, {G{\"u}nther}, {Heitzmann}, {Helling}, {Hoyer}, {Isaak}, {Kiss}, {Lam}, {Laskar}, {Lecavelier des Etangs}, {Magrin}, {Maxted}, {Mecina}, {Mordasini}, {Nascimbeni}, {Olofsson}, {Ottensamer}, {Pagano}, {Pall{\'e}}, {Peter}, {Piazza}, {Piotto}, {Pollacco}, {Queloz}, {Ragazzoni}, {Rando}, {Rauer}, {Ribas}, {Santos}, {S{\'e}gransan}, {Simon}, {Stalport}, {Szab{\'o}}, {Thomas}, {Udry}, {Van Grootel}, {Venturini},
  {Villaver}, \& {Walton}}]{Akinsanmi-2024-2}
{Akinsanmi}, B., {Barros}, S.~C.~C., {Lendl}, M., {et~al.} 2024, \aap, 685, A63, \dodoi{10.1051/0004-6361/202348502}

\bibitem[{{Alderson} {et~al.}(2023){Alderson}, {Wakeford}, {Alam}, {Batalha}, {Lothringer}, {Adams Redai}, {Barat}, {Brande}, {Damiano}, {Daylan}, {Espinoza}, {Flagg}, {Goyal}, {Grant}, {Hu}, {Inglis}, {Lee}, {Mikal-Evans}, {Ramos-Rosado}, {Roy}, {Wallack}, {Batalha}, {Bean}, {Benneke}, {Berta-Thompson}, {Carter}, {Changeat}, {Col{\'o}n}, {Crossfield}, {D{\'e}sert}, {Foreman-Mackey}, {Gibson}, {Kreidberg}, {Line}, {L{\'o}pez-Morales}, {Molaverdikhani}, {Moran}, {Morello}, {Moses}, {Mukherjee}, {Schlawin}, {Sing}, {Stevenson}, {Taylor}, {Aggarwal}, {Ahrer}, {Allen}, {Barstow}, {Bell}, {Blecic}, {Casewell}, {Chubb}, {Crouzet}, {Cubillos}, {Decin}, {Feinstein}, {Fortney}, {Harrington}, {Heng}, {Iro}, {Kempton}, {Kirk}, {Knutson}, {Krick}, {Leconte}, {Lendl}, {MacDonald}, {Mancini}, {Mansfield}, {May}, {Mayne}, {Miguel}, {Nikolov}, {Ohno}, {Palle}, {Parmentier}, {Petit dit de la Roche}, {Piaulet}, {Powell}, {Rackham}, {Redfield}, {Rogers}, {Rustamkulov}, {Tan}, {Tremblin}, {Tsai}, {Turner}, {de Val-Borro},
  {Venot}, {Welbanks}, {Wheatley}, \& {Zhang}}]{alderson23}
{Alderson}, L., {Wakeford}, H.~R., {Alam}, M.~K., {et~al.} 2023, \nat, 614, 664, \dodoi{10.1038/s41586-022-05591-3}

\bibitem[{{Altwegg} {et~al.}(2020){Altwegg}, {Balsiger}, {H{\"a}nni}, {Rubin}, {Schuhmann}, {Schroeder}, {S{\'e}mon}, {Wampfler}, {Berthelier}, {Briois}, {Combi}, {Gombosi}, {Cottin}, {De Keyser}, {Dhooghe}, {Fiethe}, \& {Fuselier}}]{Altwegg2020}
{Altwegg}, K., {Balsiger}, H., {H{\"a}nni}, N., {et~al.} 2020, Nature Astronomy, 4, 533, \dodoi{10.1038/s41550-019-0991-9}

\bibitem[{{Andrews} {et~al.}(2024){Andrews}, {Teague}, {Wirth}, {Huang}, \& {Zhu}}]{Andrews2024}
{Andrews}, S.~M., {Teague}, R., {Wirth}, C.~P., {Huang}, J., \& {Zhu}, Z. 2024, \apj, 970, 153, \dodoi{10.3847/1538-4357/ad5285}

\bibitem[{{Andrews} {et~al.}(2018){Andrews}, {Terrell}, {Tripathi}, {Ansdell}, {Williams}, \& {Wilner}}]{Andrews2018}
{Andrews}, S.~M., {Terrell}, M., {Tripathi}, A., {et~al.} 2018, \apj, 865, 157, \dodoi{10.3847/1538-4357/aadd9f}

\bibitem[{{Ansdell} {et~al.}(2017){Ansdell}, {Williams}, {Manara}, {Miotello}, {Facchini}, {van der Marel}, {Testi}, \& {van Dishoeck}}]{Ansdell2017}
{Ansdell}, M., {Williams}, J.~P., {Manara}, C.~F., {et~al.} 2017, \aj, 153, 240, \dodoi{10.3847/1538-3881/aa69c0}

\bibitem[{{Ansdell} {et~al.}(2016){Ansdell}, {Williams}, {van der Marel}, {Carpenter}, {Guidi}, {Hogerheijde}, {Mathews}, {Manara}, {Miotello}, {Natta}, {Oliveira}, {Tazzari}, {Testi}, {van Dishoeck}, \& {van Terwisga}}]{Ansdell2016}
{Ansdell}, M., {Williams}, J.~P., {van der Marel}, N., {et~al.} 2016, \apj, 828, 46, \dodoi{10.3847/0004-637X/828/1/46}

\bibitem[{{Appelgren} {et~al.}(2023){Appelgren}, {Lambrechts}, \& {van der Marel}}]{Appelgren2023}
{Appelgren}, J., {Lambrechts}, M., \& {van der Marel}, N. 2023, \aap, 673, A139, \dodoi{10.1051/0004-6361/202245252}

\bibitem[{{Arabhavi} {et~al.}(2024){Arabhavi}, {Kamp}, {Henning}, {van Dishoeck}, {Christiaens}, {Gasman}, {Perrin}, {G{\"u}del}, {Tabone}, {Kanwar}, {Waters}, {Pascucci}, {Samland}, {Perotti}, {Bettoni}, {Grant}, {Lagage}, {Ray}, {Vandenbussche}, {Absil}, {Argyriou}, {Barrado}, {Boccaletti}, {Bouwman}, {Caratti o Garatti}, {Glauser}, {Lahuis}, {Mueller}, {Olofsson}, {Pantin}, {Scheithauer}, {Morales-Calder{\'o}n}, {Franceschi}, {Jang}, {Pawellek}, {Rodgers-Lee}, {Schreiber}, {Schwarz}, {Temmink}, {Vlasblom}, {Wright}, {Colina}, \& {{\"O}stlin}}]{Arabhavi2024}
{Arabhavi}, A.~M., {Kamp}, I., {Henning}, T., {et~al.} 2024, Science, 384, 1086, \dodoi{10.1126/science.adi8147}

\bibitem[{{Arcangeli} {et~al.}(2018){Arcangeli}, {D{\'e}sert}, {Line}, {Bean}, {Parmentier}, {Stevenson}, {Kreidberg}, {Fortney}, {Mansfield}, \& {Showman}}]{arcangeli18}
{Arcangeli}, J., {D{\'e}sert}, J.-M., {Line}, M.~R., {et~al.} 2018, \apjl, 855, L30, \dodoi{10.3847/2041-8213/aab272}

\bibitem[{{Arulanantham} {et~al.}(2025){Arulanantham}, {Salyk}, {Pontoppidan}, {Banzatti}, {Zhang}, {{\"O}berg}, {Long}, {Carr}, {Najita}, {Pascucci}, {Jos{\'e} Colmenares}, {Xie}, {Huang}, {Green}, {Andrews}, {Blake}, {Bergin}, {Pinilla}, {Vioque}, {Dahl}, {Raul}, {Krijt}, \& {the JDISCS Collaboration}}]{Arulanantham2025}
{Arulanantham}, N., {Salyk}, C., {Pontoppidan}, K., {et~al.} 2025, arXiv e-prints, arXiv:2505.07562, \dodoi{10.48550/arXiv.2505.07562}

\bibitem[{{Ataiee} {et~al.}(2018){Ataiee}, {Baruteau}, {Alibert}, \& {Benz}}]{Ataiee2018}
{Ataiee}, S., {Baruteau}, C., {Alibert}, Y., \& {Benz}, W. 2018, \aap, 615, A110, \dodoi{10.1051/0004-6361/201732026}

\bibitem[{{August} {et~al.}(2023){August}, {Bean}, {Zhang}, {Lunine}, {Xue}, {Line}, \& {Smith}}]{august23}
{August}, P.~C., {Bean}, J.~L., {Zhang}, M., {et~al.} 2023, \apjl, 953, L24, \dodoi{10.3847/2041-8213/ace828}

\bibitem[{{Baehr} \& {Klahr}(2019)}]{baehr2019}
{Baehr}, H., \& {Klahr}, H. 2019, \apj, 881, 162, \dodoi{10.3847/1538-4357/ab2f85}

\bibitem[{{Banzatti} {et~al.}(2020){Banzatti}, {Pascucci}, {Bosman}, {Pinilla}, {Salyk}, {Herczeg}, {Pontoppidan}, {Vazquez}, {Watkins}, {Krijt}, {Hendler}, \& {Long}}]{Banzatti2020}
{Banzatti}, A., {Pascucci}, I., {Bosman}, A.~D., {et~al.} 2020, \apj, 903, 124, \dodoi{10.3847/1538-4357/abbc1a}

\bibitem[{{Banzatti} {et~al.}(2023){Banzatti}, {Pontoppidan}, {Carr}, {Jellison}, {Pascucci}, {Najita}, {Mu{\~n}oz-Romero}, {{\"O}berg}, {Kalyaan}, {Pinilla}, {Krijt}, {Long}, {Lambrechts}, {Rosotti}, {Herczeg}, {Salyk}, {Zhang}, {Bergin}, {Ballering}, {Meyer}, {Bruderer}, \& {Jdiscs Collaboration}}]{Banzatti2023}
{Banzatti}, A., {Pontoppidan}, K.~M., {Carr}, J.~S., {et~al.} 2023, \apjl, 957, L22, \dodoi{10.3847/2041-8213/acf5ec}

\bibitem[{{Banzatti} {et~al.}(2024){Banzatti}, {Salyk}, {Pontoppidan}, {Carr}, {Zhang}, {Arulanantham}, {Cleeves}, {Krijt}, {Najita}, {Oberg}, {Pascucci}, {Blake}, {Munoz-Romero}, {Bergin}, {Cieza}, {Pinilla}, {Long}, {Mallaney}, {Xie}, \& {the JDISCS collaboration}}]{Banzatti2024}
{Banzatti}, A., {Salyk}, C., {Pontoppidan}, K.~M., {et~al.} 2024, arXiv e-prints, arXiv:2409.16255, \dodoi{10.48550/arXiv.2409.16255}

\bibitem[{{Barat} {et~al.}(2024{\natexlab{a}}){Barat}, {D{\'e}sert}, {Vazan}, {Baeyens}, {Line}, {Fortney}, {David}, {Livingston}, {Jacobs}, {Panwar}, {Shivkumar}, {Todorov}, {Pino}, {Mraz}, \& {Petigura}}]{barat24a}
{Barat}, S., {D{\'e}sert}, J.-M., {Vazan}, A., {et~al.} 2024{\natexlab{a}}, Nature Astronomy, 8, 899, \dodoi{10.1038/s41550-024-02257-0}

\bibitem[{{Barat} {et~al.}(2024{\natexlab{b}}){Barat}, {D{\'e}sert}, {Goyal}, {Vazan}, {Kawashima}, {Fortney}, {Bean}, {Line}, {Panwar}, {Jacobs}, {Shivkumar}, {Sikora}, {Baeyens}, {Oklop{\v{c}}i{\'c}}, {David}, \& {Livingston}}]{barat24b}
{Barat}, S., {D{\'e}sert}, J.-M., {Goyal}, J.~M., {et~al.} 2024{\natexlab{b}}, \aap, 692, A198, \dodoi{10.1051/0004-6361/202451127}

\bibitem[{{Barber} {et~al.}(2024){Barber}, {Mann}, {Vanderburg}, {Krolikowski}, {Kraus}, {Ansdell}, {Pearce}, {Mace}, {Andrews}, {Boyle}, {Collins}, {De Furio}, {Dragomir}, {Espaillat}, {Feinstein}, {Fields}, {Jaffe}, {Lopez Murillo}, {Murgas}, {Newton}, {Palle}, {Sawczynec}, {Schwarz}, {Thao}, {Tofflemire}, {Watkins}, {Jenkins}, {Latham}, {Ricker}, {Seager}, {Vanderspek}, {Winn}, {Charbonneau}, {Essack}, {Rodriguez}, {Shporer}, {Twicken}, \& {Villase{\~n}or}}]{barber24}
{Barber}, M.~G., {Mann}, A.~W., {Vanderburg}, A., {et~al.} 2024, \nat, 635, 574, \dodoi{10.1038/s41586-024-08123-3}

\bibitem[{{Bardyn} {et~al.}(2017){Bardyn}, {Baklouti}, {Cottin}, {Fray}, {Briois}, {Paquette}, {Stenzel}, {Engrand}, {Fischer}, {Hornung}, {Isnard}, {Langevin}, {Lehto}, {Le Roy}, {Ligier}, {Merouane}, {Modica}, {Orthous-Daunay}, {Ryn{\"o}}, {Schulz}, {Sil{\'e}n}, {Thirkell}, {Varmuza}, {Zaprudin}, {Kissel}, \& {Hilchenbach}}]{Bardyn2017}
{Bardyn}, A., {Baklouti}, D., {Cottin}, H., {et~al.} 2017, \mnras, 469, S712, \dodoi{10.1093/mnras/stx2640}

\bibitem[{{Barenfeld} {et~al.}(2016){Barenfeld}, {Carpenter}, {Ricci}, \& {Isella}}]{Barenfeld2016}
{Barenfeld}, S.~A., {Carpenter}, J.~M., {Ricci}, L., \& {Isella}, A. 2016, \apj, 827, 142, \dodoi{10.3847/0004-637X/827/2/142}

\bibitem[{{Barrado} {et~al.}(2023){Barrado}, {Molli{\`e}re}, {Patapis}, {Min}, {Tremblin}, {Ardevol Martinez}, {Whiteford}, {Vasist}, {Argyriou}, {Samland}, {Lagage}, {Decin}, {Waters}, {Henning}, {Morales-Calder{\'o}n}, {Guedel}, {Vandenbussche}, {Absil}, {Baudoz}, {Boccaletti}, {Bouwman}, {Cossou}, {Coulais}, {Crouzet}, {Gastaud}, {Glasse}, {Glauser}, {Kamp}, {Kendrew}, {Krause}, {Lahuis}, {Mueller}, {Olofsson}, {Pye}, {Rouan}, {Royer}, {Scheithauer}, {Waldmann}, {Colina}, {van Dishoeck}, {Ray}, {{\"O}stlin}, \& {Wright}}]{Barrado2023}
{Barrado}, D., {Molli{\`e}re}, P., {Patapis}, P., {et~al.} 2023, \nat, 624, 263, \dodoi{10.1038/s41586-023-06813-y}

\bibitem[{{Barros} {et~al.}(2022){Barros}, {Akinsanmi}, {Bou{\'e}}, {Smith}, {Laskar}, {Ulmer-Moll}, {Lillo-Box}, {Queloz}, {Cameron}, {Sousa}, {Ehrenreich}, {Hooton}, {Bruno}, {Demory}, {Correia}, {Demangeon}, {Wilson}, {Bonfanti}, {Hoyer}, {Alibert}, {Alonso}, {Escud{\'e}}, {Barbato}, {B{\'a}rczy}, {Barrado}, {Baumjohann}, {Beck}, {Beck}, {Benz}, {Bergomi}, {Billot}, {Bonfils}, {Bouchy}, {Brandeker}, {Broeg}, {Cabrera}, {Cessa}, {Charnoz}, {Damme}, {Davies}, {Deleuil}, {Deline}, {Delrez}, {Erikson}, {Fortier}, {Fossati}, {Fridlund}, {Gandolfi}, {Mu{\~n}oz}, {Gillon}, {G{\"u}del}, {Isaak}, {Heng}, {Kiss}, {des Etangs}, {Lendl}, {Lovis}, {Magrin}, {Nascimbeni}, {Maxted}, {Olofsson}, {Ottensamer}, {Pagano}, {Pall{\'e}}, {Parviainen}, {Peter}, {Piotto}, {Pollacco}, {Ragazzoni}, {Rando}, {Rauer}, {Ribas}, {Santos}, {Scandariato}, {S{\'e}gransan}, {Simon}, {Steller}, {Szab{\'o}}, {Thomas}, {Udry}, {Ulmer}, {Van Grootel}, \& {Walton}}]{Barros_2022}
{Barros}, S.~C.~C., {Akinsanmi}, B., {Bou{\'e}}, G., {et~al.} 2022, \aap, 657, A52, \dodoi{10.1051/0004-6361/202142196}

\bibitem[{{Batalha} {et~al.}(2019){Batalha}, {Marley}, {Lewis}, \& {Fortney}}]{batalha19}
{Batalha}, N.~E., {Marley}, M.~S., {Lewis}, N.~K., \& {Fortney}, J.~J. 2019, \apj, 878, 70, \dodoi{10.3847/1538-4357/ab1b51}

\bibitem[{{Batygin} {et~al.}(2009){Batygin}, {Bodenheimer}, \& {Laughlin}}]{Batygin2009_firstkeccentricity}
{Batygin}, K., {Bodenheimer}, P., \& {Laughlin}, G. 2009, \apjl, 704, L49, \dodoi{10.1088/0004-637X/704/1/L49}

\bibitem[{{Baumeister} {et~al.}(2020){Baumeister}, {Padovan}, {Tosi}, {Montavon}, {Nettelmann}, {MacKenzie}, \& {Godolt}}]{Baumeister2020}
{Baumeister}, P., {Padovan}, S., {Tosi}, N., {et~al.} 2020, \apj, 889, 42, \dodoi{10.3847/1538-4357/ab5d32}

\bibitem[{Bazinet {et~al.}(2024)Bazinet, Pelletier, Benneke, Salinas, \& Mace}]{Bazinet24}
Bazinet, L., Pelletier, S., Benneke, B., Salinas, R., \& Mace, G.~N. 2024, The Astronomical Journal, 167, 206, \dodoi{10.3847/1538-3881/ad3071}

\bibitem[{{Bean} {et~al.}(2018){Bean}, {Stevenson}, {Batalha}, {Berta-Thompson}, {Kreidberg}, {Crouzet}, {Benneke}, {Line}, {Sing}, {Wakeford}, {Knutson}, {Kempton}, {D{\'e}sert}, {Crossfield}, {Batalha}, {de Wit}, {Parmentier}, {Harrington}, {Moses}, {Lopez-Morales}, {Alam}, {Blecic}, {Bruno}, {Carter}, {Chapman}, {Decin}, {Dragomir}, {Evans}, {Fortney}, {Fraine}, {Gao}, {Garc{\'\i}a Mu{\~n}oz}, {Gibson}, {Goyal}, {Heng}, {Hu}, {Kendrew}, {Kilpatrick}, {Krick}, {Lagage}, {Lendl}, {Louden}, {Madhusudhan}, {Mandell}, {Mansfield}, {May}, {Morello}, {Morley}, {Nikolov}, {Redfield}, {Roberts}, {Schlawin}, {Spake}, {Todorov}, {Tsiaras}, {Venot}, {Waalkes}, {Wheatley}, {Zellem}, {Angerhausen}, {Barrado}, {Carone}, {Casewell}, {Cubillos}, {Damiano}, {de Val-Borro}, {Drummond}, {Edwards}, {Endl}, {Espinoza}, {France}, {Gizis}, {Greene}, {Henning}, {Hong}, {Ingalls}, {Iro}, {Irwin}, {Kataria}, {Lahuis}, {Leconte}, {Lillo-Box}, {Lines}, {Lothringer}, {Mancini}, {Marchis}, {Mayne}, {Palle}, {Rauscher}, {Roudier},
  {Shkolnik}, {Southworth}, {Swain}, {Taylor}, {Teske}, {Tinetti}, {Tremblin}, {Tucker}, {van Boekel}, {Waldmann}, {Weaver}, \& {Zingales}}]{bean18}
{Bean}, J.~L., {Stevenson}, K.~B., {Batalha}, N.~M., {et~al.} 2018, \pasp, 130, 114402, \dodoi{10.1088/1538-3873/aadbf3}

\bibitem[{{Beatty} {et~al.}(2024){Beatty}, {Welbanks}, {Schlawin}, {Bell}, {Line}, {Murphy}, {Edelman}, {Greene}, {Fortney}, {Henry}, {Mukherjee}, {Ohno}, {Parmentier}, {Rauscher}, {Wiser}, \& {Arnold}}]{beatty24}
{Beatty}, T.~G., {Welbanks}, L., {Schlawin}, E., {et~al.} 2024, \apjl, 970, L10, \dodoi{10.3847/2041-8213/ad55e9}

\bibitem[{{Beiler} {et~al.}(2023){Beiler}, {Cushing}, {Kirkpatrick}, {Schneider}, {Mukherjee}, \& {Marley}}]{Beiler2023}
{Beiler}, S.~A., {Cushing}, M.~C., {Kirkpatrick}, J.~D., {et~al.} 2023, \apjl, 951, L48, \dodoi{10.3847/2041-8213/ace32c}

\bibitem[{{Bell} {et~al.}(2023){Bell}, {Welbanks}, {Schlawin}, {Line}, {Fortney}, {Greene}, {Ohno}, {Parmentier}, {Rauscher}, {Beatty}, {Mukherjee}, {Wiser}, {Boyer}, {Rieke}, \& {Stansberry}}]{bell23}
{Bell}, T.~J., {Welbanks}, L., {Schlawin}, E., {et~al.} 2023, \nat, 623, 709, \dodoi{10.1038/s41586-023-06687-0}

\bibitem[{{Ben-Yami} {et~al.}(2020){Ben-Yami}, {Madhusudhan}, {Cabot}, {Constantinou}, {Piette}, {Gandhi}, \& {Welbanks}}]{ben-yami:2020}
{Ben-Yami}, M., {Madhusudhan}, N., {Cabot}, S. H.~C., {et~al.} 2020, \apjl, 897, L5, \dodoi{10.3847/2041-8213/ab94aa}

\bibitem[{{Benneke} {et~al.}(2019){Benneke}, {Knutson}, {Lothringer}, {Crossfield}, {Moses}, {Morley}, {Kreidberg}, {Fulton}, {Dragomir}, {Howard}, {Wong}, {D{\'e}sert}, {McCullough}, {Kempton}, {Fortney}, {Gilliland}, {Deming}, \& {Kammer}}]{benneke19}
{Benneke}, B., {Knutson}, H.~A., {Lothringer}, J., {et~al.} 2019, Nature Astronomy, 3, 813, \dodoi{10.1038/s41550-019-0800-5}

\bibitem[{{Bergin} {et~al.}(2024{\natexlab{a}}){Bergin}, {Booth}, {Colmenares}, \& {Ilee}}]{Bergin2024b}
{Bergin}, E.~A., {Booth}, R.~A., {Colmenares}, M.~J., \& {Ilee}, J.~D. 2024{\natexlab{a}}, \apjl, 969, L21, \dodoi{10.3847/2041-8213/ad5839}

\bibitem[{{Bergin} {et~al.}(2016){Bergin}, {Du}, {Cleeves}, {Blake}, {Schwarz}, {Visser}, \& {Zhang}}]{Bergin2016}
{Bergin}, E.~A., {Du}, F., {Cleeves}, L.~I., {et~al.} 2016, \apj, 831, 101, \dodoi{10.3847/0004-637X/831/1/101}

\bibitem[{{Bergin} {et~al.}(2023){Bergin}, {Kempton}, {Hirschmann}, {Bastelberger}, {Teal}, {Blake}, {Ciesla}, \& {Li}}]{Bergin2023}
{Bergin}, E.~A., {Kempton}, E. M.~R., {Hirschmann}, M., {et~al.} 2023, \apjl, 949, L17, \dodoi{10.3847/2041-8213/acd377}

\bibitem[{{Bergin} {et~al.}(2013){Bergin}, {Cleeves}, {Gorti}, {Zhang}, {Blake}, {Green}, {Andrews}, {Evans}, {Henning}, {{\"O}berg}, {Pontoppidan}, {Qi}, {Salyk}, \& {van Dishoeck}}]{Bergin2013}
{Bergin}, E.~A., {Cleeves}, L.~I., {Gorti}, U., {et~al.} 2013, \nat, 493, 644, \dodoi{10.1038/nature11805}

\bibitem[{{Bergin} {et~al.}(2024{\natexlab{b}}){Bergin}, {Bosman}, {Teague}, {Calahan}, {Willacy}, {Cleeves}, {Schwarz}, {Zhang}, \& {Bruderer}}]{Bergin2024}
{Bergin}, E.~A., {Bosman}, A., {Teague}, R., {et~al.} 2024{\natexlab{b}}, \apj, 965, 147, \dodoi{10.3847/1538-4357/ad3443}

\bibitem[{{Bergner} {et~al.}(2022{\natexlab{a}}){Bergner}, {Burkhardt}, {{\"O}berg}, {Rice}, \& {Bergin}}]{Bergner2022b}
{Bergner}, J.~B., {Burkhardt}, A.~M., {{\"O}berg}, K.~I., {Rice}, T.~S., \& {Bergin}, E.~A. 2022{\natexlab{a}}, \apj, 927, 7, \dodoi{10.3847/1538-4357/ac47a2}

\bibitem[{{Bergner} {et~al.}(2022{\natexlab{b}}){Bergner}, {Shirley}, {J{\o}rgensen}, {McGuire}, {Aalto}, {Anderson}, {Chin}, {Gerin}, {Hartogh}, {Kim}, {Leisawitz}, {Najita}, {Schwarz}, {Tielens}, {Walker}, {Wilner}, \& {Wollack}}]{Bergner2022}
{Bergner}, J.~B., {Shirley}, Y.~L., {J{\o}rgensen}, J.~K., {et~al.} 2022{\natexlab{b}}, Frontiers in Astronomy and Space Sciences, 8, 246, \dodoi{10.3389/fspas.2021.793922}

\bibitem[{{Bergner} {et~al.}(2024){Bergner}, {Sturm}, {Piacentino}, {McClure}, {Oberg}, {Boogert}, {Dartois}, {Drozdovskaya}, {Fraser}, {Harsono}, {Ioppolo}, {Law}, {Lis}, {McGuire}, {Melnick}, {Noble}, {Palumbo}, {Pendleton}, {Perotti}, {Qasim}, {Rocha}, \& {van Dishoeck}}]{Bergner2024}
{Bergner}, J.~B., {Sturm}, J.~A., {Piacentino}, E.~L., {et~al.} 2024, arXiv e-prints, arXiv:2409.08117, \dodoi{10.48550/arXiv.2409.08117}

\bibitem[{{Bernab{\`o}} {et~al.}(2024){Bernab{\`o}}, {Csizmadia}, {Smith}, {Rauer}, {Hatzes}, {Esposito}, {Gandolfi}, \& {Cabrera}}]{Bernabo2024}
{Bernab{\`o}}, L.~M., {Csizmadia}, S., {Smith}, A.~M.~S., {et~al.} 2024, \aap, 684, A78, \dodoi{10.1051/0004-6361/202346852}

\bibitem[{{Bernab{\`o}} {et~al.}(2022){Bernab{\`o}}, {Turrini}, {Testi}, {Marzari}, \& {Polychroni}}]{Bernabo2022}
{Bernab{\`o}}, L.~M., {Turrini}, D., {Testi}, L., {Marzari}, F., \& {Polychroni}, D. 2022, \apjl, 927, L22, \dodoi{10.3847/2041-8213/ac574e}

\bibitem[{{Birkby}(2018)}]{birkby:2018}
{Birkby}, J.~L. 2018, arXiv e-prints.
\newblock \doarXiv{1806.04617}

\bibitem[{{Birkby} {et~al.}(2017){Birkby}, {de Kok}, {Brogi}, {Schwarz}, \& {Snellen}}]{Birkby17}
{Birkby}, J.~L., {de Kok}, R.~J., {Brogi}, M., {Schwarz}, H., \& {Snellen}, I.~A.~G. 2017, \aj, 153, 138, \dodoi{10.3847/1538-3881/aa5c87}

\bibitem[{{Birnstiel} {et~al.}(2010){Birnstiel}, {Ricci}, {Trotta}, {Dullemond}, {Natta}, {Testi}, {Dominik}, {Henning}, {Ormel}, \& {Zsom}}]{Birnstiel2010}
{Birnstiel}, T., {Ricci}, L., {Trotta}, F., {et~al.} 2010, \aap, 516, L14, \dodoi{10.1051/0004-6361/201014893}

\bibitem[{Bitsch {et~al.}(2018)Bitsch, Morbidelli, Johansen, Lega, Lambrechts, \& Crida}]{Bitsch2018}
Bitsch, B., Morbidelli, A., Johansen, A., {et~al.} 2018, Astronomy \& Astrophysics, 612, A30

\bibitem[{{Bitsch} {et~al.}(2022){Bitsch}, {Schneider}, \& {Kreidberg}}]{Bitsch2022}
{Bitsch}, B., {Schneider}, A.~D., \& {Kreidberg}, L. 2022, \aap, 665, A138, \dodoi{10.1051/0004-6361/202243345}

\bibitem[{{Blecic} {et~al.}(2017){Blecic}, {Dobbs-Dixon}, \& {Greene}}]{blecic17}
{Blecic}, J., {Dobbs-Dixon}, I., \& {Greene}, T. 2017, \apj, 848, 127, \dodoi{10.3847/1538-4357/aa8171}

\bibitem[{{Bloot} {et~al.}(2023){Bloot}, {Miguel}, {Bazot}, \& {Howard}}]{2023MNRAS.523.6282B}
{Bloot}, S., {Miguel}, Y., {Bazot}, M., \& {Howard}, S. 2023, \mnras, 523, 6282, \dodoi{10.1093/mnras/stad1873}

\bibitem[{{Bodenheimer} {et~al.}(2000){Bodenheimer}, {Hubickyj}, \& {Lissauer}}]{Bodenheimer2000}
{Bodenheimer}, P., {Hubickyj}, O., \& {Lissauer}, J.~J. 2000, \icarus, 143, 2, \dodoi{10.1006/icar.1999.6246}

\bibitem[{{Boehm} {et~al.}(2025){Boehm}, {Lewis}, {Fairman}, {Moran}, {Gasc{\'o}n}, {Wakeford}, {Alam}, {Alderson}, {Barstow}, {Batalha}, {Grant}, {L{\'o}pez-Morales}, {MacDonald}, {Marley}, \& {Ohno}}]{boehm25}
{Boehm}, V.~A., {Lewis}, N.~K., {Fairman}, C.~E., {et~al.} 2025, \aj, 169, 23, \dodoi{10.3847/1538-3881/ad8dde}

\bibitem[{{Boley} \& {Durisen}(2010)}]{boley2010}
{Boley}, A.~C., \& {Durisen}, R.~H. 2010, \apj, 724, 618, \dodoi{10.1088/0004-637X/724/1/618}

\bibitem[{{Boley} {et~al.}(2011){Boley}, {Helled}, \& {Payne}}]{Boley2011}
{Boley}, A.~C., {Helled}, R., \& {Payne}, M.~J. 2011, \apj, 735, 30, \dodoi{10.1088/0004-637X/735/1/30}

\bibitem[{{Bonnefoy} {et~al.}(2013){Bonnefoy}, {Boccaletti}, {Lagrange}, {Allard}, {Mordasini}, {Beust}, {Chauvin}, {Girard}, {Homeier}, {Apai}, {Lacour}, \& {Rouan}}]{Bonnefoy2013}
{Bonnefoy}, M., {Boccaletti}, A., {Lagrange}, A.~M., {et~al.} 2013, \aap, 555, A107, \dodoi{10.1051/0004-6361/201220838}

\bibitem[{{Bonnefoy} {et~al.}(2016){Bonnefoy}, {Zurlo}, {Baudino}, {Lucas}, {Mesa}, {Maire}, {Vigan}, {Galicher}, {Homeier}, {Marocco}, {Gratton}, {Chauvin}, {Allard}, {Desidera}, {Kasper}, {Moutou}, {Lagrange}, {Antichi}, {Baruffolo}, {Baudrand}, {Beuzit}, {Boccaletti}, {Cantalloube}, {Carbillet}, {Charton}, {Claudi}, {Costille}, {Dohlen}, {Dominik}, {Fantinel}, {Feautrier}, {Feldt}, {Fusco}, {Gigan}, {Girard}, {Gluck}, {Gry}, {Henning}, {Janson}, {Langlois}, {Madec}, {Magnard}, {Maurel}, {Mawet}, {Meyer}, {Milli}, {Moeller-Nilsson}, {Mouillet}, {Pavlov}, {Perret}, {Pujet}, {Quanz}, {Rochat}, {Rousset}, {Roux}, {Salasnich}, {Salter}, {Sauvage}, {Schmid}, {Sevin}, {Soenke}, {Stadler}, {Turatto}, {Udry}, {Vakili}, {Wahhaj}, \& {Wildi}}]{bonnefoy2016}
{Bonnefoy}, M., {Zurlo}, A., {Baudino}, J.~L., {et~al.} 2016, \aap, 587, A58, \dodoi{10.1051/0004-6361/201526906}

\bibitem[{{Booth} {et~al.}(2021){Booth}, {van der Marel}, {Leemker}, {van Dishoeck}, \& {Ohashi}}]{BoothA2021}
{Booth}, A.~S., {van der Marel}, N., {Leemker}, M., {van Dishoeck}, E.~F., \& {Ohashi}, S. 2021, \aap, 651, L6, \dodoi{10.1051/0004-6361/202141057}

\bibitem[{{Booth} \& {Clarke}(2016)}]{booth2016}
{Booth}, R.~A., \& {Clarke}, C.~J. 2016, \mnras, 458, 2676, \dodoi{10.1093/mnras/stw488}

\bibitem[{{Booth} \& {Clarke}(2019)}]{booth2019gi}
---. 2019, \mnras, 483, 3718, \dodoi{10.1093/mnras/sty3340}

\bibitem[{{Booth} {et~al.}(2017){Booth}, {Clarke}, {Madhusudhan}, \& {Ilee}}]{Booth2017}
{Booth}, R.~A., {Clarke}, C.~J., {Madhusudhan}, N., \& {Ilee}, J.~D. 2017, \mnras, 469, 3994, \dodoi{10.1093/mnras/stx1103}

\bibitem[{{Booth} \& {Ilee}(2019)}]{Booth2019}
{Booth}, R.~A., \& {Ilee}, J.~D. 2019, \mnras, 487, 3998, \dodoi{10.1093/mnras/stz1488}

\bibitem[{{Borsa} {et~al.}(2021){Borsa}, {Fossati}, {Koskinen}, {Young}, \& {Shulyak}}]{Borsa21}
{Borsa}, F., {Fossati}, L., {Koskinen}, T., {Young}, M.~E., \& {Shulyak}, D. 2021, Nature Astronomy, 6, 226, \dodoi{10.1038/s41550-021-01544-4}

\bibitem[{{Bosman} {et~al.}(2021{\natexlab{a}}){Bosman}, {Alarc{\'o}n}, {Zhang}, \& {Bergin}}]{Bosman2021b}
{Bosman}, A.~D., {Alarc{\'o}n}, F., {Zhang}, K., \& {Bergin}, E.~A. 2021{\natexlab{a}}, \apj, 910, 3, \dodoi{10.3847/1538-4357/abe127}

\bibitem[{{Bosman} {et~al.}(2019){Bosman}, {Cridland}, \& {Miguel}}]{Bosman2019}
{Bosman}, A.~D., {Cridland}, A.~J., \& {Miguel}, Y. 2019, \aap, 632, L11, \dodoi{10.1051/0004-6361/201936827}

\bibitem[{{Bosman} {et~al.}(2021{\natexlab{b}}){Bosman}, {Alarc{\'o}n}, {Bergin}, {Zhang}, {van't Hoff}, {{\"O}berg}, {Guzm{\'a}n}, {Walsh}, {Aikawa}, {Andrews}, {Bergner}, {Booth}, {Cataldi}, {Cleeves}, {Czekala}, {Furuya}, {Huang}, {Ilee}, {Law}, {Le Gal}, {Liu}, {Long}, {Loomis}, {M{\'e}nard}, {Nomura}, {Qi}, {Schwarz}, {Teague}, {Tsukagoshi}, {Yamato}, \& {Wilner}}]{Bosman2021}
{Bosman}, A.~D., {Alarc{\'o}n}, F., {Bergin}, E.~A., {et~al.} 2021{\natexlab{b}}, \apjs, 257, 7, \dodoi{10.3847/1538-4365/ac1435}

\bibitem[{{Brogi} \& {Line}(2019)}]{Brogi19}
{Brogi}, M., \& {Line}, M.~R. 2019, \aj, 157, 114, \dodoi{10.3847/1538-3881/aaffd3}

\bibitem[{{Brouwers} {et~al.}(2018){Brouwers}, {Vazan}, \& {Ormel}}]{Brouwers2018}
{Brouwers}, M.~G., {Vazan}, A., \& {Ormel}, C.~W. 2018, \aap, 611, A65, \dodoi{10.1051/0004-6361/201731824}

\bibitem[{{Br{\"u}gger} {et~al.}(2020){Br{\"u}gger}, {Burn}, {Coleman}, {Alibert}, \& {Benz}}]{Brugger2020}
{Br{\"u}gger}, N., {Burn}, R., {Coleman}, G.~A.~L., {Alibert}, Y., \& {Benz}, W. 2020, \aap, 640, A21, \dodoi{10.1051/0004-6361/202038042}

\bibitem[{{Buhler} {et~al.}(2016){Buhler}, {Knutson}, {Batygin}, {Fulton}, {Fortney}, {Burrows}, \& {Wong}}]{Buhler_2016}
{Buhler}, P.~B., {Knutson}, H.~A., {Batygin}, K., {et~al.} 2016, \apj, 821, 26, \dodoi{10.3847/0004-637X/821/1/26}

\bibitem[{{Burgasser} {et~al.}(2002){Burgasser}, {Marley}, {Ackerman}, {Saumon}, {Lodders}, {Dahn}, {Harris}, \& {Kirkpatrick}}]{burgasser02}
{Burgasser}, A.~J., {Marley}, M.~S., {Ackerman}, A.~S., {et~al.} 2002, \apjl, 571, L151, \dodoi{10.1086/341343}

\bibitem[{{Burn} \& {Mordasini}(2025)}]{Burn2025}
{Burn}, R., \& {Mordasini}, C. 2025, Planetary Population Synthesis (Cham: Springer Nature Switzerland), 1--60, \dodoi{10.1007/978-3-319-30648-3_143-2}

\bibitem[{{Burrows} \& {Liebert}(1993)}]{burrows93}
{Burrows}, A., \& {Liebert}, J. 1993, Reviews of Modern Physics, 65, 301, \dodoi{10.1103/RevModPhys.65.301}

\bibitem[{{Burrows} \& {Sharp}(1999)}]{burrows99}
{Burrows}, A., \& {Sharp}, C.~M. 1999, \apj, 512, 843, \dodoi{10.1086/306811}

\bibitem[{Cabot {et~al.}(2020)Cabot, Madhusudhan, Welbanks, Piette, \& Gandhi}]{cabot_detection_2020}
Cabot, S. H.~C., Madhusudhan, N., Welbanks, L., Piette, A., \& Gandhi, S. 2020, MNRAS, 494, 363, \dodoi{10.1093/mnras/staa748}

\bibitem[{{Cacciapuoti} {et~al.}(2024){Cacciapuoti}, {Macias}, {Gupta}, {Testi}, {Miotello}, {Espaillat}, {K{\"u}ffmeier}, {van Terwisga}, {Tobin}, {Grant}, {Manara}, {Segura-Cox}, {Wendeborn}, {Klessen}, {Maury}, {Lebreuilly}, {Hennebelle}, \& {Molinari}}]{Cacciapuoti2024}
{Cacciapuoti}, L., {Macias}, E., {Gupta}, A., {et~al.} 2024, \aap, 682, A61, \dodoi{10.1051/0004-6361/202347486}

\bibitem[{{Caldas} {et~al.}(2019){Caldas}, {Leconte}, {Selsis}, {Waldmann}, {Bord{\'e}}, {Rocchetto}, \& {Charnay}}]{caldas19}
{Caldas}, A., {Leconte}, J., {Selsis}, F., {et~al.} 2019, \aap, 623, A161, \dodoi{10.1051/0004-6361/201834384}

\bibitem[{{Carleo} {et~al.}(2022){Carleo}, {Giacobbe}, {Guilluy}, {Cubillos}, {Bonomo}, {Sozzetti}, {Brogi}, {Gandhi}, {Fossati}, {Turrini}, {Biazzo}, {Borsa}, {Lanza}, {Malavolta}, {Maggio}, {Mancini}, {Micela}, {Pino}, {Poretti}, {Rainer}, {Scandariato}, {Schisano}, {Andreuzzi}, {Bignamini}, {Cosentino}, {Fiorenzano}, {Harutyunyan}, {Molinari}, {Pedani}, {Redfield}, \& {Stoev}}]{Carleo2022}
{Carleo}, I., {Giacobbe}, P., {Guilluy}, G., {et~al.} 2022, \aj, 164, 101, \dodoi{10.3847/1538-3881/ac80bf}

\bibitem[{{Carrasco-Gonz{\'a}lez} {et~al.}(2019){Carrasco-Gonz{\'a}lez}, {Sierra}, {Flock}, {Zhu}, {Henning}, {Chandler}, {Galv{\'a}n-Madrid}, {Mac{\'\i}as}, {Anglada}, {Linz}, {Osorio}, {Rodr{\'\i}guez}, {Testi}, {Torrelles}, {P{\'e}rez}, \& {Liu}}]{Carrasco-Gonzalez2019}
{Carrasco-Gonz{\'a}lez}, C., {Sierra}, A., {Flock}, M., {et~al.} 2019, \apj, 883, 71, \dodoi{10.3847/1538-4357/ab3d33}

\bibitem[{{Carrera} \& {Simon}(2022)}]{carrera2022}
{Carrera}, D., \& {Simon}, J.~B. 2022, \apjl, 933, L10, \dodoi{10.3847/2041-8213/ac6b3e}

\bibitem[{{Carrera} {et~al.}(2021){Carrera}, {Simon}, {Li}, {Kretke}, \& {Klahr}}]{carrera2021}
{Carrera}, D., {Simon}, J.~B., {Li}, R., {Kretke}, K.~A., \& {Klahr}, H. 2021, \aj, 161, 96, \dodoi{10.3847/1538-3881/abd4d9}

\bibitem[{{Carter} {et~al.}(2024){Carter}, {May}, {Espinoza}, {Welbanks}, {Ahrer}, {Alderson}, {Brahm}, {Feinstein}, {Grant}, {Line}, {Morello}, {O'Steen}, {Radica}, {Rustamkulov}, {Stevenson}, {Turner}, {Alam}, {Anderson}, {Batalha}, {Battley}, {Bayliss}, {Bean}, {Benneke}, {Berta-Thompson}, {Brande}, {Bryant}, {Burleigh}, {Coulombe}, {Crossfield}, {Damiano}, {D{\'e}sert}, {Flagg}, {Gill}, {Inglis}, {Kirk}, {Knutson}, {Kreidberg}, {L{\'o}pez Morales}, {Mansfield}, {Moran}, {Murray}, {Nixon}, {Petit dit de la Roche}, {Rackham}, {Schlawin}, {Sing}, {Wakeford}, {Wallack}, {Wheatley}, {Zieba}, {Aggarwal}, {Barstow}, {Bell}, {Blecic}, {Caceres}, {Crouzet}, {Cubillos}, {Daylan}, {de Val-Borro}, {Decin}, {Fortney}, {Gibson}, {Heng}, {Hu}, {Kempton}, {Lagage}, {Lothringer}, {Lustig-Yaeger}, {Mancini}, {Mayne}, {Mayorga}, {Molaverdikhani}, {Nasedkin}, {Ohno}, {Parmentier}, {Powell}, {Redfield}, {Roy}, {Taylor}, \& {Zhang}}]{carter24}
{Carter}, A.~L., {May}, E.~M., {Espinoza}, N., {et~al.} 2024, Nature Astronomy, \dodoi{10.1038/s41550-024-02292-x}

\bibitem[{{Cavali{\'e}} {et~al.}(2013){Cavali{\'e}}, {Feuchtgruber}, {Lellouch}, {de Val-Borro}, {Jarchow}, {Moreno}, {Hartogh}, {Orton}, {Greathouse}, {Billebaud}, {Dobrijevic}, {Lara}, {Gonz{\'a}lez}, \& {Sagawa}}]{Cavalie2013}
{Cavali{\'e}}, T., {Feuchtgruber}, H., {Lellouch}, E., {et~al.} 2013, \aap, 553, A21, \dodoi{10.1051/0004-6361/201220797}

\bibitem[{{Chachan} {et~al.}(2023){Chachan}, {Knutson}, {Lothringer}, \& {Blake}}]{Chachan2023}
{Chachan}, Y., {Knutson}, H.~A., {Lothringer}, J., \& {Blake}, G.~A. 2023, \apj, 943, 112, \dodoi{10.3847/1538-4357/aca614}

\bibitem[{{Changeat} {et~al.}(2022){Changeat}, {Edwards}, {Al-Refaie}, {Tsiaras}, {Skinner}, {Cho}, {Yip}, {Anisman}, {Ikoma}, {Bieger}, {Venot}, {Shibata}, {Waldmann}, \& {Tinetti}}]{changeat22}
{Changeat}, Q., {Edwards}, B., {Al-Refaie}, A.~F., {et~al.} 2022, \apjs, 260, 3, \dodoi{10.3847/1538-4365/ac5cc2}

\bibitem[{{Charnay} {et~al.}(2018){Charnay}, {B{\'e}zard}, {Baudino}, {Bonnefoy}, {Boccaletti}, \& {Galicher}}]{Charnay2018}
{Charnay}, B., {B{\'e}zard}, B., {Baudino}, J.~L., {et~al.} 2018, \apj, 854, 172, \dodoi{10.3847/1538-4357/aaac7d}

\bibitem[{{Chauvin} {et~al.}(2017){Chauvin}, {Desidera}, {Lagrange}, {Vigan}, {Gratton}, {Langlois}, {Bonnefoy}, {Beuzit}, {Feldt}, {Mouillet}, {Meyer}, {Cheetham}, {Biller}, {Boccaletti}, {D'Orazi}, {Galicher}, {Hagelberg}, {Maire}, {Mesa}, {Olofsson}, {Samland}, {Schmidt}, {Sissa}, {Bonavita}, {Charnay}, {Cudel}, {Daemgen}, {Delorme}, {Janin-Potiron}, {Janson}, {Keppler}, {Le Coroller}, {Ligi}, {Marleau}, {Messina}, {Molli{\`e}re}, {Mordasini}, {M{\"u}ller}, {Peretti}, {Perrot}, {Rodet}, {Rouan}, {Zurlo}, {Dominik}, {Henning}, {Menard}, {Schmid}, {Turatto}, {Udry}, {Vakili}, {Abe}, {Antichi}, {Baruffolo}, {Baudoz}, {Baudrand}, {Blanchard}, {Bazzon}, {Buey}, {Carbillet}, {Carle}, {Charton}, {Cascone}, {Claudi}, {Costille}, {Deboulbe}, {De Caprio}, {Dohlen}, {Fantinel}, {Feautrier}, {Fusco}, {Gigan}, {Giro}, {Gisler}, {Gluck}, {Hubin}, {Hugot}, {Jaquet}, {Kasper}, {Madec}, {Magnard}, {Martinez}, {Maurel}, {Le Mignant}, {M{\"o}ller-Nilsson}, {Llored}, {Moulin}, {Orign{\'e}}, {Pavlov}, {Perret}, {Petit},
  {Pragt}, {Puget}, {Rabou}, {Ramos}, {Rigal}, {Rochat}, {Roelfsema}, {Rousset}, {Roux}, {Salasnich}, {Sauvage}, {Sevin}, {Soenke}, {Stadler}, {Suarez}, {Weber}, {Wildi}, {Antoniucci}, {Augereau}, {Baudino}, {Brandner}, {Engler}, {Girard}, {Gry}, {Kral}, {Kopytova}, {Lagadec}, {Milli}, {Moutou}, {Schlieder}, {Szul{\'a}gyi}, {Thalmann}, \& {Wahhaj}}]{Chauvin2017}
{Chauvin}, G., {Desidera}, S., {Lagrange}, A.~M., {et~al.} 2017, \aap, 605, L9, \dodoi{10.1051/0004-6361/201731152}

\bibitem[{{Chilcote} {et~al.}(2015){Chilcote}, {Barman}, {Fitzgerald}, {Graham}, {Larkin}, {Macintosh}, {Bauman}, {Burrows}, {Cardwell}, {De Rosa}, {Dillon}, {Doyon}, {Dunn}, {Erikson}, {Gavel}, {Goodsell}, {Hartung}, {Hibon}, {Ingraham}, {Kalas}, {Konopacky}, {Maire}, {Marchis}, {Marley}, {Marois}, {Millar-Blanchaer}, {Morzinski}, {Norton}, {Oppenheimer}, {Palmer}, {Patience}, {Perrin}, {Poyneer}, {Pueyo}, {Rantakyr{\"o}}, {Sadakuni}, {Saddlemyer}, {Savransky}, {Serio}, {Sivaramakrishnan}, {Song}, {Soummer}, {Thomas}, {Wallace}, {Wiktorowicz}, \& {Wolff}}]{Chilcote2015}
{Chilcote}, J., {Barman}, T., {Fitzgerald}, M.~P., {et~al.} 2015, \apjl, 798, L3, \dodoi{10.1088/2041-8205/798/1/L3}

\bibitem[{{Chilcote} {et~al.}(2017){Chilcote}, {Pueyo}, {De Rosa}, {Vargas}, {Macintosh}, {Bailey}, {Barman}, {Bauman}, {Bruzzone}, {Bulger}, {Burrows}, {Cardwell}, {Chen}, {Cotten}, {Dillon}, {Doyon}, {Draper}, {Duch{\^e}ne}, {Dunn}, {Erikson}, {Fitzgerald}, {Follette}, {Gavel}, {Goodsell}, {Graham}, {Greenbaum}, {Hartung}, {Hibon}, {Hung}, {Ingraham}, {Kalas}, {Konopacky}, {Larkin}, {Maire}, {Marchis}, {Marley}, {Marois}, {Metchev}, {Millar-Blanchaer}, {Morzinski}, {Nielsen}, {Norton}, {Oppenheimer}, {Palmer}, {Patience}, {Perrin}, {Poyneer}, {Rajan}, {Rameau}, {Rantakyr{\"o}}, {Sadakuni}, {Saddlemyer}, {Savransky}, {Schneider}, {Serio}, {Sivaramakrishnan}, {Song}, {Soummer}, {Thomas}, {Wallace}, {Wang}, {Ward-Duong}, {Wiktorowicz}, \& {Wolff}}]{Chilcote2017}
{Chilcote}, J., {Pueyo}, L., {De Rosa}, R.~J., {et~al.} 2017, \aj, 153, 182, \dodoi{10.3847/1538-3881/aa63e9}

\bibitem[{{Chomez} {et~al.}(2025){Chomez}, {Delorme}, {Lagrange}, {Gratton}, {Flasseur}, {Chauvin}, {Langlois}, {Mazoyer}, {Zurlo}, {Desidera}, {Mesa}, {Bonnefoy}, {Feldt}, {Hagelberg}, {Meyer}, {Vigan}, {Ginski}, {Kenworthy}, {Albert}, {Bergeon}, {Beuzit}, {Biller}, {Bhowmik}, {Boccaletti}, {Bonavita}, {Brandner}, {Cantalloube}, {Cheetham}, {D'Orazi}, {Dominik}, {Fontanive}, {Galicher}, {Henning}, {Janson}, {Kral}, {Lagadec}, {Lazzoni}, {Le Coroller}, {Ligi}, {Maire}, {Marleau}, {Menard}, {Messina}, {Meunier}, {Mordasini}, {Moutou}, {M{\"u}ller}, {Perrot}, {Samland}, {Schmid}, {Schmidt}, {Squicciarini}, {Sissa}, {Turatto}, {Udry}, {Abe}, {Antichi}, {Asensio-Torres}, {Baruffolo}, {Baudoz}, {Baudrand}, {Bazzon}, {Blanchard}, {Bohn}, {Brown Sevilla}, {Carbillet}, {Carle}, {Cascone}, {Charton}, {Claudi}, {Costille}, {De Caprio}, {Delboulb{\'e}}, {Dohlen}, {Engler}, {Fantinel}, {Feautrier}, {Fusco}, {Gigan}, {Girard}, {Giro}, {Gisler}, {Gluck}, {Gry}, {Hubin}, {Hugot}, {Jaquet}, {Kasper}, {Le Mignant}, {Llored},
  {Madec}, {Magnard}, {Martinez}, {Maurel}, {M{\"o}ller-Nilsson}, {Mouillet}, {Moulin}, {Orign{\'e}}, {Pavlov}, {Perret}, {Petit}, {Pragt}, {Puget}, {Rabou}, {Ramos}, {Rickman}, {Rigal}, {Rochat}, {Roelfsema}, {Rousset}, {Roux}, {Salasnich}, {Sauvage}, {Sevin}, {Soenke}, {Stadler}, {Suarez}, {Wahhaj}, {Weber}, \& {Wildi}}]{chomez25}
{Chomez}, A., {Delorme}, P., {Lagrange}, A.~M., {et~al.} 2025, \aap, 697, A99, \dodoi{10.1051/0004-6361/202451751}

\bibitem[{{Chung} {et~al.}(2024){Chung}, {Andrews}, {Gurwell}, {Wright}, {Long}, {Xu}, \& {Liu}}]{Chung2024}
{Chung}, C.-Y., {Andrews}, S.~M., {Gurwell}, M.~A., {et~al.} 2024, arXiv e-prints, arXiv:2405.19867, \dodoi{10.48550/arXiv.2405.19867}

\bibitem[{{Cieza} {et~al.}(2019){Cieza}, {Ru{\'\i}z-Rodr{\'\i}guez}, {Hales}, {Casassus}, {P{\'e}rez}, {Gonzalez-Ruilova}, {C{\'a}novas}, {Williams}, {Zurlo}, {Ansdell}, {Avenhaus}, {Bayo}, {Bertrang}, {Christiaens}, {Dent}, {Ferrero}, {Gamen}, {Olofsson}, {Orcajo}, {Pe{\~n}a Ram{\'\i}rez}, {Principe}, {Schreiber}, \& {van der Plas}}]{Cieza2019}
{Cieza}, L.~A., {Ru{\'\i}z-Rodr{\'\i}guez}, D., {Hales}, A., {et~al.} 2019, \mnras, 482, 698, \dodoi{10.1093/mnras/sty2653}

\bibitem[{{Clarke}(2009)}]{clarke2009}
{Clarke}, C.~J. 2009, \mnras, 396, 1066, \dodoi{10.1111/j.1365-2966.2009.14774.x}

\bibitem[{{Cleeves} {et~al.}(2018){Cleeves}, {{\"O}berg}, {Wilner}, {Huang}, {Loomis}, {Andrews}, \& {Guzman}}]{Cleeves2018}
{Cleeves}, L.~I., {{\"O}berg}, K.~I., {Wilner}, D.~J., {et~al.} 2018, \apj, 865, 155, \dodoi{10.3847/1538-4357/aade96}

\bibitem[{{Collings} {et~al.}(2003){Collings}, {Dever}, {Fraser}, {McCoustra}, \& {Williams}}]{Collings2003}
{Collings}, M.~P., {Dever}, J.~W., {Fraser}, H.~J., {McCoustra}, M.~R.~S., \& {Williams}, D.~A. 2003, \apj, 583, 1058, \dodoi{10.1086/345389}

\bibitem[{{Constantinou} {et~al.}(2023){Constantinou}, {Madhusudhan}, \& {Gandhi}}]{constantinou23}
{Constantinou}, S., {Madhusudhan}, N., \& {Gandhi}, S. 2023, \apjl, 943, L10, \dodoi{10.3847/2041-8213/acaead}

\bibitem[{{Coria} {et~al.}(2024){Coria}, {Hejazi}, {Crossfield}, \& {Rhem}}]{coria:2024}
{Coria}, D.~R., {Hejazi}, N., {Crossfield}, I. J.~M., \& {Rhem}, M. 2024, \apj, 974, 151, \dodoi{10.3847/1538-4357/ad7020}

\bibitem[{{Cridland} {et~al.}(2019){Cridland}, {Eistrup}, \& {van Dishoeck}}]{Cridland2019}
{Cridland}, A.~J., {Eistrup}, C., \& {van Dishoeck}, E.~F. 2019, \aap, 627, A127, \dodoi{10.1051/0004-6361/201834378}

\bibitem[{{Cridland} {et~al.}(2022){Cridland}, {Rosotti}, {Tabone}, {Tychoniec}, {McClure}, {Nazari}, \& {van Dishoeck}}]{Cridland2022}
{Cridland}, A.~J., {Rosotti}, G.~P., {Tabone}, B., {et~al.} 2022, \aap, 662, A90, \dodoi{10.1051/0004-6361/202142207}

\bibitem[{{Cridland} {et~al.}(2020){Cridland}, {van Dishoeck}, {Alessi}, \& {Pudritz}}]{Cridland2020}
{Cridland}, A.~J., {van Dishoeck}, E.~F., {Alessi}, M., \& {Pudritz}, R.~E. 2020, \aap, 642, A229, \dodoi{10.1051/0004-6361/202038767}

\bibitem[{{Crossfield}(2023)}]{crossfield23}
{Crossfield}, I. J.~M. 2023, \apjl, 952, L18, \dodoi{10.3847/2041-8213/ace35f}

\bibitem[{{Csizmadia, Sz.} {et~al.}(2019){Csizmadia, Sz.}, {Hellard, H.}, \& {Smith, A. M. S.}}]{love_number_wasp18b}
{Csizmadia, Sz.}, {Hellard, H.}, \& {Smith, A. M. S.} 2019, A\&A, 623, A45, \dodoi{10.1051/0004-6361/201834376}

\bibitem[{{Currie} {et~al.}(2023){Currie}, {Biller}, {Lagrange}, {Marois}, {Guyon}, {Nielsen}, {Bonnefoy}, \& {De Rosa}}]{Currie2023}
{Currie}, T., {Biller}, B., {Lagrange}, A., {et~al.} 2023, in Astronomical Society of the Pacific Conference Series, Vol. 534, Protostars and Planets VII, ed. S.~{Inutsuka}, Y.~{Aikawa}, T.~{Muto}, K.~{Tomida}, \& M.~{Tamura}, 799, \dodoi{10.48550/arXiv.2205.05696}

\bibitem[{{Cushing} {et~al.}(2005){Cushing}, {Rayner}, \& {Vacca}}]{cushing05}
{Cushing}, M.~C., {Rayner}, J.~T., \& {Vacca}, W.~D. 2005, \apj, 623, 1115, \dodoi{10.1086/428040}

\bibitem[{{Cuzzi} \& {Zahnle}(2004)}]{Cuzzi2004}
{Cuzzi}, J.~N., \& {Zahnle}, K.~J. 2004, \apj, 614, 490, \dodoi{10.1086/423611}

\bibitem[{{D'Alessio} {et~al.}(2006){D'Alessio}, {Calvet}, {Hartmann}, {Franco-Hern{\'a}ndez}, \& {Serv{\'\i}n}}]{DAlessio2006}
{D'Alessio}, P., {Calvet}, N., {Hartmann}, L., {Franco-Hern{\'a}ndez}, R., \& {Serv{\'\i}n}, H. 2006, \apj, 638, 314, \dodoi{10.1086/498861}

\bibitem[{{Danti} {et~al.}(2023){Danti}, {Bitsch}, \& {Mah}}]{Danti2023}
{Danti}, C., {Bitsch}, B., \& {Mah}, J. 2023, \aap, 679, L7, \dodoi{10.1051/0004-6361/202347501}

\bibitem[{{Dawson} \& {Johnson}(2018)}]{dawson2018}
{Dawson}, R.~I., \& {Johnson}, J.~A. 2018, \araa, 56, 175, \dodoi{10.1146/annurev-astro-081817-051853}

\bibitem[{{de Gruijter} {et~al.}(2025){de Gruijter}, {Tsai}, {Min}, {Waters}, {Konings}, \& {Decin}}]{deGruijter25}
{de Gruijter}, W., {Tsai}, S.-M., {Min}, M., {et~al.} 2025, \aap, 693, A132, \dodoi{10.1051/0004-6361/202450598}

\bibitem[{{de Sanctis} {et~al.}(2015){de Sanctis}, {Ammannito}, {Raponi}, {Marchi}, {McCord}, {McSween}, {Capaccioni}, {Capria}, {Carrozzo}, {Ciarniello}, {Longobardo}, {Tosi}, {Fonte}, {Formisano}, {Frigeri}, {Giardino}, {Magni}, {Palomba}, {Turrini}, {Zambon}, {Combe}, {Feldman}, {Jaumann}, {McFadden}, {Pieters}, {Prettyman}, {Toplis}, {Raymond}, \& {Russell}}]{DeSanctis2015}
{de Sanctis}, M.~C., {Ammannito}, E., {Raponi}, A., {et~al.} 2015, \nat, 528, 241, \dodoi{10.1038/nature16172}

\bibitem[{{de Wit} \& {Seager}(2013)}]{dewit13}
{de Wit}, J., \& {Seager}, S. 2013, Science, 342, 1473, \dodoi{10.1126/science.1245450}

\bibitem[{{DellaGiustina} {et~al.}(2021){DellaGiustina}, {Kaplan}, {Simon}, {Bottke}, {Avdellidou}, {Delbo}, {Ballouz}, {Golish}, {Walsh}, {Popescu}, {Campins}, {Barucci}, {Poggiali}, {Daly}, {Le Corre}, {Hamilton}, {Porter}, {Jawin}, {McCoy}, {Connolly}, {Garcia}, {Tatsumi}, {de Leon}, {Licandro}, {Fornasier}, {Daly}, {Al Asad}, {Philpott}, {Seabrook}, {Barnouin}, {Clark}, {Nolan}, {Howell}, {Binzel}, {Rizk}, {Reuter}, \& {Lauretta}}]{DellaGiustina2021}
{DellaGiustina}, D.~N., {Kaplan}, H.~H., {Simon}, A.~A., {et~al.} 2021, Nature Astronomy, 5, 31, \dodoi{10.1038/s41550-020-1195-z}

\bibitem[{{Delussu} {et~al.}(2024){Delussu}, {Birnstiel}, {Miotello}, {Pinilla}, {Rosotti}, \& {Andrews}}]{Delussu2024}
{Delussu}, L., {Birnstiel}, T., {Miotello}, A., {et~al.} 2024, arXiv e-prints, arXiv:2405.14501, \dodoi{10.48550/arXiv.2405.14501}

\bibitem[{{Deming} {et~al.}(2024){Deming}, {Fu}, {Bouwman}, {Dicken}, {Espinoza}, {Glasse}, {Greene}, {Kendrew}, {Law}, {Lustig-Yaeger}, {Garcia Marin}, \& {Schlawin}}]{deming24}
{Deming}, D., {Fu}, G., {Bouwman}, J., {et~al.} 2024, \pasp, 136, 084402, \dodoi{10.1088/1538-3873/ad6692}

\bibitem[{{Dent} {et~al.}(2019){Dent}, {Pinte}, {Cortes}, {M{\'e}nard}, {Hales}, {Fomalont}, \& {de Gregorio-Monsalvo}}]{Dent2019}
{Dent}, W.~R.~F., {Pinte}, C., {Cortes}, P.~C., {et~al.} 2019, \mnras, 482, L29, \dodoi{10.1093/mnrasl/sly181}

\bibitem[{{Dodson-Robinson} {et~al.}(2009){Dodson-Robinson}, {Willacy}, {Bodenheimer}, {Turner}, \& {Beichman}}]{Dodson2009}
{Dodson-Robinson}, S.~E., {Willacy}, K., {Bodenheimer}, P., {Turner}, N.~J., \& {Beichman}, C.~A. 2009, \icarus, 200, 672, \dodoi{10.1016/j.icarus.2008.11.023}

\bibitem[{{Dr{\k{a}}{\.z}kowska} {et~al.}(2023){Dr{\k{a}}{\.z}kowska}, {Bitsch}, {Lambrechts}, {Mulders}, {Harsono}, {Vazan}, {Liu}, {Ormel}, {Kretke}, \& {Morbidelli}}]{Drazkowska2023}
{Dr{\k{a}}{\.z}kowska}, J., {Bitsch}, B., {Lambrechts}, M., {et~al.} 2023, in Astronomical Society of the Pacific Conference Series, Vol. 534, Protostars and Planets VII, ed. S.~{Inutsuka}, Y.~{Aikawa}, T.~{Muto}, K.~{Tomida}, \& M.~{Tamura}, 717, \dodoi{10.48550/arXiv.2203.09759}

\bibitem[{{Du} {et~al.}(2015){Du}, {Bergin}, \& {Hogerheijde}}]{Du2015}
{Du}, F., {Bergin}, E.~A., \& {Hogerheijde}, M.~R. 2015, \apjl, 807, L32, \dodoi{10.1088/2041-8205/807/2/L32}

\bibitem[{{Dubrulle} {et~al.}(1995){Dubrulle}, {Morfill}, \& {Sterzik}}]{Dubrulle1995}
{Dubrulle}, B., {Morfill}, G., \& {Sterzik}, M. 1995, \icarus, 114, 237, \dodoi{10.1006/icar.1995.1058}

\bibitem[{{Duch{\^e}ne} {et~al.}(2024){Duch{\^e}ne}, {M{\'e}nard}, {Stapelfeldt}, {Villenave}, {Wolff}, {Perrin}, {Pinte}, {Tazaki}, \& {Padgett}}]{Duchene2024}
{Duch{\^e}ne}, G., {M{\'e}nard}, F., {Stapelfeldt}, K.~R., {et~al.} 2024, \aj, 167, 77, \dodoi{10.3847/1538-3881/acf9a7}

\bibitem[{{Dyrek} {et~al.}(2024){Dyrek}, {Min}, {Decin}, {Bouwman}, {Crouzet}, {Molli{\`e}re}, {Lagage}, {Konings}, {Tremblin}, {G{\"u}del}, {Pye}, {Waters}, {Henning}, {Vandenbussche}, {Ardevol Martinez}, {Argyriou}, {Ducrot}, {Heinke}, {van Looveren}, {Absil}, {Barrado}, {Baudoz}, {Boccaletti}, {Cossou}, {Coulais}, {Edwards}, {Gastaud}, {Glasse}, {Glauser}, {Greene}, {Kendrew}, {Krause}, {Lahuis}, {Mueller}, {Olofsson}, {Patapis}, {Rouan}, {Royer}, {Scheithauer}, {Waldmann}, {Whiteford}, {Colina}, {van Dishoeck}, {{\"O}stlin}, {Ray}, \& {Wright}}]{dyrek24}
{Dyrek}, A., {Min}, M., {Decin}, L., {et~al.} 2024, \nat, 625, 51, \dodoi{10.1038/s41586-023-06849-0}

\bibitem[{{Eberlein} {et~al.}(2024){Eberlein}, {Bitsch}, \& {Helled}}]{Eberlein2024}
{Eberlein}, M., {Bitsch}, B., \& {Helled}, R. 2024, \aap, 691, A50, \dodoi{10.1051/0004-6361/202449840}

\bibitem[{{Ehrenreich} {et~al.}(2014){Ehrenreich}, {Bonfils}, {Lovis}, {Delfosse}, {Forveille}, {Mayor}, {Neves}, {Santos}, {Udry}, \& {S{\'e}gransan}}]{ehrenreich14}
{Ehrenreich}, D., {Bonfils}, X., {Lovis}, C., {et~al.} 2014, \aap, 570, A89, \dodoi{10.1051/0004-6361/201423809}

\bibitem[{{Eistrup} {et~al.}(2018){Eistrup}, {Walsh}, \& {van Dishoeck}}]{Eistrup2018}
{Eistrup}, C., {Walsh}, C., \& {van Dishoeck}, E.~F. 2018, \aap, 613, A14, \dodoi{10.1051/0004-6361/201731302}

\bibitem[{{Emsenhuber} {et~al.}(2021){Emsenhuber}, {Mordasini}, {Burn}, {Alibert}, {Benz}, \& {Asphaug}}]{emsenhuber2021}
{Emsenhuber}, A., {Mordasini}, C., {Burn}, R., {et~al.} 2021, \aap, 656, A70, \dodoi{10.1051/0004-6361/202038863}

\bibitem[{{Eriksson} {et~al.}(2022){Eriksson}, {Ronnet}, {Johansen}, {Helled}, {Valletta}, \& {Petit}}]{Eriksson2022}
{Eriksson}, L. E.~J., {Ronnet}, T., {Johansen}, A., {et~al.} 2022, \aap, 661, A73, \dodoi{10.1051/0004-6361/202142391}

\bibitem[{{Estrada} {et~al.}(2016){Estrada}, {Cuzzi}, \& {Morgan}}]{Estrada2016}
{Estrada}, P.~R., {Cuzzi}, J.~N., \& {Morgan}, D.~A. 2016, \apj, 818, 200, \dodoi{10.3847/0004-637X/818/2/200}

\bibitem[{{Facchini} {et~al.}(2019){Facchini}, {van Dishoeck}, {Manara}, {Tazzari}, {Maud}, {Cazzoletti}, {Rosotti}, {van der Marel}, {Pinilla}, \& {Clarke}}]{Facchini2019}
{Facchini}, S., {van Dishoeck}, E.~F., {Manara}, C.~F., {et~al.} 2019, \aap, 626, L2, \dodoi{10.1051/0004-6361/201935496}

\bibitem[{{Fegley} \& {Schaefer}(2010)}]{Fegley2010}
{Fegley}, B., \& {Schaefer}, L. 2010, in Astrophysics and Space Science Proceedings, Vol.~16, Principles and Perspectives in Cosmochemistry, ed. A.~{Goswami} \& B.~E. {Reddy}, 347, \dodoi{10.1007/978-3-642-10352-0_7}

\bibitem[{{Feinstein} {et~al.}(2024){Feinstein}, {Welbanks}, {Ahrer}, {Alderson}, {Barat}, {Brande}, {Crossfield}, {Desert}, {Duvvuri}, {Espinoza}, {France}, {Gao}, {Guzman Caloca}, {Levine}, {Livingston}, {Lunine}, {Luque}, {Mann}, {Mukherjee}, {Murray}, {Owen}, {Rackham}, {Radica}, {Rockcliffe}, {Rogers}, {Seager}, {Seligman}, {Shapiro}, {Thao}, \& {Vissapragada}}]{feinstein24}
{Feinstein}, A., {Welbanks}, L., {Ahrer}, E.-M., {et~al.} 2024, {KRONOS: Keys to Revealing the Origin and Nature Of sub-neptune Systems}, JWST Proposal. Cycle 3, ID. \#5959

\bibitem[{{Feinstein} {et~al.}(2025){Feinstein}, {Bergner}, {Barber}, {Booth}, {Espaillat}, {Mann}, {Penzlin}, {Qi}, {Seligman}, {Thao}, \& {Welbanks}}]{feinstein25}
{Feinstein}, A., {Bergner}, J., {Barber}, M., {et~al.} 2025, {Measuring the Bulk Properties of a 3 Myr Transiting Exoplanet and its Original Protoplanetary Disk}, JWST Proposal. Cycle 4, ID. \#8597

\bibitem[{{Feinstein} {et~al.}(2023){Feinstein}, {Radica}, {Welbanks}, {Murray}, {Ohno}, {Coulombe}, {Espinoza}, {Bean}, {Teske}, {Benneke}, {Line}, {Rustamkulov}, {Saba}, {Tsiaras}, {Barstow}, {Fortney}, {Gao}, {Knutson}, {MacDonald}, {Mikal-Evans}, {Rackham}, {Taylor}, {Parmentier}, {Batalha}, {Berta-Thompson}, {Carter}, {Changeat}, {dos Santos}, {Gibson}, {Goyal}, {Kreidberg}, {L{\'o}pez-Morales}, {Lothringer}, {Miguel}, {Molaverdikhani}, {Moran}, {Morello}, {Mukherjee}, {Sing}, {Stevenson}, {Wakeford}, {Ahrer}, {Alam}, {Alderson}, {Allen}, {Batalha}, {Bell}, {Blecic}, {Brande}, {Caceres}, {Casewell}, {Chubb}, {Crossfield}, {Crouzet}, {Cubillos}, {Decin}, {D{\'e}sert}, {Harrington}, {Heng}, {Henning}, {Iro}, {Kempton}, {Kendrew}, {Kirk}, {Krick}, {Lagage}, {Lendl}, {Mancini}, {Mansfield}, {May}, {Mayne}, {Nikolov}, {Palle}, {Petit dit de la Roche}, {Piaulet}, {Powell}, {Redfield}, {Rogers}, {Roman}, {Roy}, {Nixon}, {Schlawin}, {Tan}, {Tremblin}, {Turner}, {Venot}, {Waalkes}, {Wheatley}, \&
  {Zhang}}]{feinstein23}
{Feinstein}, A.~D., {Radica}, M., {Welbanks}, L., {et~al.} 2023, \nat, 614, 670, \dodoi{10.1038/s41586-022-05674-1}

\bibitem[{{Feng} {et~al.}(2016){Feng}, {Line}, {Fortney}, {Stevenson}, {Bean}, {Kreidberg}, \& {Parmentier}}]{feng16}
{Feng}, Y.~K., {Line}, M.~R., {Fortney}, J.~J., {et~al.} 2016, \apj, 829, 52, \dodoi{10.3847/0004-637X/829/1/52}

\bibitem[{{Flagg} {et~al.}(2016){Flagg}, {Weinberger}, \& {Matthews}}]{Flagg2016}
{Flagg}, L., {Weinberger}, A.~J., \& {Matthews}, K. 2016, \icarus, 264, 1, \dodoi{10.1016/j.icarus.2015.08.024}

\bibitem[{{Flagg} {et~al.}(2023){Flagg}, {Turner}, {Deibert}, {Ridden-Harper}, {de Mooij}, {MacDonald}, {Jayawardhana}, {Gibson}, {Langeveld}, \& {Sing}}]{flagg23}
{Flagg}, L., {Turner}, J.~D., {Deibert}, E., {et~al.} 2023, \apjl, 953, L19, \dodoi{10.3847/2041-8213/ace529}

\bibitem[{{Fletcher} {et~al.}(2014){Fletcher}, {Greathouse}, {Orton}, {Irwin}, {Mousis}, {Sinclair}, \& {Giles}}]{Fletcher2014}
{Fletcher}, L.~N., {Greathouse}, T.~K., {Orton}, G.~S., {et~al.} 2014, \icarus, 238, 170, \dodoi{10.1016/j.icarus.2014.05.007}

\bibitem[{{Fonte} {et~al.}(2023){Fonte}, {Turrini}, {Pacetti}, {Schisano}, {Molinari}, {Polychroni}, {Politi}, \& {Changeat}}]{fonte23}
{Fonte}, S., {Turrini}, D., {Pacetti}, E., {et~al.} 2023, \mnras, 520, 4683, \dodoi{10.1093/mnras/stad245}

\bibitem[{{Forgan} \& {Rice}(2013)}]{forgan2013}
{Forgan}, D., \& {Rice}, K. 2013, \mnras, 432, 3168, \dodoi{10.1093/mnras/stt672}

\bibitem[{{Fortney} {et~al.}(2008){Fortney}, {Marley}, {Saumon}, \& {Lodders}}]{Fortney2008}
{Fortney}, J.~J., {Marley}, M.~S., {Saumon}, D., \& {Lodders}, K. 2008, \apj, 683, 1104, \dodoi{10.1086/589942}

\bibitem[{{Fournier-Tondreau} {et~al.}(2024){Fournier-Tondreau}, {MacDonald}, {Radica}, {Lafreni{\`e}re}, {Welbanks}, {Piaulet}, {Coulombe}, {Allart}, {Morel}, {Artigau}, {Albert}, {Lim}, {Doyon}, {Benneke}, {Rowe}, {Darveau-Bernier}, {Cowan}, {Lewis}, {Cook}, {Flagg}, {Genest}, {Pelletier}, {Johnstone}, {Dang}, {Kaltenegger}, {Taylor}, \& {Turner}}]{fournier24}
{Fournier-Tondreau}, M., {MacDonald}, R.~J., {Radica}, M., {et~al.} 2024, \mnras, 528, 3354, \dodoi{10.1093/mnras/stad3813}

\bibitem[{{Fu} {et~al.}(2017){Fu}, {Deming}, {Knutson}, {Madhusudhan}, {Mandell}, \& {Fraine}}]{fu17}
{Fu}, G., {Deming}, D., {Knutson}, H., {et~al.} 2017, \apjl, 847, L22, \dodoi{10.3847/2041-8213/aa8e40}

\bibitem[{{Fu} {et~al.}(2025){Fu}, {Stevenson}, {Sing}, {Mukherjee}, {Welbanks}, {Thorngren}, {Tsai}, {Gao}, {Lothringer}, {Beatty}, {Gapp}, {Evans-Soma}, {Allart}, {Pelletier}, {Thao}, \& {Mann}}]{fu25}
{Fu}, G., {Stevenson}, K.~B., {Sing}, D.~K., {et~al.} 2025, arXiv e-prints, arXiv:2501.02081, \dodoi{10.48550/arXiv.2501.02081}

\bibitem[{{Furuya} {et~al.}(2022){Furuya}, {Lee}, \& {Nomura}}]{Furuya2022}
{Furuya}, K., {Lee}, S., \& {Nomura}, H. 2022, \apj, 938, 29, \dodoi{10.3847/1538-4357/ac9233}

\bibitem[{{Gammie}(2001)}]{gammie2001}
{Gammie}, C.~F. 2001, \apj, 553, 174, \dodoi{10.1086/320631}

\bibitem[{{Gandhi} {et~al.}(2023{\natexlab{a}}){Gandhi}, {de Regt}, {Snellen}, {Zhang}, {Rugers}, {van Leur}, \& {Bosschaart}}]{Gandhi2023}
{Gandhi}, S., {de Regt}, S., {Snellen}, I., {et~al.} 2023{\natexlab{a}}, \apjl, 957, L36, \dodoi{10.3847/2041-8213/ad07e2}

\bibitem[{{Gandhi} {et~al.}(2023{\natexlab{b}}){Gandhi}, {Kesseli}, {Zhang}, {Louca}, {Snellen}, {Brogi}, {Miguel}, {Casasayas-Barris}, {Pelletier}, {Landman}, {Maguire}, \& {Gibson}}]{Gandhi2023B}
{Gandhi}, S., {Kesseli}, A., {Zhang}, Y., {et~al.} 2023{\natexlab{b}}, \aj, 165, 242, \dodoi{10.3847/1538-3881/accd65}

\bibitem[{{Gao} {et~al.}(2020){Gao}, {Thorngren}, {Lee}, {Fortney}, {Morley}, {Wakeford}, {Powell}, {Stevenson}, \& {Zhang}}]{gao20}
{Gao}, P., {Thorngren}, D.~P., {Lee}, E. K.~H., {et~al.} 2020, Nature Astronomy, 4, 951, \dodoi{10.1038/s41550-020-1114-3}

\bibitem[{{Gasman} {et~al.}(2025){Gasman}, {Temmink}, {van Dishoeck}, {Kurtovic}, {Grant}, {Sellek}, {Tabone}, {Henning}, {Kamp}, {G{\"u}del}, {Barrado}, {Caratti o Garatti}, {Glauser}, {Waters}, {Arabhavi}, {Jang}, {Kanwar}, {Lienert}, {Perotti}, {Schwarz}, \& {Vlasblom}}]{Gasman2025}
{Gasman}, D., {Temmink}, M., {van Dishoeck}, E.~F., {et~al.} 2025, \aap, 694, A147, \dodoi{10.1051/0004-6361/202452152}

\bibitem[{Gibson {et~al.}(2022)Gibson, Nugroho, Lothringer, Maguire, \& Sing}]{Gibson22}
Gibson, N.~P., Nugroho, S.~K., Lothringer, J., Maguire, C., \& Sing, D.~K. 2022, Monthly Notices of the Royal Astronomical Society, 512, 4618, \dodoi{10.1093/mnras/stac091}

\bibitem[{{Gibson} {et~al.}(2020){Gibson}, {Merritt}, {Nugroho}, {Cubillos}, {de Mooij}, {Mikal-Evans}, {Fossati}, {Lothringer}, {Nikolov}, {Sing}, {Spake}, {Watson}, \& {Wilson}}]{Gibson20}
{Gibson}, N.~P., {Merritt}, S., {Nugroho}, S.~K., {et~al.} 2020, \mnras, 493, 2215, \dodoi{10.1093/mnras/staa228}

\bibitem[{Gibson {et~al.}(2020)Gibson, Merritt, Nugroho, Cubillos, de~Mooij, Mikal-Evans, Fossati, Lothringer, Nikolov, Sing, Spake, Watson, \& Wilson}]{gibson_detection_2020}
Gibson, N.~P., Merritt, S., Nugroho, S.~K., {et~al.} 2020, Monthly Notices of the Royal Astronomical Society, 493, 2215, \dodoi{10.1093/mnras/staa228}

\bibitem[{{Ginsburg} {et~al.}(2019){Ginsburg}, {McGuire}, {Plambeck}, {Bally}, {Goddi}, \& {Wright}}]{Ginsburg2019}
{Ginsburg}, A., {McGuire}, B., {Plambeck}, R., {et~al.} 2019, \apj, 872, 54, \dodoi{10.3847/1538-4357/aafb71}

\bibitem[{{Goldreich} {et~al.}(2004){Goldreich}, {Lithwick}, \& {Sari}}]{Goldreich2004}
{Goldreich}, P., {Lithwick}, Y., \& {Sari}, R. 2004, \araa, 42, 549, \dodoi{10.1146/annurev.astro.42.053102.134004}

\bibitem[{{Grant} {et~al.}(2023){Grant}, {van Dishoeck}, {Tabone}, {Gasman}, {Henning}, {Kamp}, {G{\"u}del}, {Lagage}, {Bettoni}, {Perotti}, {Christiaens}, {Samland}, {Arabhavi}, {Argyriou}, {Abergel}, {Absil}, {Barrado}, {Boccaletti}, {Bouwman}, {Caratti o Garatti}, {Geers}, {Glauser}, {Guadarrama}, {Jang}, {Kanwar}, {Lahuis}, {Morales-Calder{\'o}n}, {Mueller}, {Nehm{\'e}}, {Olofsson}, {Pantin}, {Pawellek}, {Ray}, {Rodgers-Lee}, {Scheithauer}, {Schreiber}, {Schwarz}, {Temmink}, {Vandenbussche}, {Vlasblom}, {Waters}, {Wright}, {Colina}, {Greve}, {Justannont}, \& {{\"O}stlin}}]{Grant2023}
{Grant}, S.~L., {van Dishoeck}, E.~F., {Tabone}, B., {et~al.} 2023, \apjl, 947, L6, \dodoi{10.3847/2041-8213/acc44b}

\bibitem[{{GRAVITY Collaboration} {et~al.}(2020){GRAVITY Collaboration}, {Nowak}, {Lacour}, {Molli{\`e}re}, {Wang}, {Charnay}, {van Dishoeck}, {Abuter}, {Amorim}, {Berger}, {Beust}, {Bonnefoy}, {Bonnet}, {Brandner}, {Buron}, {Cantalloube}, {Collin}, {Chapron}, {Cl{\'e}net}, {Coud{\'e} Du Foresto}, {de Zeeuw}, {Dembet}, {Dexter}, {Duvert}, {Eckart}, {Eisenhauer}, {F{\"o}rster Schreiber}, {F{\'e}dou}, {Garcia Lopez}, {Gao}, {Gendron}, {Genzel}, {Gillessen}, {Hau{\ss}mann}, {Henning}, {Hippler}, {Hubert}, {Jocou}, {Kervella}, {Lagrange}, {Lapeyr{\`e}re}, {Le Bouquin}, {L{\'e}na}, {Maire}, {Ott}, {Paumard}, {Paladini}, {Perraut}, {Perrin}, {Pueyo}, {Pfuhl}, {Rabien}, {Rau}, {Rodr{\'\i}guez-Coira}, {Rousset}, {Scheithauer}, {Shangguan}, {Straub}, {Straubmeier}, {Sturm}, {Tacconi}, {Vincent}, {Widmann}, {Wieprecht}, {Wiezorrek}, {Woillez}, {Yazici}, \& {Ziegler}}]{GRAVITYCollaboration2020}
{GRAVITY Collaboration}, {Nowak}, M., {Lacour}, S., {et~al.} 2020, \aap, 633, A110, \dodoi{10.1051/0004-6361/201936898}

\bibitem[{{Guidi} {et~al.}(2022){Guidi}, {Isella}, {Testi}, {Chandler}, {Liu}, {Schmid}, {Rosotti}, {Meng}, {Jennings}, {Williams}, {Carpenter}, {de Gregorio-Monsalvo}, {Li}, {Liu}, {Ortolani}, {Quanz}, {Ricci}, \& {Tazzari}}]{Guidi2022}
{Guidi}, G., {Isella}, A., {Testi}, L., {et~al.} 2022, \aap, 664, A137, \dodoi{10.1051/0004-6361/202142303}

\bibitem[{{Guillot} {et~al.}(1994){Guillot}, {Gautier}, {Chabrier}, \& {Mosser}}]{1994Icar..112..337G}
{Guillot}, T., {Gautier}, D., {Chabrier}, G., \& {Mosser}, B. 1994, \icarus, 112, 337, \dodoi{10.1006/icar.1994.1188}

\bibitem[{{Guillot} \& {Morel}(1995)}]{1995A&AS..109..109G}
{Guillot}, T., \& {Morel}, P. 1995, \aaps, 109, 109

\bibitem[{{Guilluy} {et~al.}(2022){Guilluy}, {Giacobbe}, {Carleo}, {Cubillos}, {Sozzetti}, {Bonomo}, {Brogi}, {Gandhi}, {Fossati}, {Nascimbeni}, {Turrini}, {Schisano}, {Borsa}, {Lanza}, {Mancini}, {Maggio}, {Malavolta}, {Micela}, {Pino}, {Rainer}, {Bignamini}, {Claudi}, {Cosentino}, {Covino}, {Desidera}, {Fiorenzano}, {Harutyunyan}, {Lorenzi}, {Knapic}, {Molinari}, {Pacetti}, {Pagano}, {Pedani}, {Piotto}, \& {Poretti}}]{Guilluy2022}
{Guilluy}, G., {Giacobbe}, P., {Carleo}, I., {et~al.} 2022, \aap, 665, A104, \dodoi{10.1051/0004-6361/202243854}

\bibitem[{{Gupta} {et~al.}(2024){Gupta}, {Miotello}, {Williams}, {Birnstiel}, {Kuffmeier}, \& {Yen}}]{Gupta2024}
{Gupta}, A., {Miotello}, A., {Williams}, J.~P., {et~al.} 2024, \aap, 683, A133, \dodoi{10.1051/0004-6361/202348007}

\bibitem[{{Hardy} {et~al.}(2017){Hardy}, {Harrington}, {Hardin}, {Madhusudhan}, {Loredo}, {Challener}, {Foster}, {Cubillos}, \& {Blecic}}]{Hardy_2017}
{Hardy}, R.~A., {Harrington}, J., {Hardin}, M.~R., {et~al.} 2017, \apj, 836, 143, \dodoi{10.3847/1538-4357/836/1/143}

\bibitem[{{Heays} {et~al.}(2014){Heays}, {Visser}, {Gredel}, {Ubachs}, {Lewis}, {Gibson}, \& {van Dishoeck}}]{Heays2014}
{Heays}, A.~N., {Visser}, R., {Gredel}, R., {et~al.} 2014, \aap, 562, A61, \dodoi{10.1051/0004-6361/201322832}

\bibitem[{{Hellard} {et~al.}(2019){Hellard}, {Csizmadia}, {Padovan}, {Rauer}, {Cabrera}, {Sohl}, {Spohn}, \& {Breuer}}]{Hellard_2019}
{Hellard}, H., {Csizmadia}, S., {Padovan}, S., {et~al.} 2019, \apj, 878, 119, \dodoi{10.3847/1538-4357/ab2048}

\bibitem[{{Heng}(2016)}]{heng16}
{Heng}, K. 2016, \apjl, 826, L16, \dodoi{10.3847/2041-8205/826/1/L16}

\bibitem[{{Henning} {et~al.}(2024){Henning}, {Kamp}, {Samland}, {Arabhavi}, {Kanwar}, {van Dishoeck}, {G{\"u}del}, {Lagage}, {Waelkens}, {Abergel}, {Absil}, {Barrado}, {Boccaletti}, {Bouwman}, {Caratti o Garatti}, {Geers}, {Glauser}, {Lahuis}, {Mueller}, {Nehm{\'e}}, {Olofsson}, {Pantin}, {Ray}, {Scheithauer}, {Vandenbussche}, {Waters}, {Wright}, {Argyriou}, {Christiaens}, {Franceschi}, {Gasman}, {Grant}, {Guadarrama}, {Jang}, {Morales-Calder{\'o}n}, {Pawellek}, {Perotti}, {Rodgers-Lee}, {Schreiber}, {Schwarz}, {Tabone}, {Temmink}, {Vlasblom}, {Colina}, {Greve}, \& {{\"O}stlin}}]{Henning2024}
{Henning}, T., {Kamp}, I., {Samland}, M., {et~al.} 2024, \pasp, 136, 054302, \dodoi{10.1088/1538-3873/ad3455}

\bibitem[{Hobbs {et~al.}(2021)Hobbs, Rimmer, Shorttle, \& Madhusudhan}]{hobbs21}
Hobbs, R., Rimmer, P.~B., Shorttle, O., \& Madhusudhan, N. 2021, Monthly Notices of the Royal Astronomical Society, 506, 3186, \dodoi{10.1093/mnras/stab1839}

\bibitem[{{Hobbs} {et~al.}(2022){Hobbs}, {Shorttle}, \& {Madhusudhan}}]{Hobbs2022}
{Hobbs}, R., {Shorttle}, O., \& {Madhusudhan}, N. 2022, \mnras, 516, 1032, \dodoi{10.1093/mnras/stac2106}

\bibitem[{{Hoch} {et~al.}(2022){Hoch}, {Konopacky}, {Barman}, {Theissen}, {Brock}, {Perrin}, {Ruffio}, {Macintosh}, \& {Marois}}]{hoch22}
{Hoch}, K. K.~W., {Konopacky}, Q.~M., {Barman}, T.~S., {et~al.} 2022, \aj, 164, 155, \dodoi{10.3847/1538-3881/ac84d4}

\bibitem[{{Hoch} {et~al.}(2023){Hoch}, {Konopacky}, {Theissen}, {Ruffio}, {Barman}, {Rickman}, {Perrin}, {Macintosh}, \& {Marois}}]{hoch23}
{Hoch}, K. K.~W., {Konopacky}, Q.~M., {Theissen}, C.~A., {et~al.} 2023, \aj, 166, 85, \dodoi{10.3847/1538-3881/ace442}

\bibitem[{{Hoeijmakers} {et~al.}(2020){Hoeijmakers}, {Seidel}, {Pino}, {Kitzmann}, {Sindel}, {Ehrenreich}, {Oza}, {Bourrier}, {Allart}, {Gebek}, {Lovis}, {Yurchenko}, {Astudillo-Defru}, {Bayliss}, {Cegla}, {Lavie}, {Lendl}, {Melo}, {Murgas}, {Nascimbeni}, {Pepe}, {S{\'e}gransan}, {Udry}, {Wyttenbach}, \& {Heng}}]{hoeijmakers:2020}
{Hoeijmakers}, H.~J., {Seidel}, J.~V., {Pino}, L., {et~al.} 2020, \aap, 641, A123, \dodoi{10.1051/0004-6361/202038365}

\bibitem[{{Hoeijmakers} {et~al.}(2024){Hoeijmakers}, {Kitzmann}, {Morris}, {Prinoth}, {Borsato}, {Thorsbro}, {Pino}, {Lee}, {Ak{\i}n}, {Seidel}, {Birkby}, {Allart}, \& {Heng}}]{hoeijmakers:2024}
{Hoeijmakers}, H.~J., {Kitzmann}, D., {Morris}, B.~M., {et~al.} 2024, \aap, 685, A139, \dodoi{10.1051/0004-6361/202244968}

\bibitem[{{Hood} {et~al.}(2024){Hood}, {Mukherjee}, {Fortney}, {Line}, {Faherty}, {Merchan}, {Burningham}, {Su{\'a}rez}, {Kiman}, {Gagn{\'e}}, {Beichman}, {Vos}, {Bardalez Gagliuffi}, {Meisner}, \& {Gonzales}}]{Hood2024}
{Hood}, C.~E., {Mukherjee}, S., {Fortney}, J.~J., {et~al.} 2024, arXiv e-prints, arXiv:2402.05345, \dodoi{10.48550/arXiv.2402.05345}

\bibitem[{{Houge} {et~al.}(2025){Houge}, {Krijt}, {Banzatti}, {Blake}, {Pinilla}, {Pontoppidan}, {Trapman}, {Williams}, \& {Zhang}}]{Houge2025}
{Houge}, A., {Krijt}, S., {Banzatti}, A., {et~al.} 2025, \mnras, 537, 691, \dodoi{10.1093/mnras/staf057}

\bibitem[{{Hsu} {et~al.}(2024){Hsu}, {Wang}, {Blake}, {Xuan}, {Zhang}, {Ruffio}, {Horstman}, {Cronin}, {Sappey}, {Xin}, {Finnerty}, {Echeverri}, {Mawet}, {Jovanovic}, {Do {\'O}}, {Baker}, {Bartos}, {Calvin}, {Cetre}, {Delorme}, {Doppmann}, {Fitzgerald}, {Liberman}, {L{\'o}pez}, {Morris}, {Pezzato-Rovner}, {Schofield}, {Skemer}, {Wallace}, \& {Wang}}]{hsu24}
{Hsu}, C.-C., {Wang}, J.~J., {Blake}, G.~A., {et~al.} 2024, \apjl, 977, L47, \dodoi{10.3847/2041-8213/ad95e8}

\bibitem[{{Huang} {et~al.}(2023){Huang}, {Bergin}, {Bae}, {Benisty}, \& {Andrews}}]{Huang2023}
{Huang}, J., {Bergin}, E.~A., {Bae}, J., {Benisty}, M., \& {Andrews}, S.~M. 2023, \apj, 943, 107, \dodoi{10.3847/1538-4357/aca89c}

\bibitem[{{Huang} {et~al.}(2021){Huang}, {Bergin}, {{\"O}berg}, {Andrews}, {Teague}, {Law}, {Kalas}, {Aikawa}, {Bae}, {Bergner}, {Booth}, {Bosman}, {Calahan}, {Cataldi}, {Cleeves}, {Czekala}, {Ilee}, {Le Gal}, {Guzm{\'a}n}, {Long}, {Loomis}, {M{\'e}nard}, {Nomura}, {Qi}, {Schwarz}, {Tsukagoshi}, {van't Hoff}, {Walsh}, {Wilner}, {Yamato}, \& {Zhang}}]{Huang2021}
{Huang}, J., {Bergin}, E.~A., {{\"O}berg}, K.~I., {et~al.} 2021, \apjs, 257, 19, \dodoi{10.3847/1538-4365/ac143e}

\bibitem[{{Hueso} {et~al.}(2018){Hueso}, {Delcroix}, {S{\'a}nchez-Lavega}, {Pedranghelu}, {Kernbauer}, {McKeon}, {Fleckstein}, {Wesley}, {G{\'o}mez-Forrellad}, {Rojas}, \& {Juaristi}}]{Hueso2018}
{Hueso}, R., {Delcroix}, M., {S{\'a}nchez-Lavega}, A., {et~al.} 2018, \aap, 617, A68, \dodoi{10.1051/0004-6361/201832689}

\bibitem[{{Ilee} {et~al.}(2017){Ilee}, {Forgan}, {Evans}, {Hall}, {Booth}, {Clarke}, {Rice}, {Boley}, {Caselli}, {Hartquist}, \& {Rawlings}}]{ilee2017}
{Ilee}, J.~D., {Forgan}, D.~H., {Evans}, M.~G., {et~al.} 2017, \mnras, 472, 189, \dodoi{10.1093/mnras/stx1966}

\bibitem[{{Inglis} {et~al.}(2024){Inglis}, {Batalha}, {Lewis}, {Kataria}, {Knutson}, {Kilpatrick}, {Gagnebin}, {Mukherjee}, {Pettyjohn}, {Crossfield}, {Foote}, {Grant}, {Henry}, {Lally}, {McKemmish}, {Sing}, {Wakeford}, {Zapata Trujillo}, \& {Zellem}}]{inglis24}
{Inglis}, J., {Batalha}, N.~E., {Lewis}, N.~K., {et~al.} 2024, \apjl, 973, L41, \dodoi{10.3847/2041-8213/ad725e}

\bibitem[{{Izidoro} {et~al.}(2017){Izidoro}, {Ogihara}, {Raymond}, {Morbidelli}, {Pierens}, {Bitsch}, {Cossou}, \& {Hersant}}]{izidoro2017}
{Izidoro}, A., {Ogihara}, M., {Raymond}, S.~N., {et~al.} 2017, \mnras, 470, 1750, \dodoi{10.1093/mnras/stx1232}

\bibitem[{{Izidoro} {et~al.}(2022){Izidoro}, {Schlichting}, {Isella}, {Dasgupta}, {Zimmermann}, \& {Bitsch}}]{Izidoro2022}
{Izidoro}, A., {Schlichting}, H.~E., {Isella}, A., {et~al.} 2022, \apjl, 939, L19, \dodoi{10.3847/2041-8213/ac990d}

\bibitem[{{Jang} {et~al.}(2024){Jang}, {Waters}, {Kaeufer}, {Tamanai}, {Perotti}, {Christiaens}, {Kamp}, {Henning}, {Min}, {Arabhavi}, {Barrado}, {van Dishoeck}, {Gasman}, {Grant}, {G{\"u}del}, {Lagage}, {Lahuis}, {Schwarz}, {Tabone}, \& {Temmink}}]{Jang2024}
{Jang}, H., {Waters}, R., {Kaeufer}, T., {et~al.} 2024, \aap, 691, A148, \dodoi{10.1051/0004-6361/202451589}

\bibitem[{{Johansen} \& {Lambrechts}(2017)}]{Johansen2017}
{Johansen}, A., \& {Lambrechts}, M. 2017, Annual Review of Earth and Planetary Sciences, 45, 359, \dodoi{10.1146/annurev-earth-063016-020226}

\bibitem[{{Johansen} {et~al.}(2021){Johansen}, {Ronnet}, {Bizzarro}, {Schiller}, {Lambrechts}, {Nordlund}, \& {Lammer}}]{Johansen2021}
{Johansen}, A., {Ronnet}, T., {Bizzarro}, M., {et~al.} 2021, Science Advances, 7, eabc0444, \dodoi{10.1126/sciadv.abc0444}

\bibitem[{{JWST Transiting Exoplanet Community Early Release Science Team} {et~al.}(2023){JWST Transiting Exoplanet Community Early Release Science Team}, {Ahrer}, {Alderson}, {Batalha}, {Batalha}, {Bean}, {Beatty}, {Bell}, {Benneke}, {Berta-Thompson}, {Carter}, {Crossfield}, {Espinoza}, {Feinstein}, {Fortney}, {Gibson}, {Goyal}, {Kempton}, {Kirk}, {Kreidberg}, {L{\'o}pez-Morales}, {Line}, {Lothringer}, {Moran}, {Mukherjee}, {Ohno}, {Parmentier}, {Piaulet}, {Rustamkulov}, {Schlawin}, {Sing}, {Stevenson}, {Wakeford}, {Allen}, {Birkmann}, {Brande}, {Crouzet}, {Cubillos}, {Damiano}, {D{\'e}sert}, {Gao}, {Harrington}, {Hu}, {Kendrew}, {Knutson}, {Lagage}, {Leconte}, {Lendl}, {MacDonald}, {May}, {Miguel}, {Molaverdikhani}, {Moses}, {Murray}, {Nehring}, {Nikolov}, {Petit dit de la Roche}, {Radica}, {Roy}, {Stassun}, {Taylor}, {Waalkes}, {Wachiraphan}, {Welbanks}, {Wheatley}, {Aggarwal}, {Alam}, {Banerjee}, {Barstow}, {Blecic}, {Casewell}, {Changeat}, {Chubb}, {Col{\'o}n}, {Coulombe}, {Daylan}, {de Val-Borro},
  {Decin}, {Dos Santos}, {Flagg}, {France}, {Fu}, {Garc{\'\i}a Mu{\~n}oz}, {Gizis}, {Glidden}, {Grant}, {Heng}, {Henning}, {Hong}, {Inglis}, {Iro}, {Kataria}, {Komacek}, {Krick}, {Lee}, {Lewis}, {Lillo-Box}, {Lustig-Yaeger}, {Mancini}, {Mandell}, {Mansfield}, {Marley}, {Mikal-Evans}, {Morello}, {Nixon}, {Ortiz Ceballos}, {Piette}, {Powell}, {Rackham}, {Ramos-Rosado}, {Rauscher}, {Redfield}, {Rogers}, {Roman}, {Roudier}, {Scarsdale}, {Shkolnik}, {Southworth}, {Spake}, {Steinrueck}, {Tan}, {Teske}, {Tremblin}, {Tsai}, {Tucker}, {Turner}, {Valenti}, {Venot}, {Waldmann}, {Wallack}, {Zhang}, \& {Zieba}}]{jtec23}
{JWST Transiting Exoplanet Community Early Release Science Team}, {Ahrer}, E.-M., {Alderson}, L., {et~al.} 2023, \nat, 614, 649, \dodoi{10.1038/s41586-022-05269-w}

\bibitem[{{Kalyaan} {et~al.}(2021){Kalyaan}, {Pinilla}, {Krijt}, {Mulders}, \& {Banzatti}}]{Kalyaan2021}
{Kalyaan}, A., {Pinilla}, P., {Krijt}, S., {Mulders}, G.~D., \& {Banzatti}, A. 2021, \apj, 921, 84, \dodoi{10.3847/1538-4357/ac1e96}

\bibitem[{{Kama} {et~al.}(2019){Kama}, {Shorttle}, {Jermyn}, {Folsom}, {Furuya}, {Bergin}, {Walsh}, \& {Keller}}]{Kama2019}
{Kama}, M., {Shorttle}, O., {Jermyn}, A.~S., {et~al.} 2019, \apj, 885, 114, \dodoi{10.3847/1538-4357/ab45f8}

\bibitem[{{Kamp} {et~al.}(2021){Kamp}, {Honda}, {Nomura}, {Audard}, {Fedele}, {Waters}, {Aikawa}, {Banzatti}, {Bowey}, {Bradford}, {Dominik}, {Furuya}, {Habart}, {Ishihara}, {Johnstone}, {Kennedy}, {Kim}, {Kral}, {Lai}, {Larsson}, {McClure}, {Miotello}, {Momose}, {Nakagawa}, {Naylor}, {Nisini}, {Notsu}, {Onaka}, {Pantin}, {Podio}, {Riviere Marichalar}, {Rocha}, {Roelfsema}, {Shimonishi}, {Tang}, {Takami}, {Tazaki}, {Wolf}, {Wyatt}, \& {Ysard}}]{Kamp2021}
{Kamp}, I., {Honda}, M., {Nomura}, H., {et~al.} 2021, \pasa, 38, e055, \dodoi{10.1017/pasa.2021.31}

\bibitem[{{Kanwar} {et~al.}(2024){Kanwar}, {Kamp}, {Jang}, {Waters}, {van Dishoeck}, {Christiaens}, {Arabhavi}, {Henning}, {G{\"u}del}, {Woitke}, {Absil}, {Barrado}, {Garatti}, {Glauser}, {Lahuis}, {Scheithauer}, {Vandenbussche}, {Gasman}, {Grant}, {Kurtovic}, {Perotti}, {Tabone}, \& {Temmink}}]{Kanwar2024}
{Kanwar}, J., {Kamp}, I., {Jang}, H., {et~al.} 2024, arXiv e-prints, arXiv:2407.14362.
\newblock \doarXiv{2407.14362}

\bibitem[{{Kataoka} {et~al.}(2016){Kataoka}, {Tsukagoshi}, {Momose}, {Nagai}, {Muto}, {Dullemond}, {Pohl}, {Fukagawa}, {Shibai}, {Hanawa}, \& {Murakawa}}]{Kataoka2016}
{Kataoka}, A., {Tsukagoshi}, T., {Momose}, M., {et~al.} 2016, \apjl, 831, L12, \dodoi{10.3847/2041-8205/831/2/L12}

\bibitem[{{Kebukawa} {et~al.}(2020){Kebukawa}, {Zolensky}, {Ito}, {Ogawa}, {Takano}, {Ohkouchi}, {Nakato}, {Suga}, {Takeichi}, {Takahashi}, \& {Kobayashi}}]{Kebukawa2020}
{Kebukawa}, Y., {Zolensky}, M.~E., {Ito}, M., {et~al.} 2020, \gca, 271, 61, \dodoi{10.1016/j.gca.2019.12.012}

\bibitem[{{Keppler} {et~al.}(2018){Keppler}, {Benisty}, {M{\"u}ller}, {Henning}, {van Boekel}, {Cantalloube}, {Ginski}, {van Holstein}, {Maire}, {Pohl}, {Samland}, {Avenhaus}, {Baudino}, {Boccaletti}, {de Boer}, {Bonnefoy}, {Chauvin}, {Desidera}, {Langlois}, {Lazzoni}, {Marleau}, {Mordasini}, {Pawellek}, {Stolker}, {Vigan}, {Zurlo}, {Birnstiel}, {Brandner}, {Feldt}, {Flock}, {Girard}, {Gratton}, {Hagelberg}, {Isella}, {Janson}, {Juhasz}, {Kemmer}, {Kral}, {Lagrange}, {Launhardt}, {Matter}, {M{\'e}nard}, {Milli}, {Molli{\`e}re}, {Olofsson}, {P{\'e}rez}, {Pinilla}, {Pinte}, {Quanz}, {Schmidt}, {Udry}, {Wahhaj}, {Williams}, {Buenzli}, {Cudel}, {Dominik}, {Galicher}, {Kasper}, {Lannier}, {Mesa}, {Mouillet}, {Peretti}, {Perrot}, {Salter}, {Sissa}, {Wildi}, {Abe}, {Antichi}, {Augereau}, {Baruffolo}, {Baudoz}, {Bazzon}, {Beuzit}, {Blanchard}, {Brems}, {Buey}, {De Caprio}, {Carbillet}, {Carle}, {Cascone}, {Cheetham}, {Claudi}, {Costille}, {Delboulb{\'e}}, {Dohlen}, {Fantinel}, {Feautrier}, {Fusco}, {Giro}, {Gluck},
  {Gry}, {Hubin}, {Hugot}, {Jaquet}, {Le Mignant}, {Llored}, {Madec}, {Magnard}, {Martinez}, {Maurel}, {Meyer}, {M{\"o}ller-Nilsson}, {Moulin}, {Mugnier}, {Orign{\'e}}, {Pavlov}, {Perret}, {Petit}, {Pragt}, {Puget}, {Rabou}, {Ramos}, {Rigal}, {Rochat}, {Roelfsema}, {Rousset}, {Roux}, {Salasnich}, {Sauvage}, {Sevin}, {Soenke}, {Stadler}, {Suarez}, {Turatto}, \& {Weber}}]{keppler18}
{Keppler}, M., {Benisty}, M., {M{\"u}ller}, A., {et~al.} 2018, \aap, 617, A44, \dodoi{10.1051/0004-6361/201832957}

\bibitem[{{Keyte} {et~al.}(2024){Keyte}, {Kama}, {Chuang}, {Cleeves}, {Drozdovskaya}, {Furuya}, {Rawlings}, \& {Shorttle}}]{Keyte2024}
{Keyte}, L., {Kama}, M., {Chuang}, K.-J., {et~al.} 2024, \mnras, 528, 388, \dodoi{10.1093/mnras/stae019}

\bibitem[{Khorshid {et~al.}(2023)Khorshid, Min, \& D\'esert}]{Khorshid23}
Khorshid, N., Min, M., \& D\'esert, J.~M. 2023, Astronomy \& Astrophysics, 675, A95, \dodoi{10.1051/0004-6361/202245469}

\bibitem[{{Kirk} {et~al.}(2024){Kirk}, {Ahrer}, {Penzlin}, {Owen}, {Booth}, {Alderson}, {Christie}, {Claringbold}, {Esparza-Borges}, {Fisher}, {L{\'o}pez-Morales}, {Mayne}, {McCormack}, {Meech}, {Panwar}, {Powell}, {Sergeev}, {Taylor}, {Tsai}, {Valentine}, {Wakeford}, {Wheatley}, \& {Zamyatina}}]{kirk2024}
{Kirk}, J., {Ahrer}, E.-M., {Penzlin}, A. B.~T., {et~al.} 2024, RAS Techniques and Instruments, 3, 691, \dodoi{10.1093/rasti/rzae043}

\bibitem[{{Kirk} {et~al.}(2025){Kirk}, {Ahrer}, {Claringbold}, {Zamyatina}, {Fisher}, {McCormack}, {Panwar}, {Powell}, {Taylor}, {Thorngren}, {Christie}, {Esparza-Borges}, {Tsai}, {Alderson}, {Booth}, {Fairman}, {L{\'o}pez-Morales}, {Mayne}, {Meech}, {Molli{\`e}re}, {Owen}, {Penzlin}, {Sergeev}, {Valentine}, {Wakeford}, \& {Wheatley}}]{kirk2025}
{Kirk}, J., {Ahrer}, E.-M., {Claringbold}, A.~B., {et~al.} 2025, \mnras, 537, 3027, \dodoi{10.1093/mnras/staf208}

\bibitem[{{Kitzmann} {et~al.}(2018){Kitzmann}, {Heng}, {Rimmer}, {Hoeijmakers}, {Tsai}, {Malik}, {Lendl}, {Deitrick}, \& {Demory}}]{Kitzmann18}
{Kitzmann}, D., {Heng}, K., {Rimmer}, P.~B., {et~al.} 2018, \apj, 863, 183, \dodoi{10.3847/1538-4357/aace5a}

\bibitem[{{Knierim} \& {Helled}(2024)}]{2024ApJ...977..227K}
{Knierim}, H., \& {Helled}, R. 2024, \apj, 977, 227, \dodoi{10.3847/1538-4357/ad8dd0}

\bibitem[{{Knierim} {et~al.}(2022){Knierim}, {Shibata}, \& {Helled}}]{2022A&A...665L...5K}
{Knierim}, H., {Shibata}, S., \& {Helled}, R. 2022, \aap, 665, L5, \dodoi{10.1051/0004-6361/202244516}

\bibitem[{{Knutson} {et~al.}(2014){Knutson}, {Benneke}, {Deming}, \& {Homeier}}]{knutson14}
{Knutson}, H.~A., {Benneke}, B., {Deming}, D., \& {Homeier}, D. 2014, \nat, 505, 66.
\newblock \doarXiv{1401.3350}

\bibitem[{{Konopacky} {et~al.}(2013){Konopacky}, {Barman}, {Macintosh}, \& {Marois}}]{Konopacky2013}
{Konopacky}, Q.~M., {Barman}, T.~S., {Macintosh}, B.~A., \& {Marois}, C. 2013, Science, 339, 1398, \dodoi{10.1126/science.1232003}

\bibitem[{{Kratter} \& {Lodato}(2016)}]{Kratter2016}
{Kratter}, K., \& {Lodato}, G. 2016, \araa, 54, 271, \dodoi{10.1146/annurev-astro-081915-023307}

\bibitem[{{Kreidberg} {et~al.}(2018){Kreidberg}, {Line}, {Thorngren}, {Morley}, \& {Stevenson}}]{kreidberg18}
{Kreidberg}, L., {Line}, M.~R., {Thorngren}, D., {Morley}, C.~V., \& {Stevenson}, K.~B. 2018, \apjl, 858, L6, \dodoi{10.3847/2041-8213/aabfce}

\bibitem[{{Kreidberg} {et~al.}(2014){Kreidberg}, {Bean}, {D{\'e}sert}, {Benneke}, {Deming}, {Stevenson}, {Seager}, {Berta-Thompson}, {Seifahrt}, \& {Homeier}}]{kreidberg14}
{Kreidberg}, L., {Bean}, J.~L., {D{\'e}sert}, J.-M., {et~al.} 2014, \nat, 505, 69, \dodoi{10.1038/nature12888}

\bibitem[{{Krijt} {et~al.}(2020){Krijt}, {Bosman}, {Zhang}, {Schwarz}, {Ciesla}, \& {Bergin}}]{Krijt2020}
{Krijt}, S., {Bosman}, A.~D., {Zhang}, K., {et~al.} 2020, \apj, 899, 134, \dodoi{10.3847/1538-4357/aba75d}

\bibitem[{{Kruczkiewicz} {et~al.}(2021){Kruczkiewicz}, {Vitorino}, {Congiu}, {Theul{\'e}}, \& {Dulieu}}]{Kruczkiewicz2021}
{Kruczkiewicz}, F., {Vitorino}, J., {Congiu}, E., {Theul{\'e}}, P., \& {Dulieu}, F. 2021, \aap, 652, A29, \dodoi{10.1051/0004-6361/202140579}

\bibitem[{{Kuffmeier} {et~al.}(2023){Kuffmeier}, {Jensen}, \& {Haugb{\o}lle}}]{Kuffmeier2023}
{Kuffmeier}, M., {Jensen}, S.~S., \& {Haugb{\o}lle}, T. 2023, European Physical Journal Plus, 138, 272, \dodoi{10.1140/epjp/s13360-023-03880-y}

\bibitem[{{K{\"u}hnle} {et~al.}(2025){K{\"u}hnle}, {Patapis}, {Molli{\`e}re}, {Tremblin}, {Matthews}, {Glauser}, {Whiteford}, {Vasist}, {Absil}, {Barrado}, {Min}, {Lagage}, {Waters}, {Guedel}, {Henning}, {Vandenbussche}, {Baudoz}, {Decin}, {Pye}, {Royer}, {van Dishoeck}, {{\"O}stlin}, {Ray}, \& {Wright}}]{kuehnle2025}
{K{\"u}hnle}, H., {Patapis}, P., {Molli{\`e}re}, P., {et~al.} 2025, \aap, 695, A224, \dodoi{10.1051/0004-6361/202452547}

\bibitem[{{Lambrechts} {et~al.}(2014){Lambrechts}, {Johansen}, \& {Morbidelli}}]{Lambrechts2014}
{Lambrechts}, M., {Johansen}, A., \& {Morbidelli}, A. 2014, \aap, 572, A35, \dodoi{10.1051/0004-6361/201423814}

\bibitem[{{Langeveld} {et~al.}(2022){Langeveld}, {Madhusudhan}, \& {Cabot}}]{Langeveld22}
{Langeveld}, A.~B., {Madhusudhan}, N., \& {Cabot}, S. H.~C. 2022, \mnras, 514, 5192, \dodoi{10.1093/mnras/stac1539}

\bibitem[{{Laskar} \& {Petit}(2017)}]{Laskar2017}
{Laskar}, J., \& {Petit}, A.~C. 2017, \aap, 605, A72, \dodoi{10.1051/0004-6361/201630022}

\bibitem[{{Le Gal} {et~al.}(2021){Le Gal}, {{\"O}berg}, {Teague}, {Loomis}, {Law}, {Walsh}, {Bergin}, {M{\'e}nard}, {Wilner}, {Andrews}, {Aikawa}, {Booth}, {Cataldi}, {Bergner}, {Bosman}, {Cleeves}, {Czekala}, {Furuya}, {Guzm{\'a}n}, {Huang}, {Ilee}, {Nomura}, {Qi}, {Schwarz}, {Tsukagoshi}, {Yamato}, \& {Zhang}}]{LeGal2021}
{Le Gal}, R., {{\"O}berg}, K.~I., {Teague}, R., {et~al.} 2021, \apjs, 257, 12, \dodoi{10.3847/1538-4365/ac2583}

\bibitem[{{Lecavelier Des Etangs} {et~al.}(2008{\natexlab{a}}){Lecavelier Des Etangs}, {Pont}, {Vidal-Madjar}, \& {Sing}}]{lecavelier08b}
{Lecavelier Des Etangs}, A., {Pont}, F., {Vidal-Madjar}, A., \& {Sing}, D. 2008{\natexlab{a}}, \aap, 481, L83, \dodoi{10.1051/0004-6361:200809388}

\bibitem[{{Lecavelier Des Etangs} {et~al.}(2008{\natexlab{b}}){Lecavelier Des Etangs}, {Vidal-Madjar}, {D{\'e}sert}, \& {Sing}}]{lecavelier08a}
{Lecavelier Des Etangs}, A., {Vidal-Madjar}, A., {D{\'e}sert}, J.~M., \& {Sing}, D. 2008{\natexlab{b}}, \aap, 485, 865, \dodoi{10.1051/0004-6361:200809704}

\bibitem[{{Lee}(2024)}]{Lee2024}
{Lee}, E.~J. 2024, \apjl, 970, L15, \dodoi{10.3847/2041-8213/ad5d8e}

\bibitem[{{Lee} \& {Chiang}(2015)}]{Lee2015}
{Lee}, E.~J., \& {Chiang}, E. 2015, \apj, 811, 41, \dodoi{10.1088/0004-637X/811/1/41}

\bibitem[{{Leemker} {et~al.}(2023){Leemker}, {Booth}, {van Dishoeck}, {van der Marel}, {Tabone}, {Ligterink}, {Brunken}, \& {Hogerheijde}}]{Leemker2023}
{Leemker}, M., {Booth}, A.~S., {van Dishoeck}, E.~F., {et~al.} 2023, \aap, 673, A7, \dodoi{10.1051/0004-6361/202245662}

\bibitem[{{Lew} {et~al.}(2024){Lew}, {Roellig}, {Batalha}, {Line}, {Greene}, {Murkherjee}, {Freedman}, {Meyer}, {Beichman}, {Alves de Oliveira}, {De Furio}, {Johnstone}, {Greenbaum}, {Marley}, {Fortney}, {Young}, {Leisenring}, {Boyer}, {Hodapp}, {Misselt}, {Stansberry}, \& {Rieke}}]{Lew2024}
{Lew}, B. W.~P., {Roellig}, T., {Batalha}, N.~E., {et~al.} 2024, \aj, 167, 237, \dodoi{10.3847/1538-3881/ad3425}

\bibitem[{{Li} {et~al.}(2021){Li}, {Bergin}, {Blake}, {Ciesla}, \& {Hirschmann}}]{Li2021}
{Li}, J., {Bergin}, E.~A., {Blake}, G.~A., {Ciesla}, F.~J., \& {Hirschmann}, M.~M. 2021, Science Advances, 7, eabd3632, \dodoi{10.1126/sciadv.abd3632}

\bibitem[{{Limbach} \& {Turner}(2015)}]{Limbach2015}
{Limbach}, M.~A., \& {Turner}, E.~L. 2015, Proceedings of the National Academy of Science, 112, 20, \dodoi{10.1073/pnas.1406545111}

\bibitem[{{Line} \& {Parmentier}(2016)}]{line16}
{Line}, M.~R., \& {Parmentier}, V. 2016, \apj, 820, 78, \dodoi{10.3847/0004-637X/820/1/78}

\bibitem[{{Line} {et~al.}(2021){Line}, {Brogi}, {Bean}, {Gandhi}, {Zalesky}, {Parmentier}, {Smith}, {Mace}, {Mansfield}, {Kempton}, {Fortney}, {Shkolnik}, {Patience}, {Rauscher}, {D{\'e}sert}, \& {Wardenier}}]{line21}
{Line}, M.~R., {Brogi}, M., {Bean}, J.~L., {et~al.} 2021, \nat, 598, 580, \dodoi{10.1038/s41586-021-03912-6}

\bibitem[{{Lissauer}(1987)}]{Lissauer1987}
{Lissauer}, J.~J. 1987, \icarus, 69, 249, \dodoi{10.1016/0019-1035(87)90104-7}

\bibitem[{{Liu} {et~al.}(2024){Liu}, {Roussel}, {Linz}, {Fang}, {Wolf}, {Kirchschlager}, {Henning}, {Yang}, {Du}, {Flock}, \& {Wang}}]{LiuY2024}
{Liu}, Y., {Roussel}, H., {Linz}, H., {et~al.} 2024, \aap, 692, A148, \dodoi{10.1051/0004-6361/202451981}

\bibitem[{{Lodders}(2004)}]{Lodders2004}
{Lodders}, K. 2004, \apj, 611, 587, \dodoi{10.1086/421970}

\bibitem[{{Lodders}(2010)}]{Lodders2010}
{Lodders}, K. 2010, in Astrophysics and Space Science Proceedings, Vol.~16, Principles and Perspectives in Cosmochemistry, ed. A.~{Goswami} \& B.~E. {Reddy}, 379, \dodoi{10.1007/978-3-642-10352-0_8}

\bibitem[{{Lodders} {et~al.}(2009){Lodders}, {Palme}, \& {Gail}}]{Lodders2009}
{Lodders}, K., {Palme}, H., \& {Gail}, H.~P. 2009, Landolt B{\"o}rnstein, 4B, 712, \dodoi{10.1007/978-3-540-88055-4_34}

\bibitem[{{Lothringer} {et~al.}(2018){Lothringer}, {Barman}, \& {Koskinen}}]{Lothringer18}
{Lothringer}, J.~D., {Barman}, T., \& {Koskinen}, T. 2018, \apj, 866, 27, \dodoi{10.3847/1538-4357/aadd9e}

\bibitem[{{Lothringer} {et~al.}(2020){Lothringer}, {Fu}, {Sing}, \& {Barman}}]{Lothringer20}
{Lothringer}, J.~D., {Fu}, G., {Sing}, D.~K., \& {Barman}, T.~S. 2020, \apjl, 898, L14, \dodoi{10.3847/2041-8213/aba265}

\bibitem[{{Lothringer} {et~al.}(2021){Lothringer}, {Rustamkulov}, {Sing}, {Gibson}, {Wilson}, \& {Schlaufman}}]{Lothringer21}
{Lothringer}, J.~D., {Rustamkulov}, Z., {Sing}, D.~K., {et~al.} 2021, \apj, 914, 12, \dodoi{10.3847/1538-4357/abf8a9}

\bibitem[{{Lothringer} {et~al.}(2022){Lothringer}, {Sing}, {Rustamkulov}, {Wakeford}, {Stevenson}, {Nikolov}, {Lavvas}, {Spake}, \& {Winch}}]{Lothringer22}
{Lothringer}, J.~D., {Sing}, D.~K., {Rustamkulov}, Z., {et~al.} 2022, \nat, 604, 49, \dodoi{10.1038/s41586-022-04453-2}

\bibitem[{{Louca} {et~al.}(2023){Louca}, {Miguel}, \& {Kubyshkina}}]{Louca2023}
{Louca}, A.~J., {Miguel}, Y., \& {Kubyshkina}, D. 2023, \apjl, 956, L19, \dodoi{10.3847/2041-8213/acfaec}

\bibitem[{{Luhman}(2014)}]{Luhman2014}
{Luhman}, K.~L. 2014, \apjl, 786, L18, \dodoi{10.1088/2041-8205/786/2/L18}

\bibitem[{{Mac{\'\i}as} {et~al.}(2021){Mac{\'\i}as}, {Guerra-Alvarado}, {Carrasco-Gonz{\'a}lez}, {Ribas}, {Espaillat}, {Huang}, \& {Andrews}}]{Macias2021}
{Mac{\'\i}as}, E., {Guerra-Alvarado}, O., {Carrasco-Gonz{\'a}lez}, C., {et~al.} 2021, \aap, 648, A33, \dodoi{10.1051/0004-6361/202039812}

\bibitem[{{Macintosh} {et~al.}(2015){Macintosh}, {Graham}, {Barman}, {De Rosa}, {Konopacky}, {Marley}, {Marois}, {Nielsen}, {Pueyo}, {Rajan}, {Rameau}, {Saumon}, {Wang}, {Patience}, {Ammons}, {Arriaga}, {Artigau}, {Beckwith}, {Brewster}, {Bruzzone}, {Bulger}, {Burningham}, {Burrows}, {Chen}, {Chiang}, {Chilcote}, {Dawson}, {Dong}, {Doyon}, {Draper}, {Duch{\^e}ne}, {Esposito}, {Fabrycky}, {Fitzgerald}, {Follette}, {Fortney}, {Gerard}, {Goodsell}, {Greenbaum}, {Hibon}, {Hinkley}, {Cotten}, {Hung}, {Ingraham}, {Johnson-Groh}, {Kalas}, {Lafreniere}, {Larkin}, {Lee}, {Line}, {Long}, {Maire}, {Marchis}, {Matthews}, {Max}, {Metchev}, {Millar-Blanchaer}, {Mittal}, {Morley}, {Morzinski}, {Murray-Clay}, {Oppenheimer}, {Palmer}, {Patel}, {Perrin}, {Poyneer}, {Rafikov}, {Rantakyr{\"o}}, {Rice}, {Rojo}, {Rudy}, {Ruffio}, {Ruiz}, {Sadakuni}, {Saddlemyer}, {Salama}, {Savransky}, {Schneider}, {Sivaramakrishnan}, {Song}, {Soummer}, {Thomas}, {Vasisht}, {Wallace}, {Ward-Duong}, {Wiktorowicz}, {Wolff}, \&
  {Zuckerman}}]{Macintosh2015}
{Macintosh}, B., {Graham}, J.~R., {Barman}, T., {et~al.} 2015, Science, 350, 64, \dodoi{10.1126/science.aac5891}

\bibitem[{{Madhusudhan}(2019)}]{madhusudhan19}
{Madhusudhan}, N. 2019, \araa, 57, 617, \dodoi{10.1146/annurev-astro-081817-051846}

\bibitem[{{Madhusudhan} {et~al.}(2014){Madhusudhan}, {Amin}, \& {Kennedy}}]{madhusudhan2014}
{Madhusudhan}, N., {Amin}, M.~A., \& {Kennedy}, G.~M. 2014, \apjl, 794, L12, \dodoi{10.1088/2041-8205/794/1/L12}

\bibitem[{{Madhusudhan} {et~al.}(2017){Madhusudhan}, {Bitsch}, {Johansen}, \& {Eriksson}}]{madhu2017}
{Madhusudhan}, N., {Bitsch}, B., {Johansen}, A., \& {Eriksson}, L. 2017, \mnras, 469, 4102, \dodoi{10.1093/mnras/stx1139}

\bibitem[{{Madhusudhan} \& {Seager}(2009)}]{madhusudhan09}
{Madhusudhan}, N., \& {Seager}, S. 2009, \apj, 707, 24, \dodoi{10.1088/0004-637X/707/1/24}

\bibitem[{{Madhusudhan} {et~al.}(2011){Madhusudhan}, {Harrington}, {Stevenson}, {Nymeyer}, {Campo}, {Wheatley}, {Deming}, {Blecic}, {Hardy}, {Lust}, {Anderson}, {Collier-Cameron}, {Britt}, {Bowman}, {Hebb}, {Hellier}, {Maxted}, {Pollacco}, \& {West}}]{Madhusudhan2011}
{Madhusudhan}, N., {Harrington}, J., {Stevenson}, K.~B., {et~al.} 2011, \nat, 469, 64, \dodoi{10.1038/nature09602}

\bibitem[{Maguire {et~al.}(2023)Maguire, Gibson, Nugroho, Ramkumar, Fortune, Merritt, \& de Mooij}]{maguire_high-resolution_2023}
Maguire, C., Gibson, N.~P., Nugroho, S.~K., {et~al.} 2023, Monthly Notices of the Royal Astronomical Society, 519, 1030, \dodoi{10.1093/mnras/stac3388}

\bibitem[{{Mah} {et~al.}(2023){Mah}, {Bitsch}, {Pascucci}, \& {Henning}}]{Mah2023}
{Mah}, J., {Bitsch}, B., {Pascucci}, I., \& {Henning}, T. 2023, \aap, 677, L7, \dodoi{10.1051/0004-6361/202347169}

\bibitem[{{Mah} {et~al.}(2024){Mah}, {Savvidou}, \& {Bitsch}}]{Mah2024}
{Mah}, J., {Savvidou}, S., \& {Bitsch}, B. 2024, \aap, 686, L17, \dodoi{10.1051/0004-6361/202450322}

\bibitem[{{Males} {et~al.}(2014){Males}, {Close}, {Morzinski}, {Wahhaj}, {Liu}, {Skemer}, {Kopon}, {Follette}, {Puglisi}, {Esposito}, {Riccardi}, {Pinna}, {Xompero}, {Briguglio}, {Biller}, {Nielsen}, {Hinz}, {Rodigas}, {Hayward}, {Chun}, {Ftaclas}, {Toomey}, \& {Wu}}]{Males2014}
{Males}, J.~R., {Close}, L.~M., {Morzinski}, K.~M., {et~al.} 2014, \apj, 786, 32, \dodoi{10.1088/0004-637X/786/1/32}

\bibitem[{{Malsky} {et~al.}(2023){Malsky}, {Rogers}, {Kempton}, \& {Marounina}}]{malsky23}
{Malsky}, I., {Rogers}, L., {Kempton}, E. M.~R., \& {Marounina}, N. 2023, Nature Astronomy, 7, 57, \dodoi{10.1038/s41550-022-01823-8}

\bibitem[{{Mamajek} \& {Bell}(2014)}]{mamajek14}
{Mamajek}, E.~E., \& {Bell}, C. P.~M. 2014, \mnras, 445, 2169, \dodoi{10.1093/mnras/stu1894}

\bibitem[{{Manara} {et~al.}(2023){Manara}, {Ansdell}, {Rosotti}, {Hughes}, {Armitage}, {Lodato}, \& {Williams}}]{Manara2023}
{Manara}, C.~F., {Ansdell}, M., {Rosotti}, G.~P., {et~al.} 2023, in Astronomical Society of the Pacific Conference Series, Vol. 534, Protostars and Planets VII, ed. S.~{Inutsuka}, Y.~{Aikawa}, T.~{Muto}, K.~{Tomida}, \& M.~{Tamura}, 539, \dodoi{10.48550/arXiv.2203.09930}

\bibitem[{{Manara} {et~al.}(2018){Manara}, {Morbidelli}, \& {Guillot}}]{Manara2018}
{Manara}, C.~F., {Morbidelli}, A., \& {Guillot}, T. 2018, \aap, 618, L3, \dodoi{10.1051/0004-6361/201834076}

\bibitem[{{Mankovich} \& {Fuller}(2021)}]{2021NatAs...5.1103M}
{Mankovich}, C.~R., \& {Fuller}, J. 2021, Nature Astronomy, 5, 1103, \dodoi{10.1038/s41550-021-01448-3}

\bibitem[{{Mansfield} {et~al.}(2022){Mansfield}, {Wiser}, {Stevenson}, {Smith}, {Line}, {Bean}, {Fortney}, {Parmentier}, {Kempton}, {Arcangeli}, {D{\'e}sert}, {Kilpatrick}, {Kreidberg}, \& {Malik}}]{mansfield22}
{Mansfield}, M., {Wiser}, L., {Stevenson}, K.~B., {et~al.} 2022, \aj, 163, 261, \dodoi{10.3847/1538-3881/ac658f}

\bibitem[{{Marley} {et~al.}(2007){Marley}, {Fortney}, {Hubickyj}, {Bodenheimer}, \& {Lissauer}}]{Marley2007}
{Marley}, M.~S., {Fortney}, J.~J., {Hubickyj}, O., {Bodenheimer}, P., \& {Lissauer}, J.~J. 2007, \apj, 655, 541, \dodoi{10.1086/509759}

\bibitem[{{McClure} {et~al.}(2023){McClure}, {Rocha}, {Pontoppidan}, {Crouzet}, {Chu}, {Dartois}, {Lamberts}, {Noble}, {Pendleton}, {Perotti}, {Qasim}, {Rachid}, {Smith}, {Sun}, {Beck}, {Boogert}, {Brown}, {Caselli}, {Charnley}, {Cuppen}, {Dickinson}, {Drozdovskaya}, {Egami}, {Erkal}, {Fraser}, {Garrod}, {Harsono}, {Ioppolo}, {Jim{\'e}nez-Serra}, {Jin}, {J{\o}rgensen}, {Kristensen}, {Lis}, {McCoustra}, {McGuire}, {Melnick}, {{\~A}-berg}, {Palumbo}, {Shimonishi}, {Sturm}, {van Dishoeck}, \& {Linnartz}}]{McClure2023}
{McClure}, M.~K., {Rocha}, W.~R.~M., {Pontoppidan}, K.~M., {et~al.} 2023, Nature Astronomy, 7, 431, \dodoi{10.1038/s41550-022-01875-w}

\bibitem[{{McCord} {et~al.}(2012){McCord}, {Li}, {Combe}, {McSween}, {Jaumann}, {Reddy}, {Tosi}, {Williams}, {Blewett}, {Turrini}, {Palomba}, {Pieters}, {de Sanctis}, {Ammannito}, {Capria}, {Le Corre}, {Longobardo}, {Nathues}, {Mittlefehldt}, {Schr{\"o}der}, {Hiesinger}, {Beck}, {Capaccioni}, {Carsenty}, {Keller}, {Denevi}, {Sunshine}, {Raymond}, \& {Russell}}]{McCord2012}
{McCord}, T.~B., {Li}, J.~Y., {Combe}, J.~P., {et~al.} 2012, \nat, 491, 83, \dodoi{10.1038/nature11561}

\bibitem[{Merritt {et~al.}(2020)Merritt, Gibson, Nugroho, Mooij, Hooton, Matthews, McKemmish, Mikal-Evans, Nikolov, Sing, Spake, \& Watson}]{merritt_non-detection_2020}
Merritt, S.~R., Gibson, N.~P., Nugroho, S.~K., {et~al.} 2020, Astronomy \& Astrophysics, 636, A117, \dodoi{10.1051/0004-6361/201937409}

\bibitem[{{Meru} \& {Bate}(2011)}]{meru2011}
{Meru}, F., \& {Bate}, M.~R. 2011, \mnras, 411, L1, \dodoi{10.1111/j.1745-3933.2010.00978.x}

\bibitem[{{Miguel} {et~al.}(2016){Miguel}, {Guillot}, \& {Fayon}}]{2016A&A...596A.114M}
{Miguel}, Y., {Guillot}, T., \& {Fayon}, L. 2016, \aap, 596, A114, \dodoi{10.1051/0004-6361/201629732}

\bibitem[{{Miguel} \& {Vazan}(2023)}]{2023RemS...15..681M}
{Miguel}, Y., \& {Vazan}, A. 2023, Remote Sensing, 15, 681, \dodoi{10.3390/rs15030681}

\bibitem[{{Miguel} {et~al.}(2022){Miguel}, {Bazot}, {Guillot}, {Howard}, {Galanti}, {Kaspi}, {Hubbard}, {Militzer}, {Helled}, {Atreya}, {Connerney}, {Durante}, {Kulowski}, {Lunine}, {Stevenson}, \& {Bolton}}]{2022A&A...662A..18M}
{Miguel}, Y., {Bazot}, M., {Guillot}, T., {et~al.} 2022, \aap, 662, A18, \dodoi{10.1051/0004-6361/202243207}

\bibitem[{{Mikal-Evans} {et~al.}(2019){Mikal-Evans}, {Sing}, {Goyal}, {Drummond}, {Carter}, {Henry}, {Wakeford}, {Lewis}, {Marley}, {Tremblin}, {Nikolov}, {Kataria}, {Deming}, \& {Ballester}}]{mikalevans19}
{Mikal-Evans}, T., {Sing}, D.~K., {Goyal}, J.~M., {et~al.} 2019, \mnras, 488, 2222, \dodoi{10.1093/mnras/stz1753}

\bibitem[{{Miles} {et~al.}(2023){Miles}, {Biller}, {Patapis}, {Worthen}, {Rickman}, {Hoch}, {Skemer}, {Perrin}, {Whiteford}, {Chen}, {Sargent}, {Mukherjee}, {Morley}, {Moran}, {Bonnefoy}, {Petrus}, {Carter}, {Choquet}, {Hinkley}, {Ward-Duong}, {Leisenring}, {Millar-Blanchaer}, {Pueyo}, {Ray}, {Sallum}, {Stapelfeldt}, {Stone}, {Wang}, {Absil}, {Balmer}, {Boccaletti}, {Bonavita}, {Booth}, {Bowler}, {Chauvin}, {Christiaens}, {Currie}, {Danielski}, {Fortney}, {Girard}, {Grady}, {Greenbaum}, {Henning}, {Hines}, {Janson}, {Kalas}, {Kammerer}, {Kennedy}, {Kenworthy}, {Kervella}, {Lagage}, {Lew}, {Liu}, {Macintosh}, {Marino}, {Marley}, {Marois}, {Matthews}, {Matthews}, {Mawet}, {McElwain}, {Metchev}, {Meyer}, {Molliere}, {Pantin}, {Quirrenbach}, {Rebollido}, {Ren}, {Schneider}, {Vasist}, {Wyatt}, {Zhou}, {Briesemeister}, {Bryan}, {Calissendorff}, {Cantalloube}, {Cugno}, {De Furio}, {Dupuy}, {Factor}, {Faherty}, {Fitzgerald}, {Franson}, {Gonzales}, {Hood}, {Howe}, {Kraus}, {Kuzuhara}, {Lagrange}, {Lawson}, {Lazzoni},
  {Liu}, {Llop-Sayson}, {Lloyd}, {Martinez}, {Mazoyer}, {Quanz}, {Redai}, {Samland}, {Schlieder}, {Tamura}, {Tan}, {Uyama}, {Vigan}, {Vos}, {Wagner}, {Wolff}, {Ygouf}, {Zhang}, {Zhang}, \& {Zhang}}]{Miles2023}
{Miles}, B.~E., {Biller}, B.~A., {Patapis}, P., {et~al.} 2023, \apjl, 946, L6, \dodoi{10.3847/2041-8213/acb04a}

\bibitem[{{Militzer} {et~al.}(2022){Militzer}, {Hubbard}, {Wahl}, {Lunine}, {Galanti}, {Kaspi}, {Miguel}, {Guillot}, {Moore}, {Parisi}, {Connerney}, {Helled}, {Cao}, {Mankovich}, {Stevenson}, {Park}, {Wong}, {Atreya}, {Anderson}, \& {Bolton}}]{2022PSJ.....3..185M}
{Militzer}, B., {Hubbard}, W.~B., {Wahl}, S., {et~al.} 2022, \psj, 3, 185, \dodoi{10.3847/PSJ/ac7ec8}

\bibitem[{{Millar} {et~al.}(1989){Millar}, {Bennett}, \& {Herbst}}]{Millar1989}
{Millar}, T.~J., {Bennett}, A., \& {Herbst}, E. 1989, \apj, 340, 906, \dodoi{10.1086/167444}

\bibitem[{{Miller} \& {Fortney}(2011)}]{miller11}
{Miller}, N., \& {Fortney}, J.~J. 2011, \apjl, 736, L29, \dodoi{10.1088/2041-8205/736/2/L29}

\bibitem[{{Minissale} {et~al.}(2022){Minissale}, {Aikawa}, {Bergin}, {Bertin}, {Brown}, {Cazaux}, {Charnley}, {Coutens}, {Cuppen}, {Guzman}, {Linnartz}, {McCoustra}, {Rimola}, {Schrauwen}, {Toubin}, {Ugliengo}, {Watanabe}, {Wakelam}, \& {Dulieu}}]{Minissale2022}
{Minissale}, M., {Aikawa}, Y., {Bergin}, E., {et~al.} 2022, ACS Earth and Space Chemistry, 6, 597, \dodoi{10.1021/acsearthspacechem.1c00357}

\bibitem[{{Miotello} {et~al.}(2023){Miotello}, {Kamp}, {Birnstiel}, {Cleeves}, \& {Kataoka}}]{Miotello2023}
{Miotello}, A., {Kamp}, I., {Birnstiel}, T., {Cleeves}, L.~C., \& {Kataoka}, A. 2023, in Astronomical Society of the Pacific Conference Series, Vol. 534, Protostars and Planets VII, ed. S.~{Inutsuka}, Y.~{Aikawa}, T.~{Muto}, K.~{Tomida}, \& M.~{Tamura}, 501, \dodoi{10.48550/arXiv.2203.09818}

\bibitem[{{Miotello} {et~al.}(2019){Miotello}, {Facchini}, {van Dishoeck}, {Cazzoletti}, {Testi}, {Williams}, {Ansdell}, {van Terwisga}, \& {van der Marel}}]{Miotello2019}
{Miotello}, A., {Facchini}, S., {van Dishoeck}, E.~F., {et~al.} 2019, \aap, 631, A69, \dodoi{10.1051/0004-6361/201935441}

\bibitem[{{Mizuno}(1980)}]{Mizuno1980}
{Mizuno}, H. 1980, Progress of Theoretical Physics, 64, 544, \dodoi{10.1143/PTP.64.544}

\bibitem[{{Mol Lous} \& {Miguel}(2020)}]{2020MNRAS.495.2994M}
{Mol Lous}, M., \& {Miguel}, Y. 2020, \mnras, 495, 2994, \dodoi{10.1093/mnras/staa1405}

\bibitem[{{Molli{\`e}re} \& {Snellen}(2019)}]{Molliere2019}
{Molli{\`e}re}, P., \& {Snellen}, I.~A.~G. 2019, \aap, 622, A139, \dodoi{10.1051/0004-6361/201834169}

\bibitem[{{Molli{\`e}re} {et~al.}(2022){Molli{\`e}re}, {Molyarova}, {Bitsch}, {Henning}, {Schneider}, {Kreidberg}, {Eistrup}, {Burn}, {Nasedkin}, {Semenov}, {Mordasini}, {Schlecker}, {Schwarz}, {Lacour}, {Nowak}, \& {Schulik}}]{molliere2022}
{Molli{\`e}re}, P., {Molyarova}, T., {Bitsch}, B., {et~al.} 2022, \apj, 934, 74, \dodoi{10.3847/1538-4357/ac6a56}

\bibitem[{{Molyarova} {et~al.}(2021){Molyarova}, {Vorobyov}, {Akimkin}, {Skliarevskii}, {Wiebe}, \& {G{\"u}del}}]{molyarova2021}
{Molyarova}, T., {Vorobyov}, E.~I., {Akimkin}, V., {et~al.} 2021, \apj, 910, 153, \dodoi{10.3847/1538-4357/abe2b0}

\bibitem[{{Morbidelli} {et~al.}(2023){Morbidelli}, {Batygin}, \& {Lega}}]{Morbidelli2023}
{Morbidelli}, A., {Batygin}, K., \& {Lega}, E. 2023, \aap, 675, A75, \dodoi{10.1051/0004-6361/202346868}

\bibitem[{{Mordasini} \& {Burn}(2024)}]{Mordasini2024}
{Mordasini}, C., \& {Burn}, R. 2024, Reviews in Mineralogy and Geochemistry, 90, 55, \dodoi{10.2138/rmg.2024.90.03}

\bibitem[{{Mori} \& {Kataoka}(2021)}]{Mori2021}
{Mori}, T., \& {Kataoka}, A. 2021, \apj, 908, 153, \dodoi{10.3847/1538-4357/abd08a}

\bibitem[{{Morley} {et~al.}(2019){Morley}, {Skemer}, {Miles}, {Line}, {Lopez}, {Brogi}, {Freedman}, \& {Marley}}]{Morley2019}
{Morley}, C.~V., {Skemer}, A.~J., {Miles}, B.~E., {et~al.} 2019, \apjl, 882, L29, \dodoi{10.3847/2041-8213/ab3c65}

\bibitem[{{Mulders} {et~al.}(2021){Mulders}, {Pascucci}, {Ciesla}, \& {Fernandes}}]{Mulders2021}
{Mulders}, G.~D., {Pascucci}, I., {Ciesla}, F.~J., \& {Fernandes}, R.~B. 2021, \apj, 920, 66, \dodoi{10.3847/1538-4357/ac178e}

\bibitem[{{M{\"u}ller} {et~al.}(2024){M{\"u}ller}, {Bitsch}, \& {Schneider}}]{Muller2024}
{M{\"u}ller}, J., {Bitsch}, B., \& {Schneider}, A.~D. 2024, \aap, 688, A139, \dodoi{10.1051/0004-6361/202346748}

\bibitem[{{Nasedkin} {et~al.}(2024){Nasedkin}, {Molli{\`e}re}, {Lacour}, {Nowak}, {Kreidberg}, {Stolker}, {Wang}, {Balmer}, {Kammerer}, {Shangguan}, {Abuter}, {Amorim}, {Asensio-Torres}, {Benisty}, {Berger}, {Beust}, {Blunt}, {Boccaletti}, {Bonnefoy}, {Bonnet}, {Bordoni}, {Bourdarot}, {Brandner}, {Cantalloube}, {Caselli}, {Charnay}, {Chauvin}, {Chavez}, {Choquet}, {Christiaens}, {Cl{\'e}net}, {Coud{\'e} Du Foresto}, {Cridland}, {Davies}, {Dembet}, {Dexter}, {Drescher}, {Duvert}, {Eckart}, {Eisenhauer}, {F{\"o}rster Schreiber}, {Garcia}, {Garcia Lopez}, {Gendron}, {Genzel}, {Gillessen}, {Girard}, {Grant}, {Haubois}, {Hei{\ss}el}, {Henning}, {Hinkley}, {Hippler}, {Houll{\'e}}, {Hubert}, {Jocou}, {Keppler}, {Kervella}, {Kurtovic}, {Lagrange}, {Lapeyr{\`e}re}, {Le Bouquin}, {Lutz}, {Maire}, {Mang}, {Marleau}, {M{\'e}rand}, {Monnier}, {Mordasini}, {Ott}, {Otten}, {Paladini}, {Paumard}, {Perraut}, {Perrin}, {Pfuhl}, {Pourr{\'e}}, {Pueyo}, {Ribeiro}, {Rickman}, {Ruffio}, {Rustamkulov}, {Shimizu}, {Sing}, {Stadler},
  {Straub}, {Straubmeier}, {Sturm}, {Tacconi}, {van Dishoeck}, {Vigan}, {Vincent}, {von Fellenberg}, {Widmann}, {Winterhalder}, {Woillez}, {Yazici}, \& {Gravity Collaboration}}]{Nasedkin2024}
{Nasedkin}, E., {Molli{\`e}re}, P., {Lacour}, S., {et~al.} 2024, \aap, 687, A298, \dodoi{10.1051/0004-6361/202449328}

\bibitem[{{Nayakshin}(2010)}]{nayakshin2010}
{Nayakshin}, S. 2010, \mnras, 408, L36, \dodoi{10.1111/j.1745-3933.2010.00923.x}

\bibitem[{{Nayakshin}(2015)}]{nayakshin2015}
---. 2015, \mnras, 446, 459, \dodoi{10.1093/mnras/stu2074}

\bibitem[{{Nikolov} {et~al.}(2022){Nikolov}, {Sing}, {Spake}, {Smalley}, {Goyal}, {Mikal-Evans}, {Wakeford}, {Rustamkulov}, {Deming}, {Fortney}, {Carter}, {Gibson}, \& {Mayne}}]{nikolov22}
{Nikolov}, N.~K., {Sing}, D.~K., {Spake}, J.~J., {et~al.} 2022, \mnras, 515, 3037, \dodoi{10.1093/mnras/stac1530}

\bibitem[{{Nittler} {et~al.}(2019){Nittler}, {Stroud}, {Trigo-Rodr{\'\i}guez}, {De Gregorio}, {Alexander}, {Davidson}, {Moyano-Cambero}, \& {Tanbakouei}}]{Nittler2019}
{Nittler}, L.~R., {Stroud}, R.~M., {Trigo-Rodr{\'\i}guez}, J.~M., {et~al.} 2019, Nature Astronomy, 3, 659, \dodoi{10.1038/s41550-019-0737-8}

\bibitem[{{Nixon} \& {Madhusudhan}(2021)}]{nixon21}
{Nixon}, M.~C., \& {Madhusudhan}, N. 2021, \mnras, 505, 3414, \dodoi{10.1093/mnras/stab1500}

\bibitem[{{Nomura} {et~al.}(2022){Nomura}, {Furuya}, {Cordiner}, {Charnley}, {Alexander}, {Nixon}, {Guzman}, {Yurimoto}, {Tsukagoshi}, \& {Iino}}]{Nomura2022}
{Nomura}, H., {Furuya}, K., {Cordiner}, M.~A., {et~al.} 2022, arXiv e-prints, arXiv:2203.10863, \dodoi{10.48550/arXiv.2203.10863}

\bibitem[{{{\"O}berg} \& {Bergin}(2016)}]{Oberg2016}
{{\"O}berg}, K.~I., \& {Bergin}, E.~A. 2016, \apjl, 831, L19, \dodoi{10.3847/2041-8205/831/2/L19}

\bibitem[{{{\"O}berg} \& {Bergin}(2021)}]{Oberg2021}
---. 2021, \physrep, 893, 1, \dodoi{10.1016/j.physrep.2020.09.004}

\bibitem[{{{\"O}berg} {et~al.}(2011){{\"O}berg}, {Murray-Clay}, \& {Bergin}}]{Oberg2011}
{{\"O}berg}, K.~I., {Murray-Clay}, R., \& {Bergin}, E.~A. 2011, \apjl, 743, L16, \dodoi{10.1088/2041-8205/743/1/L16}

\bibitem[{{{\"O}berg} \& {Wordsworth}(2019)}]{Oberg2019}
{{\"O}berg}, K.~I., \& {Wordsworth}, R. 2019, \aj, 158, 194, \dodoi{10.3847/1538-3881/ab46a8}

\bibitem[{Ouyang {et~al.}(2023)Ouyang, Wang, Zhai, Chen, Rojo, Liu, Zhao, Huang, \& Zhao}]{ouyang_detection_2023}
Ouyang, Q., Wang, W., Zhai, M., {et~al.} 2023, Research in Astronomy and Astrophysics, 23, 065010, \dodoi{10.1088/1674-4527/accbb2}

\bibitem[{{Owen} {et~al.}(1999){Owen}, {Mahaffy}, {Niemann}, {Atreya}, {Donahue}, {Bar-Nun}, \& {de Pater}}]{Owen1999}
{Owen}, T., {Mahaffy}, P., {Niemann}, H.~B., {et~al.} 1999, \nat, 402, 269, \dodoi{10.1038/46232}

\bibitem[{{Paardekooper}(2012)}]{paardekooper2012}
{Paardekooper}, S.-J. 2012, \mnras, 421, 3286, \dodoi{10.1111/j.1365-2966.2012.20553.x}

\bibitem[{{Pacetti} {et~al.}(2022){Pacetti}, {Turrini}, {Schisano}, {Molinari}, {Fonte}, {Politi}, {Hennebelle}, {Klessen}, {Testi}, \& {Lebreuilly}}]{Pacetti2022}
{Pacetti}, E., {Turrini}, D., {Schisano}, E., {et~al.} 2022, \apj, 937, 36, \dodoi{10.3847/1538-4357/ac8b11}

\bibitem[{{Palme} {et~al.}(2014){Palme}, {Lodders}, \& {Jones}}]{Palme2014}
{Palme}, H., {Lodders}, K., \& {Jones}, A. 2014, in Planets, Asteriods, Comets and The Solar System, ed. A.~M. {Davis}, Vol.~2, 15--36

\bibitem[{{Parker} {et~al.}(2024){Parker}, {Birkby}, {Landman}, {Wardenier}, {Young}, {Vaughan}, {van Sluijs}, {Brogi}, {Parmentier}, \& {Line}}]{Parker2024}
{Parker}, L.~T., {Birkby}, J.~L., {Landman}, R., {et~al.} 2024, \mnras, 531, 2356, \dodoi{10.1093/mnras/stae1277}

\bibitem[{{Parmentier} {et~al.}(2018){Parmentier}, {Line}, {Bean}, {Mansfield}, {Kreidberg}, {Lupu}, {Visscher}, {D{\'e}sert}, {Fortney}, {Deleuil}, {Arcangeli}, {Showman}, \& {Marley}}]{Parmentier18}
{Parmentier}, V., {Line}, M.~R., {Bean}, J.~L., {et~al.} 2018, \aap, 617, A110, \dodoi{10.1051/0004-6361/201833059}

\bibitem[{{Pascucci} {et~al.}(2009){Pascucci}, {Apai}, {Luhman}, {Henning}, {Bouwman}, {Meyer}, {Lahuis}, \& {Natta}}]{Pascucci2009}
{Pascucci}, I., {Apai}, D., {Luhman}, K., {et~al.} 2009, \apj, 696, 143, \dodoi{10.1088/0004-637X/696/1/143}

\bibitem[{{Pascucci} {et~al.}(2016){Pascucci}, {Testi}, {Herczeg}, {Long}, {Manara}, {Hendler}, {Mulders}, {Krijt}, {Ciesla}, {Henning}, {Mohanty}, {Drabek-Maunder}, {Apai}, {Sz{\H{u}}cs}, {Sacco}, \& {Olofsson}}]{Pascucci2016}
{Pascucci}, I., {Testi}, L., {Herczeg}, G.~J., {et~al.} 2016, \apj, 831, 125, \dodoi{10.3847/0004-637X/831/2/125}

\bibitem[{{Paxton} {et~al.}(2013){Paxton}, {Cantiello}, {Arras}, {Bildsten}, {Brown}, {Dotter}, {Mankovich}, {Montgomery}, {Stello}, {Timmes}, \& {Townsend}}]{2013ApJS..208....4P}
{Paxton}, B., {Cantiello}, M., {Arras}, P., {et~al.} 2013, \apjs, 208, 4, \dodoi{10.1088/0067-0049/208/1/4}

\bibitem[{Pelletier {et~al.}(2023)Pelletier, Benneke, Ali-Dib, Prinoth, Kasper, Seifahrt, Bean, Debras, Klein, Bazinet, Hoeijmakers, Kesseli, Lim, Carmona, Pino, Casasayas-Barris, Hood, \& Stürmer}]{Pelletier23}
Pelletier, S., Benneke, B., Ali-Dib, M., {et~al.} 2023, Nature, 619, 491, \dodoi{10.1038/s41586-023-06134-0}

\bibitem[{{Pelletier} {et~al.}(2024){Pelletier}, {Benneke}, {Chachan}, {Bazinet}, {Allart}, {Hoeijmakers}, {Lavail}, {Prinoth}, {Coulombe}, {Lothringer}, {Parmentier}, {Smith}, {Borsato}, \& {Thorsbro}}]{Pelletier24}
{Pelletier}, S., {Benneke}, B., {Chachan}, Y., {et~al.} 2024, arXiv e-prints, arXiv:2410.18183, \dodoi{10.48550/arXiv.2410.18183}

\bibitem[{{Penzlin} {et~al.}(2024){Penzlin}, {Booth}, {Kirk}, {Owen}, {Ahrer}, {Christie}, {Claringbold}, {Esparza-Borges}, {L{\'o}pez-Morales}, {Mayne}, {McCormack}, {Meech}, {Panwar}, {Powell}, {Sergeev}, {Taylor}, {Wheatley}, \& {Zamyatina}}]{Penzlin2024}
{Penzlin}, A. B.~T., {Booth}, R.~A., {Kirk}, J., {et~al.} 2024, arXiv e-prints, arXiv:2407.03199, \dodoi{10.48550/arXiv.2407.03199}

\bibitem[{{Perotti} {et~al.}(2023){Perotti}, {Christiaens}, {Henning}, {Tabone}, {Waters}, {Kamp}, {Olofsson}, {Grant}, {Gasman}, {Bouwman}, {Samland}, {Franceschi}, {van Dishoeck}, {Schwarz}, {G{\"u}del}, {Lagage}, {Ray}, {Vandenbussche}, {Abergel}, {Absil}, {Arabhavi}, {Argyriou}, {Barrado}, {Boccaletti}, {Caratti o Garatti}, {Geers}, {Glauser}, {Justannont}, {Lahuis}, {Mueller}, {Nehm{\'e}}, {Pantin}, {Scheithauer}, {Waelkens}, {Guadarrama}, {Jang}, {Kanwar}, {Morales-Calder{\'o}n}, {Pawellek}, {Rodgers-Lee}, {Schreiber}, {Colina}, {Greve}, {{\"O}stlin}, \& {Wright}}]{Perotti2023}
{Perotti}, G., {Christiaens}, V., {Henning}, T., {et~al.} 2023, \nat, 620, 516, \dodoi{10.1038/s41586-023-06317-9}

\bibitem[{{Perotti} {et~al.}(2025){Perotti}, {Kurtovic}, {Henning}, {Olofsson}, {Arabhavi}, {Schwarz}, {Kanwar}, {van Boekel}, {Kamp}, {Pascucci}, {van Dishoeck}, {G{\"u}del}, {Lagage}, {Barrado}, {Garatti}, {Glauser}, {Lahuis}, {Christiaens}, {Franceschi}, {Gasman}, {Grant}, {Jang}, {Kaeufer}, {Morales-Calder{\'o}n}, {Temmink}, \& {Vlasblom}}]{Perotti2025}
{Perotti}, G., {Kurtovic}, N.~T., {Henning}, T., {et~al.} 2025, arXiv e-prints, arXiv:2504.11424, \dodoi{10.48550/arXiv.2504.11424}

\bibitem[{Petrovic {et~al.}(2024)Petrovic, Booth, \& Clarke}]{Petrovic2024}
Petrovic, H.~J., Booth, R.~A., \& Clarke, C.~J. 2024, Material Transport in Protoplanetary Discs with Massive Embedded Planets.
\newblock \doarXiv{2409.16245}

\bibitem[{{Petrus} {et~al.}(2024){Petrus}, {Whiteford}, {Patapis}, {Biller}, {Skemer}, {Hinkley}, {Su{\'a}rez}, {Palma-Bifani}, {Morley}, {Tremblin}, {Charnay}, {Vos}, {Wang}, {Stone}, {Bonnefoy}, {Chauvin}, {Miles}, {Carter}, {Lueber}, {Helling}, {Sutlieff}, {Janson}, {Gonzales}, {Hoch}, {Absil}, {Balmer}, {Boccaletti}, {Bonavita}, {Booth}, {Bowler}, {Briesemeister}, {Bryan}, {Calissendorff}, {Cantalloube}, {Chen}, {Choquet}, {Christiaens}, {Cugno}, {Currie}, {Danielski}, {De Furio}, {Dupuy}, {Factor}, {Faherty}, {Fitzgerald}, {Fortney}, {Franson}, {Girard}, {Grady}, {Henning}, {Hines}, {Hood}, {Howe}, {Kalas}, {Kammerer}, {Kennedy}, {Kenworthy}, {Kervella}, {Kim}, {Kitzmann}, {Kraus}, {Kuzuhara}, {Lagage}, {Lagrange}, {Lawson}, {Lazzoni}, {Leisenring}, {Lew}, {Liu}, {Liu}, {Llop-Sayson}, {Lloyd}, {Macintosh}, {M{\^a}lin}, {Manjavacas}, {Marino}, {Marley}, {Marois}, {Martinez}, {Matthews}, {Matthews}, {Mawet}, {Mazoyer}, {McElwain}, {Metchev}, {Meyer}, {Millar-Blanchaer}, {Molli{\`e}re}, {Moran},
  {Mukherjee}, {Pantin}, {Perrin}, {Pueyo}, {Quanz}, {Quirrenbach}, {Ray}, {Rebollido}, {Adams Redai}, {Ren}, {Rickman}, {Sallum}, {Samland}, {Sargent}, {Schlieder}, {Stapelfeldt}, {Tamura}, {Tan}, {Theissen}, {Uyama}, {Vasist}, {Vigan}, {Wagner}, {Ward-Duong}, {Wolff}, {Worthen}, {Wyatt}, {Ygouf}, {Zurlo}, {Zhang}, {Zhang}, {Zhang}, \& {Zhou}}]{Petrus2024}
{Petrus}, S., {Whiteford}, N., {Patapis}, P., {et~al.} 2024, \apjl, 966, L11, \dodoi{10.3847/2041-8213/ad3e7c}

\bibitem[{{Piette} \& {Madhusudhan}(2020)}]{piette20}
{Piette}, A. A.~A., \& {Madhusudhan}, N. 2020, \apj, 904, 154, \dodoi{10.3847/1538-4357/abbfb1}

\bibitem[{{Pineda} {et~al.}(2023){Pineda}, {Arzoumanian}, {Andre}, {Friesen}, {Zavagno}, {Clarke}, {Inoue}, {Chen}, {Lee}, {Soler}, \& {Kuffmeier}}]{Pineda2023}
{Pineda}, J.~E., {Arzoumanian}, D., {Andre}, P., {et~al.} 2023, in Astronomical Society of the Pacific Conference Series, Vol. 534, Protostars and Planets VII, ed. S.~{Inutsuka}, Y.~{Aikawa}, T.~{Muto}, K.~{Tomida}, \& M.~{Tamura}, 233, \dodoi{10.48550/arXiv.2205.03935}

\bibitem[{{Pinhas} {et~al.}(2018){Pinhas}, {Rackham}, {Madhusudhan}, \& {Apai}}]{pinhas18}
{Pinhas}, A., {Rackham}, B.~V., {Madhusudhan}, N., \& {Apai}, D. 2018, \mnras, 480, 5314, \dodoi{10.1093/mnras/sty2209}

\bibitem[{{Pinilla} {et~al.}(2024){Pinilla}, {Benisty}, {Waters}, {Bae}, \& {Facchini}}]{Pinilla2024}
{Pinilla}, P., {Benisty}, M., {Waters}, R., {Bae}, J., \& {Facchini}, S. 2024, \aap, 686, A135, \dodoi{10.1051/0004-6361/202348707}

\bibitem[{{Pinilla} {et~al.}(2012){Pinilla}, {Birnstiel}, {Ricci}, {Dullemond}, {Uribe}, {Testi}, \& {Natta}}]{Pinilla2012}
{Pinilla}, P., {Birnstiel}, T., {Ricci}, L., {et~al.} 2012, \aap, 538, A114, \dodoi{10.1051/0004-6361/201118204}

\bibitem[{{Pinte} {et~al.}(2016){Pinte}, {Dent}, {M{\'e}nard}, {Hales}, {Hill}, {Cortes}, \& {de Gregorio-Monsalvo}}]{Pinte2016}
{Pinte}, C., {Dent}, W.~R.~F., {M{\'e}nard}, F., {et~al.} 2016, \apj, 816, 25, \dodoi{10.3847/0004-637X/816/1/25}

\bibitem[{{Pirani} {et~al.}(2019){Pirani}, {Johansen}, {Bitsch}, {Mustill}, \& {Turrini}}]{pirani2019}
{Pirani}, S., {Johansen}, A., {Bitsch}, B., {Mustill}, A.~J., \& {Turrini}, D. 2019, \aap, 623, A169, \dodoi{10.1051/0004-6361/201833713}

\bibitem[{{Piso} {et~al.}(2015){Piso}, {{\"O}berg}, {Birnstiel}, \& {Murray-Clay}}]{Piso2015}
{Piso}, A.-M.~A., {{\"O}berg}, K.~I., {Birnstiel}, T., \& {Murray-Clay}, R.~A. 2015, \apj, 815, 109, \dodoi{10.1088/0004-637X/815/2/109}

\bibitem[{{Pizzati} {et~al.}(2023){Pizzati}, {Rosotti}, \& {Tabone}}]{Pizzati2023}
{Pizzati}, E., {Rosotti}, G.~P., \& {Tabone}, B. 2023, \mnras, 524, 3184, \dodoi{10.1093/mnras/stad2057}

\bibitem[{{Poch} {et~al.}(2020){Poch}, {Istiqomah}, {Quirico}, {Beck}, {Schmitt}, {Theul{\'e}}, {Faure}, {Hily-Blant}, {Bonal}, {Raponi}, {Ciarniello}, {Rousseau}, {Potin}, {Brissaud}, {Flandinet}, {Filacchione}, {Pommerol}, {Thomas}, {Kappel}, {Mennella}, {Moroz}, {Vinogradoff}, {Arnold}, {Erard}, {Bockel{\'e}e-Morvan}, {Leyrat}, {Capaccioni}, {De Sanctis}, {Longobardo}, {Mancarella}, {Palomba}, \& {Tosi}}]{Poch2020}
{Poch}, O., {Istiqomah}, I., {Quirico}, E., {et~al.} 2020, Science, 367, aaw7462, \dodoi{10.1126/science.aaw7462}

\bibitem[{{Polanski} {et~al.}(2022){Polanski}, {Crossfield}, {Howard}, {Isaacson}, \& {Rice}}]{polanski22}
{Polanski}, A.~S., {Crossfield}, I. J.~M., {Howard}, A.~W., {Isaacson}, H., \& {Rice}, M. 2022, Research Notes of the American Astronomical Society, 6, 155, \dodoi{10.3847/2515-5172/ac8676}

\bibitem[{{Pollack} {et~al.}(1996){Pollack}, {Hubickyj}, {Bodenheimer}, {Lissauer}, {Podolak}, \& {Greenzweig}}]{Pollack1996}
{Pollack}, J.~B., {Hubickyj}, O., {Bodenheimer}, P., {et~al.} 1996, \icarus, 124, 62, \dodoi{10.1006/icar.1996.0190}

\bibitem[{{Polman} \& {Mordasini}(2024)}]{Polman2024}
{Polman}, J., \& {Mordasini}, C. 2024, \aap, 692, A202, \dodoi{10.1051/0004-6361/202451897}

\bibitem[{Polman {et~al.}(2023)Polman, Waters, Min, Miguel, \& Khorshid}]{polman23}
Polman, J., Waters, L. B. F.~M., Min, M., Miguel, Y., \& Khorshid, N. 2023, Astronomy \& Astrophysics, 670, A161, \dodoi{10.1051/0004-6361/202244647}

\bibitem[{{Polychroni} {et~al.}(2025){Polychroni}, {Turrini}, {Ivanovski}, {Marzari}, {Testi}, {Politi}, {Sozzetti, A.}, {Trigo-Rodriguez, J. M.}, {Desidera, S.}, {Drozdovskaya, M. N.}, {Fonte, S.}, {Molinari, S.}, {Naponiello, L.}, {Pacetti, E.}, {Schisano, E.}, {Simonetti, P.}, \& {Zusi, M.}}]{Polychroni2025}
{Polychroni}, D., {Turrini}, D., {Ivanovski}, S., {et~al.} 2025, \aap, 697, A158, \dodoi{10.1051/0004-6361/202453097}

\bibitem[{{Powell} {et~al.}(2024){Powell}, {Feinstein}, {Lee}, {Zhang}, {Tsai}, {Taylor}, {Kirk}, {Bell}, {Barstow}, {Gao}, {Bean}, {Blecic}, {Chubb}, {Crossfield}, {Jordan}, {Kitzmann}, {Moran}, {Morello}, {Moses}, {Welbanks}, {Yang}, {Zhang}, {Ahrer}, {Bello-Arufe}, {Brande}, {Casewell}, {Crouzet}, {Cubillos}, {Demory}, {Dyrek}, {Flagg}, {Hu}, {Inglis}, {Jones}, {Kreidberg}, {L{\'o}pez-Morales}, {Lagage}, {Meier Vald{\'e}s}, {Miguel}, {Parmentier}, {Piette}, {Rackham}, {Radica}, {Redfield}, {Stevenson}, {Wakeford}, {Aggarwal}, {Alam}, {Batalha}, {Batalha}, {Benneke}, {Berta-Thompson}, {Brady}, {Caceres}, {Carter}, {D{\'e}sert}, {Harrington}, {Iro}, {Line}, {Lothringer}, {MacDonald}, {Mancini}, {Molaverdikhani}, {Mukherjee}, {Nixon}, {Oza}, {Palle}, {Rustamkulov}, {Sing}, {Steinrueck}, {Venot}, {Wheatley}, \& {Yurchenko}}]{powell24}
{Powell}, D., {Feinstein}, A.~D., {Lee}, E. K.~H., {et~al.} 2024, \nat, 626, 979, \dodoi{10.1038/s41586-024-07040-9}

\bibitem[{{Rackham} {et~al.}(2018){Rackham}, {Apai}, \& {Giampapa}}]{rackham18}
{Rackham}, B.~V., {Apai}, D., \& {Giampapa}, M.~S. 2018, \apj, 853, 122, \dodoi{10.3847/1538-4357/aaa08c}

\bibitem[{{Radica} {et~al.}(2023){Radica}, {Welbanks}, {Espinoza}, {Taylor}, {Coulombe}, {Feinstein}, {Goyal}, {Scarsdale}, {Albert}, {Baghel}, {Bean}, {Blecic}, {Lafreni{\`e}re}, {MacDonald}, {Zamyatina}, {Allart1}, {Artigau}, {Batalha}, {Cook}, {Cowan}, {Dang}, {Doyon}, {Fournier-Tondreau}, {Johnstone}, {Line}, {Moran}, {Mukherjee}, {Pelletier}, {Roy}, {Talens}, {Filippazzo}, {Pontoppidan}, \& {Volk}}]{radica23}
{Radica}, M., {Welbanks}, L., {Espinoza}, N., {et~al.} 2023, \mnras, 524, 835, \dodoi{10.1093/mnras/stad1762}

\bibitem[{{Rafikov}(2005)}]{rafikov2005gi}
{Rafikov}, R.~R. 2005, \apjl, 621, L69, \dodoi{10.1086/428899}

\bibitem[{{Rafikov}(2006)}]{Rafikov2006}
---. 2006, \apj, 648, 666, \dodoi{10.1086/505695}

\bibitem[{{Ragozzine} \& {Wolf}(2009)}]{Ragozzine_2009}
{Ragozzine}, D., \& {Wolf}, A.~S. 2009, \apj, 698, 1778, \dodoi{10.1088/0004-637X/698/2/1778}

\bibitem[{{Raymond} \& {Izidoro}(2017)}]{Raymond2017}
{Raymond}, S.~N., \& {Izidoro}, A. 2017, \icarus, 297, 134, \dodoi{10.1016/j.icarus.2017.06.030}

\bibitem[{{Reboussin} {et~al.}(2015){Reboussin}, {Wakelam}, {Guilloteau}, {Hersant}, \& {Dutrey}}]{Reboussin2015}
{Reboussin}, L., {Wakelam}, V., {Guilloteau}, S., {Hersant}, F., \& {Dutrey}, A. 2015, \aap, 579, A82, \dodoi{10.1051/0004-6361/201525885}

\bibitem[{{Reggiani} {et~al.}(2022){Reggiani}, {Schlaufman}, {Healy}, {Lothringer}, \& {Sing}}]{Reggiani22}
{Reggiani}, H., {Schlaufman}, K.~C., {Healy}, B.~F., {Lothringer}, J.~D., \& {Sing}, D.~K. 2022, \aj, 163, 159, \dodoi{10.3847/1538-3881/ac4d9f}

\bibitem[{{Ricci} {et~al.}(2010){Ricci}, {Testi}, {Natta}, \& {Brooks}}]{Ricci2010}
{Ricci}, L., {Testi}, L., {Natta}, A., \& {Brooks}, K.~J. 2010, \aap, 521, A66, \dodoi{10.1051/0004-6361/201015039}

\bibitem[{{Rice} {et~al.}(2003){Rice}, {Armitage}, {Bate}, \& {Bonnell}}]{rice2003}
{Rice}, W.~K.~M., {Armitage}, P.~J., {Bate}, M.~R., \& {Bonnell}, I.~A. 2003, \mnras, 339, 1025, \dodoi{10.1046/j.1365-8711.2003.06253.x}

\bibitem[{{Rice} {et~al.}(2005){Rice}, {Lodato}, \& {Armitage}}]{rice2005}
{Rice}, W.~K.~M., {Lodato}, G., \& {Armitage}, P.~J. 2005, \mnras, 364, L56, \dodoi{10.1111/j.1745-3933.2005.00105.x}

\bibitem[{{Rice} {et~al.}(2004){Rice}, {Lodato}, {Pringle}, {Armitage}, \& {Bonnell}}]{rice2004}
{Rice}, W.~K.~M., {Lodato}, G., {Pringle}, J.~E., {Armitage}, P.~J., \& {Bonnell}, I.~A. 2004, \mnras, 355, 543, \dodoi{10.1111/j.1365-2966.2004.08339.x}

\bibitem[{{Rimmer} {et~al.}(2020){Rimmer}, {Ferus}, {Waldmann}, {Kn{\'\i}{\v{z}}ek}, {Kalvaitis}, {Ivanek}, {Kubel{\'\i}k}, {Yurchenko}, {Burian}, {Dost{\'a}l}, {Juha}, {Dud{\v{z}}{\'a}k}, {Kr{\r{u}}s}, {Tennyson}, {Civi{\v{s}}}, {Archibald}, \& {Granville-Willett}}]{rimmer20}
{Rimmer}, P.~B., {Ferus}, M., {Waldmann}, I.~P., {et~al.} 2020, \apj, 888, 21, \dodoi{10.3847/1538-4357/ab55e8}

\bibitem[{{Rizzuto} {et~al.}(2020){Rizzuto}, {Newton}, {Mann}, {Tofflemire}, {Vanderburg}, {Kraus}, {Wood}, {Quinn}, {Zhou}, {Thao}, {Law}, {Ziegler}, \& {Brice{\~n}o}}]{rizzuto20}
{Rizzuto}, A.~C., {Newton}, E.~R., {Mann}, A.~W., {et~al.} 2020, \aj, 160, 33, \dodoi{10.3847/1538-3881/ab94b7}

\bibitem[{{Rodler} {et~al.}(2012){Rodler}, {Lopez-Morales}, \& {Ribas}}]{rodler12}
{Rodler}, F., {Lopez-Morales}, M., \& {Ribas}, I. 2012, \apjl, 753, L25, \dodoi{10.1088/2041-8205/753/1/L25}

\bibitem[{{Roellig} {et~al.}(2004){Roellig}, {Van Cleve}, {Sloan}, {Wilson}, {Saumon}, {Leggett}, {Marley}, {Cushing}, {Kirkpatrick}, {Mainzer}, \& {Houck}}]{Roellig2004}
{Roellig}, T.~L., {Van Cleve}, J.~E., {Sloan}, G.~C., {et~al.} 2004, \apjs, 154, 418, \dodoi{10.1086/421978}

\bibitem[{{Rowland} {et~al.}(2024){Rowland}, {Morley}, {Miles}, {Suarez}, {Faherty}, {Skemer}, {Beiler}, {Line}, {Bjoraker}, {Fortney}, {Vos}, {Alejandro Merchan}, {Marley}, {Burningham}, {Freedman}, {Gharib-Nezhad}, {Batalha}, {Lupu}, {Visscher}, {Schneider}, {Geballe}, {Carter}, {Allers}, {Mang}, {Apai}, {Limbach}, \& {Wilson}}]{Rowland2024}
{Rowland}, M.~J., {Morley}, C.~V., {Miles}, B.~E., {et~al.} 2024, \apjl, 977, L49, \dodoi{10.3847/2041-8213/ad9744}

\bibitem[{{Rustamkulov} {et~al.}(2023){Rustamkulov}, {Sing}, {Mukherjee}, {May}, {Kirk}, {Schlawin}, {Line}, {Piaulet}, {Carter}, {Batalha}, {Goyal}, {L{\'o}pez-Morales}, {Lothringer}, {MacDonald}, {Moran}, {Stevenson}, {Wakeford}, {Espinoza}, {Bean}, {Batalha}, {Benneke}, {Berta-Thompson}, {Crossfield}, {Gao}, {Kreidberg}, {Powell}, {Cubillos}, {Gibson}, {Leconte}, {Molaverdikhani}, {Nikolov}, {Parmentier}, {Roy}, {Taylor}, {Turner}, {Wheatley}, {Aggarwal}, {Ahrer}, {Alam}, {Alderson}, {Allen}, {Banerjee}, {Barat}, {Barrado}, {Barstow}, {Bell}, {Blecic}, {Brande}, {Casewell}, {Changeat}, {Chubb}, {Crouzet}, {Daylan}, {Decin}, {D{\'e}sert}, {Mikal-Evans}, {Feinstein}, {Flagg}, {Fortney}, {Harrington}, {Heng}, {Hong}, {Hu}, {Iro}, {Kataria}, {Kempton}, {Krick}, {Lendl}, {Lillo-Box}, {Louca}, {Lustig-Yaeger}, {Mancini}, {Mansfield}, {Mayne}, {Miguel}, {Morello}, {Ohno}, {Palle}, {Petit dit de la Roche}, {Rackham}, {Radica}, {Ramos-Rosado}, {Redfield}, {Rogers}, {Shkolnik}, {Southworth}, {Teske}, {Tremblin},
  {Tucker}, {Venot}, {Waalkes}, {Welbanks}, {Zhang}, \& {Zieba}}]{rustamkulov23}
{Rustamkulov}, Z., {Sing}, D.~K., {Mukherjee}, S., {et~al.} 2023, \nat, 614, 659, \dodoi{10.1038/s41586-022-05677-y}

\bibitem[{{Safronov}(1960)}]{safronov1960}
{Safronov}, V.~S. 1960, Annales d'Astrophysique, 23, 979

\bibitem[{{Sainsbury-Martinez} \& {Walsh}(2024)}]{Sainsbury-Martinez2024}
{Sainsbury-Martinez}, F., \& {Walsh}, C. 2024, \apj, 966, 39, \dodoi{10.3847/1538-4357/ad28b3}

\bibitem[{{Sanchis} {et~al.}(2021){Sanchis}, {Testi}, {Natta}, {Facchini}, {Manara}, {Miotello}, {Ercolano}, {Henning}, {Preibisch}, {Carpenter}, {de Gregorio-Monsalvo}, {Jayawardhana}, {Lopez}, {Mu{\v{z}}i{\'c}}, {Pascucci}, {Santamar{\'\i}a-Miranda}, {van Terwisga}, \& {Williams}}]{Sanchis2021}
{Sanchis}, E., {Testi}, L., {Natta}, A., {et~al.} 2021, \aap, 649, A19, \dodoi{10.1051/0004-6361/202039733}

\bibitem[{Savel {et~al.}(2024)Savel, M.-R.~Kempton, Beltz, \& Malsky}]{savel:zenodo}
Savel, A., M.-R.~Kempton, E., Beltz, H., \& Malsky, I. 2024, arjunsavel/hires-literature, v1.0.5,  Zenodo, \dodoi{10.5281/zenodo.13306679}

\bibitem[{{Savvidou} \& {Bitsch}(2023)}]{Savvidou2023}
{Savvidou}, S., \& {Bitsch}, B. 2023, \aap, 679, A42, \dodoi{10.1051/0004-6361/202245793}

\bibitem[{{Savvidou} \& {Bitsch}(2024)}]{Savvidou2024}
---. 2024, arXiv e-prints, arXiv:2407.08533, \dodoi{10.48550/arXiv.2407.08533}

\bibitem[{{Schneider} \& {Bitsch}(2021{\natexlab{a}})}]{Schneider2021b}
{Schneider}, A.~D., \& {Bitsch}, B. 2021{\natexlab{a}}, \aap, 654, A72, \dodoi{10.1051/0004-6361/202141096}

\bibitem[{{Schneider} \& {Bitsch}(2021{\natexlab{b}})}]{Schneider2021}
---. 2021{\natexlab{b}}, \aap, 654, A71, \dodoi{10.1051/0004-6361/202039640}

\bibitem[{{Schwarz} {et~al.}(2018){Schwarz}, {Bergin}, {Cleeves}, {Zhang}, {{\"O}berg}, {Blake}, \& {Anderson}}]{Schwarz2018}
{Schwarz}, K.~R., {Bergin}, E.~A., {Cleeves}, L.~I., {et~al.} 2018, \apj, 856, 85, \dodoi{10.3847/1538-4357/aaae08}

\bibitem[{{Schwarz} {et~al.}(2019){Schwarz}, {Bergin}, {Cleeves}, {Zhang}, {{\"O}berg}, {Blake}, \& {Anderson}}]{Schwarz2019}
---. 2019, \apj, 877, 131, \dodoi{10.3847/1538-4357/ab1c5e}

\bibitem[{{Schwarz} {et~al.}(2024){Schwarz}, {Tielens}, {Najita}, {Bergner}, {Kral}, {Anderson}, {Chin}, {Leisawitz}, {Wilner}, {Roelfsema}, {van der Tak}, {Young}, \& {Walker}}]{Schwarz2024}
{Schwarz}, K.~R., {Tielens}, A., {Najita}, J., {et~al.} 2024, Journal of Astronomical Telescopes, Instruments, and Systems, 10, 042307, \dodoi{10.1117/1.JATIS.10.4.042307}

\bibitem[{{Seager} \& {Sasselov}(1998)}]{seager98}
{Seager}, S., \& {Sasselov}, D.~D. 1998, \apjl, 502, L157, \dodoi{10.1086/311498}

\bibitem[{{Seligman} {et~al.}(2022){Seligman}, {Becker}, {Adams}, {Feinstein}, \& {Rogers}}]{seligman2022_comets}
{Seligman}, D.~Z., {Becker}, J., {Adams}, F.~C., {Feinstein}, A.~D., \& {Rogers}, L.~A. 2022, arXiv e-prints, arXiv:2204.12653.
\newblock \doarXiv{2204.12653}

\bibitem[{{Sellek} {et~al.}(2020){Sellek}, {Booth}, \& {Clarke}}]{Sellek2020}
{Sellek}, A.~D., {Booth}, R.~A., \& {Clarke}, C.~J. 2020, \mnras, 498, 2845, \dodoi{10.1093/mnras/staa2519}

\bibitem[{{Semenov} {et~al.}(2018){Semenov}, {Favre}, {Fedele}, {Guilloteau}, {Teague}, {Henning}, {Dutrey}, {Chapillon}, {Hersant}, \& {Pi{\'e}tu}}]{Semenov2018}
{Semenov}, D., {Favre}, C., {Fedele}, D., {et~al.} 2018, \aap, 617, A28, \dodoi{10.1051/0004-6361/201832980}

\bibitem[{{Shibata} {et~al.}(2020){Shibata}, {Helled}, \& {Ikoma}}]{Shibata2020}
{Shibata}, S., {Helled}, R., \& {Ikoma}, M. 2020, \aap, 633, A33, \dodoi{10.1051/0004-6361/201936700}

\bibitem[{{Siebenaler} {et~al.}(2025){Siebenaler}, {Miguel}, {de Regt}, \& {Guillot}}]{2025A&A...693A.308S}
{Siebenaler}, L., {Miguel}, Y., {de Regt}, S., \& {Guillot}, T. 2025, \aap, 693, A308, \dodoi{10.1051/0004-6361/202452860}

\bibitem[{{Sierra} {et~al.}(2021){Sierra}, {P{\'e}rez}, {Zhang}, {Law}, {Guzm{\'a}n}, {Qi}, {Bosman}, {{\"O}berg}, {Andrews}, {Long}, {Teague}, {Booth}, {Walsh}, {Wilner}, {M{\'e}nard}, {Cataldi}, {Czekala}, {Bae}, {Huang}, {Bergner}, {Ilee}, {Benisty}, {Le Gal}, {Loomis}, {Tsukagoshi}, {Liu}, {Yamato}, \& {Aikawa}}]{Sierra2021}
{Sierra}, A., {P{\'e}rez}, L.~M., {Zhang}, K., {et~al.} 2021, \apjs, 257, 14, \dodoi{10.3847/1538-4365/ac1431}

\bibitem[{Silva {et~al.}(2022)Silva, Demangeon, Santos, Allart, Borsa, Cristo, Esparza-Borges, Seidel, Palle, Sousa, Tabernero, Osorio, Cristiani, Pepe, Rebolo, Adibekyan, Alibert, Barros, Bouchy, Bourrier, Curto, Marcantonio, D’Odorico, Ehrenreich, Figueira, Hernández, Lovis, Martins, Mehner, Micela, Molaro, Mounzer, Nunes, Sozzetti, Mascareño, \& Udry}]{silva_detection_2022}
Silva, T.~A., Demangeon, O. D.~S., Santos, N.~C., {et~al.} 2022, Astronomy \& Astrophysics, 666, L10, \dodoi{10.1051/0004-6361/202244489}

\bibitem[{{Simon} {et~al.}(2023){Simon}, {Rajappan}, \& {{\"O}berg}}]{Simon2023}
{Simon}, A., {Rajappan}, M., \& {{\"O}berg}, K.~I. 2023, \apj, 955, 5, \dodoi{10.3847/1538-4357/aceaf8}

\bibitem[{{Sing} {et~al.}(2016){Sing}, {Fortney}, {Nikolov}, {Wakeford}, {Kataria}, {Evans}, {Aigrain}, {Ballester}, {Burrows}, {Deming}, {D{\'e}sert}, {Gibson}, {Henry}, {Huitson}, {Knutson}, {Lecavelier Des Etangs}, {Pont}, {Showman}, {Vidal-Madjar}, {Williamson}, \& {Wilson}}]{sing16}
{Sing}, D.~K., {Fortney}, J.~J., {Nikolov}, N., {et~al.} 2016, \nat, 529, 59, \dodoi{10.1038/nature16068}

\bibitem[{{Sing} {et~al.}(2019){Sing}, {Lavvas}, {Ballester}, {Lecavelier des Etangs}, {Marley}, {Nikolov}, {Ben-Jaffel}, {Bourrier}, {Buchhave}, {Deming}, {Ehrenreich}, {Mikal-Evans}, {Kataria}, {Lewis}, {L{\'o}pez-Morales}, {Garc{\'\i}a Mu{\~n}oz}, {Henry}, {Sanz-Forcada}, {Spake}, {Wakeford}, \& {PanCET Collaboration}}]{Sing19}
{Sing}, D.~K., {Lavvas}, P., {Ballester}, G.~E., {et~al.} 2019, \aj, 158, 91, \dodoi{10.3847/1538-3881/ab2986}

\bibitem[{{Sing} {et~al.}(2024){Sing}, {Rustamkulov}, {Thorngren}, {Barstow}, {Tremblin}, {Alves de Oliveira}, {Beck}, {Birkmann}, {Challener}, {Crouzet}, {Espinoza}, {Ferruit}, {Giardino}, {Gressier}, {Lee}, {Lewis}, {Maiolino}, {Manjavacas}, {Rauscher}, {Sirianni}, \& {Valenti}}]{sing24}
{Sing}, D.~K., {Rustamkulov}, Z., {Thorngren}, D.~P., {et~al.} 2024, \nat, 630, 831, \dodoi{10.1038/s41586-024-07395-z}

\bibitem[{{Smith} \& {Terrile}(1984)}]{Smith1984}
{Smith}, B.~A., \& {Terrile}, R.~J. 1984, Science, 226, 1421, \dodoi{10.1126/science.226.4681.1421}

\bibitem[{{Smith} {et~al.}(2024{\natexlab{a}}){Smith}, {Line}, {Bean}, {Brogi}, {August}, {Welbanks}, {Desert}, {Lunine}, {Sanchez}, {Mansfield}, {Pino}, {Rauscher}, {Kempton}, {Zalesky}, \& {Fowler}}]{smith24}
{Smith}, P. C.~B., {Line}, M.~R., {Bean}, J.~L., {et~al.} 2024{\natexlab{a}}, \aj, 167, 110, \dodoi{10.3847/1538-3881/ad17bf}

\bibitem[{{Smith} {et~al.}(2024{\natexlab{b}}){Smith}, {Sanchez}, {Line}, {Rauscher}, {Weiner Mansfield}, {Kempton}, {Savel}, {Wardenier}, {Pino}, {Bean}, {Beltz}, {Panwar}, {Brogi}, {Malsky}, {Fortney}, {Desert}, {Pelletier}, {Parmentier}, {Kanumalla}, {Welbanks}, {Meyer}, \& {Monnier}}]{Smith24b}
{Smith}, P. C.~B., {Sanchez}, J.~A., {Line}, M.~R., {et~al.} 2024{\natexlab{b}}, arXiv e-prints, arXiv:2410.19017, \dodoi{10.48550/arXiv.2410.19017}

\bibitem[{{Snellen}(2025)}]{Snellen2025}
{Snellen}, I. 2025, arXiv e-prints, arXiv:2505.08926, \dodoi{10.48550/arXiv.2505.08926}

\bibitem[{{Snellen} {et~al.}(2010){Snellen}, {de Kok}, {de Mooij}, \& {Albrecht}}]{snellen10}
{Snellen}, I. A.~G., {de Kok}, R.~J., {de Mooij}, E. J.~W., \& {Albrecht}, S. 2010, \nat, 465, 1049, \dodoi{10.1038/nature09111}

\bibitem[{{Spiegel} \& {Burrows}(2012)}]{Spiegel2012}
{Spiegel}, D.~S., \& {Burrows}, A. 2012, \apj, 745, 174, \dodoi{10.1088/0004-637X/745/2/174}

\bibitem[{{Stammler} {et~al.}(2023){Stammler}, {Lichtenberg}, {Dr{\k{a}}{\.z}kowska}, \& {Birnstiel}}]{Stammler2023}
{Stammler}, S.~M., {Lichtenberg}, T., {Dr{\k{a}}{\.z}kowska}, J., \& {Birnstiel}, T. 2023, \aap, 670, L5, \dodoi{10.1051/0004-6361/202245512}

\bibitem[{{Steinmeyer} {et~al.}(2023){Steinmeyer}, {Woitke}, \& {Johansen}}]{Steinmeyer2023}
{Steinmeyer}, M.-L., {Woitke}, P., \& {Johansen}, A. 2023, \aap, 677, A181, \dodoi{10.1051/0004-6361/202245636}

\bibitem[{{Stevenson}(1982)}]{Stevenson1982}
{Stevenson}, D.~J. 1982, \planss, 30, 755, \dodoi{10.1016/0032-0633(82)90108-8}

\bibitem[{{Stevenson} {et~al.}(2016){Stevenson}, {Lewis}, {Bean}, {Beichman}, {Fraine}, {Kilpatrick}, {Krick}, {Lothringer}, {Mandell}, {Valenti}, {Agol}, {Angerhausen}, {Barstow}, {Birkmann}, {Burrows}, {Charbonneau}, {Cowan}, {Crouzet}, {Cubillos}, {Curry}, {Dalba}, {de Wit}, {Deming}, {D{\'e}sert}, {Doyon}, {Dragomir}, {Ehrenreich}, {Fortney}, {Garc{\'\i}a Mu{\~n}oz}, {Gibson}, {Gizis}, {Greene}, {Harrington}, {Heng}, {Kataria}, {Kempton}, {Knutson}, {Kreidberg}, {Lafreni{\`e}re}, {Lagage}, {Line}, {Lopez-Morales}, {Madhusudhan}, {Morley}, {Rocchetto}, {Schlawin}, {Shkolnik}, {Shporer}, {Sing}, {Todorov}, {Tucker}, \& {Wakeford}}]{stevenson16}
{Stevenson}, K.~B., {Lewis}, N.~K., {Bean}, J.~L., {et~al.} 2016, \pasp, 128, 094401, \dodoi{10.1088/1538-3873/128/967/094401}

\bibitem[{{Sturm} {et~al.}(2023){Sturm}, {McClure}, {Bergner}, {Harsono}, {Dartois}, {Drozdovskaya}, {Ioppolo}, {{\"O}berg}, {Law}, {Palumbo}, {Pendleton}, {Rocha}, {Terada}, \& {Urso}}]{Sturm2023}
{Sturm}, J.~A., {McClure}, M.~K., {Bergner}, J.~B., {et~al.} 2023, \aap, 677, A18, \dodoi{10.1051/0004-6361/202346053}

\bibitem[{{Szul{\'a}gyi} {et~al.}(2022){Szul{\'a}gyi}, {Binkert}, \& {Surville}}]{Szulagyi2022}
{Szul{\'a}gyi}, J., {Binkert}, F., \& {Surville}, C. 2022, \apj, 924, 1, \dodoi{10.3847/1538-4357/ac32d1}

\bibitem[{{Tabone} {et~al.}(2023){Tabone}, {Bettoni}, {van Dishoeck}, {Arabhavi}, {Grant}, {Gasman}, {Henning}, {Kamp}, {G{\"u}del}, {Lagage}, {Ray}, {Vandenbussche}, {Abergel}, {Absil}, {Argyriou}, {Barrado}, {Boccaletti}, {Bouwman}, {Caratti o Garatti}, {Geers}, {Glauser}, {Justannont}, {Lahuis}, {Mueller}, {Nehm{\'e}}, {Olofsson}, {Pantin}, {Scheithauer}, {Waelkens}, {Waters}, {Black}, {Christiaens}, {Guadarrama}, {Morales-Calder{\'o}n}, {Jang}, {Kanwar}, {Pawellek}, {Perotti}, {Perrin}, {Rodgers-Lee}, {Samland}, {Schreiber}, {Schwarz}, {Colina}, {{\"O}stlin}, \& {Wright}}]{Tabone2023}
{Tabone}, B., {Bettoni}, G., {van Dishoeck}, E.~F., {et~al.} 2023, Nature Astronomy, 7, 805, \dodoi{10.1038/s41550-023-01965-3}

\bibitem[{{Takeuchi} \& {Lin}(2002)}]{Takeuchi2002}
{Takeuchi}, T., \& {Lin}, D.~N.~C. 2002, \apj, 581, 1344, \dodoi{10.1086/344437}

\bibitem[{{Tanaka} \& {Ida}(1999)}]{Tanaka1999}
{Tanaka}, H., \& {Ida}, S. 1999, \icarus, 139, 350, \dodoi{10.1006/icar.1999.6107}

\bibitem[{{Tatsumi} {et~al.}(2021){Tatsumi}, {Sugimoto}, {Riu}, {Sugita}, {Nakamura}, {Hiroi}, {Morota}, {Popescu}, {Michikami}, {Kitazato}, {Matsuoka}, {Kameda}, {Honda}, {Yamada}, {Sakatani}, {Kouyama}, {Yokota}, {Honda}, {Suzuki}, {Cho}, {Ogawa}, {Hayakawa}, {Sawada}, {Yoshioka}, {Pilorget}, {Ishida}, {Domingue}, {Hirata}, {Sasaki}, {de Le{\'o}n}, {Barucci}, {Michel}, {Suemitsu}, {Saiki}, {Tanaka}, {Terui}, {Nakazawa}, {Kikuchi}, {Yamaguchi}, {Ogawa}, {Ono}, {Mimasu}, {Yoshikawa}, {Takahashi}, {Takei}, {Fujii}, {Yamamoto}, {Okada}, {Hirose}, {Hosoda}, {Mori}, {Shimada}, {Soldini}, {Tsukizaki}, {Mizuno}, {Iwata}, {Yano}, {Ozaki}, {Abe}, {Ohtake}, {Namiki}, {Tachibana}, {Arakawa}, {Ikeda}, {Ishiguro}, {Wada}, {Yabuta}, {Takeuchi}, {Shimaki}, {Shirai}, {Hirata}, {Iijima}, {Tsuda}, {Watanabe}, \& {Yoshikawa}}]{Tatsumi2021}
{Tatsumi}, E., {Sugimoto}, C., {Riu}, L., {et~al.} 2021, Nature Astronomy, 5, 39, \dodoi{10.1038/s41550-020-1179-z}

\bibitem[{{Taylor} {et~al.}(2004){Taylor}, {Atreya}, {Encrenaz}, {Hunten}, {Irwin}, \& {Owen}}]{Taylor2004}
{Taylor}, F.~W., {Atreya}, S.~K., {Encrenaz}, T., {et~al.} 2004, in Jupiter. The Planet, Satellites and Magnetosphere, ed. F.~{Bagenal}, T.~E. {Dowling}, \& W.~B. {McKinnon}, Vol.~1 (Cambridge University Press), 59--78

\bibitem[{{Taylor} {et~al.}(2023){Taylor}, {Radica}, {Welbanks}, {MacDonald}, {Blecic}, {Zamyatina}, {Roth}, {Bean}, {Parmentier}, {Coulombe}, {Feinstein}, {Espinoza}, {Benneke}, {Lafreni{\`e}re}, {Doyon}, \& {Ahrer}}]{taylor23}
{Taylor}, J., {Radica}, M., {Welbanks}, L., {et~al.} 2023, \mnras, 524, 817, \dodoi{10.1093/mnras/stad1547}

\bibitem[{{Tazzari} {et~al.}(2021){Tazzari}, {Testi}, {Natta}, {Williams}, {Ansdell}, {Carpenter}, {Facchini}, {Guidi}, {Hogherheijde}, {Manara}, {Miotello}, \& {van der Marel}}]{Tazzari2021}
{Tazzari}, M., {Testi}, L., {Natta}, A., {et~al.} 2021, \mnras, 506, 5117, \dodoi{10.1093/mnras/stab1912}

\bibitem[{{Testi} {et~al.}(2014){Testi}, {Birnstiel}, {Ricci}, {Andrews}, {Blum}, {Carpenter}, {Dominik}, {Isella}, {Natta}, {Williams}, \& {Wilner}}]{Testi2014}
{Testi}, L., {Birnstiel}, T., {Ricci}, L., {et~al.} 2014, in Protostars and Planets VI, ed. H.~{Beuther}, R.~S. {Klessen}, C.~P. {Dullemond}, \& T.~{Henning}, 339--361, \dodoi{10.2458/azu_uapress_9780816531240-ch015}

\bibitem[{{Testi} {et~al.}(2022){Testi}, {Natta}, {Manara}, {de Gregorio Monsalvo}, {Lodato}, {Lopez}, {Muzic}, {Pascucci}, {Sanchis}, {Miranda}, {Scholz}, {De Simone}, \& {Williams}}]{Testi2022}
{Testi}, L., {Natta}, A., {Manara}, C.~F., {et~al.} 2022, \aap, 663, A98, \dodoi{10.1051/0004-6361/202141380}

\bibitem[{{Thao} {et~al.}(2024){Thao}, {Mann}, {Feinstein}, {Gao}, {Thorngren}, {Rotman}, {Welbanks}, {Brown}, {Duvvuri}, {France}, {Longo}, {Sandoval}, {Schneider}, {Wilson}, {Youngblood}, {Vanderburg}, {Barber}, {Wood}, {Batalha}, {Kraus}, {Murray}, {Newton}, {Rizzuto}, {Tofflemire}, {Tsai}, {Bean}, {Berta-Thompson}, {Evans-Soma}, {Froning}, {Kempton}, {Miguel}, \& {Pineda}}]{thao24}
{Thao}, P.~C., {Mann}, A.~W., {Feinstein}, A.~D., {et~al.} 2024, \aj, 168, 297, \dodoi{10.3847/1538-3881/ad81d7}

\bibitem[{{Thorngren}(2024)}]{Thorngren2024}
{Thorngren}, D.~P. 2024, arXiv e-prints, arXiv:2405.05307, \dodoi{10.48550/arXiv.2405.05307}

\bibitem[{{Thorngren} {et~al.}(2016){Thorngren}, {Fortney}, {Murray-Clay}, \& {Lopez}}]{thorngren16}
{Thorngren}, D.~P., {Fortney}, J.~J., {Murray-Clay}, R.~A., \& {Lopez}, E.~D. 2016, \apj, 831, 64, \dodoi{10.3847/0004-637X/831/1/64}

\bibitem[{{Tobin} {et~al.}(2023){Tobin}, {van't Hoff}, {Leemker}, {van Dishoeck}, {Paneque-Carre{\~n}o}, {Furuya}, {Harsono}, {Persson}, {Cleeves}, {Sheehan}, \& {Cieza}}]{Tobin2023}
{Tobin}, J.~J., {van't Hoff}, M. L.~R., {Leemker}, M., {et~al.} 2023, \nat, 615, 227, \dodoi{10.1038/s41586-022-05676-z}

\bibitem[{{Toci} {et~al.}(2021){Toci}, {Rosotti}, {Lodato}, {Testi}, \& {Trapman}}]{Toci2021}
{Toci}, C., {Rosotti}, G., {Lodato}, G., {Testi}, L., \& {Trapman}, L. 2021, \mnras, 507, 818, \dodoi{10.1093/mnras/stab2112}

\bibitem[{{Toomre}(1964)}]{toomre1964}
{Toomre}, A. 1964, \apj, 139, 1217, \dodoi{10.1086/147861}

\bibitem[{{Trapman} {et~al.}(2020){Trapman}, {Ansdell}, {Hogerheijde}, {Facchini}, {Manara}, {Miotello}, {Williams}, \& {Bruderer}}]{Trapman2020}
{Trapman}, L., {Ansdell}, M., {Hogerheijde}, M.~R., {et~al.} 2020, \aap, 638, A38, \dodoi{10.1051/0004-6361/201834537}

\bibitem[{{Trapman} {et~al.}(2019){Trapman}, {Facchini}, {Hogerheijde}, {van Dishoeck}, \& {Bruderer}}]{Trapman2019}
{Trapman}, L., {Facchini}, S., {Hogerheijde}, M.~R., {van Dishoeck}, E.~F., \& {Bruderer}, S. 2019, \aap, 629, A79, \dodoi{10.1051/0004-6361/201834723}

\bibitem[{{Trapman} {et~al.}(2022){Trapman}, {Zhang}, {van't Hoff}, {Hogerheijde}, \& {Bergin}}]{Trapman2022}
{Trapman}, L., {Zhang}, K., {van't Hoff}, M. L.~R., {Hogerheijde}, M.~R., \& {Bergin}, E.~A. 2022, \apjl, 926, L2, \dodoi{10.3847/2041-8213/ac4f47}

\bibitem[{{Tremblin} {et~al.}(2015){Tremblin}, {Amundsen}, {Mourier}, {Baraffe}, {Chabrier}, {Drummond}, {Homeier}, \& {Venot}}]{Tremblin2015}
{Tremblin}, P., {Amundsen}, D.~S., {Mourier}, P., {et~al.} 2015, \apjl, 804, L17, \dodoi{10.1088/2041-8205/804/1/L17}

\bibitem[{{Tsai} {et~al.}(2023){Tsai}, {Lee}, {Powell}, {Gao}, {Zhang}, {Moses}, {H{\'e}brard}, {Venot}, {Parmentier}, {Jordan}, {Hu}, {Alam}, {Alderson}, {Batalha}, {Bean}, {Benneke}, {Bierson}, {Brady}, {Carone}, {Carter}, {Chubb}, {Inglis}, {Leconte}, {Line}, {L{\'o}pez-Morales}, {Miguel}, {Molaverdikhani}, {Rustamkulov}, {Sing}, {Stevenson}, {Wakeford}, {Yang}, {Aggarwal}, {Baeyens}, {Barat}, {de Val-Borro}, {Daylan}, {Fortney}, {France}, {Goyal}, {Grant}, {Kirk}, {Kreidberg}, {Louca}, {Moran}, {Mukherjee}, {Nasedkin}, {Ohno}, {Rackham}, {Redfield}, {Taylor}, {Tremblin}, {Visscher}, {Wallack}, {Welbanks}, {Youngblood}, {Ahrer}, {Batalha}, {Behr}, {Berta-Thompson}, {Blecic}, {Casewell}, {Crossfield}, {Crouzet}, {Cubillos}, {Decin}, {D{\'e}sert}, {Feinstein}, {Gibson}, {Harrington}, {Heng}, {Henning}, {Kempton}, {Krick}, {Lagage}, {Lendl}, {Lothringer}, {Mansfield}, {Mayne}, {Mikal-Evans}, {Palle}, {Schlawin}, {Shorttle}, {Wheatley}, \& {Yurchenko}}]{tsai23}
{Tsai}, S.-M., {Lee}, E. K.~H., {Powell}, D., {et~al.} 2023, \nat, 617, 483, \dodoi{10.1038/s41586-023-05902-2}

\bibitem[{{Turrini} {et~al.}(2012){Turrini}, {Coradini}, \& {Magni}}]{Turrini2012}
{Turrini}, D., {Coradini}, A., \& {Magni}, G. 2012, \apj, 750, 8, \dodoi{10.1088/0004-637X/750/1/8}

\bibitem[{{Turrini} {et~al.}(2011){Turrini}, {Magni}, \& {Coradini}}]{Turrini2011}
{Turrini}, D., {Magni}, G., \& {Coradini}, A. 2011, Monthly Notices of the Royal Astronomical Society, 413, 2439, \dodoi{10.1111/j.1365-2966.2011.18316.x}

\bibitem[{{Turrini} {et~al.}(2019){Turrini}, {Marzari}, {Polychroni}, \& {Testi}}]{Turrini2019}
{Turrini}, D., {Marzari}, F., {Polychroni}, D., \& {Testi}, L. 2019, \apj, 877, 50, \dodoi{10.3847/1538-4357/ab18f5}

\bibitem[{{Turrini} {et~al.}(2015){Turrini}, {Nelson}, \& {Barbieri}}]{Turrini2015}
{Turrini}, D., {Nelson}, R.~P., \& {Barbieri}, M. 2015, Experimental Astronomy, 40, 501, \dodoi{10.1007/s10686-014-9401-6}

\bibitem[{{Turrini} \& {Svetsov}(2014)}]{Turrini2014b}
{Turrini}, D., \& {Svetsov}, V. 2014, Life, 4, 4, \dodoi{10.3390/life4010004}

\bibitem[{{Turrini} {et~al.}(2018){Turrini}, {Svetsov}, {Consolmagno}, {Sirono}, \& {Jutzi}}]{Turrini2018b}
{Turrini}, D., {Svetsov}, V., {Consolmagno}, G., {Sirono}, S., \& {Jutzi}, M. 2018, \icarus, 311, 224, \dodoi{10.1016/j.icarus.2018.04.004}

\bibitem[{{Turrini} {et~al.}(2016){Turrini}, {Svetsov}, {Consolmagno}, {Sirono}, \& {Pirani}}]{Turrini2016}
{Turrini}, D., {Svetsov}, V., {Consolmagno}, G., {Sirono}, S., \& {Pirani}, S. 2016, \icarus, 280, 328, \dodoi{10.1016/j.icarus.2016.07.009}

\bibitem[{{Turrini} {et~al.}(2020){Turrini}, {Zinzi}, \& {Belinchon}}]{Turrini2020}
{Turrini}, D., {Zinzi}, A., \& {Belinchon}, J.~A. 2020, \aap, 636, A53, \dodoi{10.1051/0004-6361/201936301}

\bibitem[{{Turrini} {et~al.}(2014){Turrini}, {Combe}, {McCord}, {Oklay}, {Vincent}, {Prettyman}, {McSween}, {Consolmagno}, {De Sanctis}, {Le Corre}, {Longobardo}, {Palomba}, \& {Russell}}]{Turrini2014}
{Turrini}, D., {Combe}, J.~P., {McCord}, T.~B., {et~al.} 2014, \icarus, 240, 86, \dodoi{10.1016/j.icarus.2014.02.021}

\bibitem[{{Turrini} {et~al.}(2021){Turrini}, {Schisano}, {Fonte}, {Molinari}, {Politi}, {Fedele}, {Pani{\'c}}, {Kama}, {Changeat}, \& {Tinetti}}]{Turrini2021}
{Turrini}, D., {Schisano}, E., {Fonte}, S., {et~al.} 2021, \apj, 909, 40, \dodoi{10.3847/1538-4357/abd6e5}

\bibitem[{{Turrini} {et~al.}(2023){Turrini}, {Marzari}, {Polychroni}, {Claudi}, {Desidera}, {Mesa}, {Pinamonti}, {Sozzetti}, {Su{\'a}rez Mascare{\~n}o}, {Damasso}, {Benatti}, {Malavolta}, {Micela}, {Zinzi}, {B{\'e}jar}, {Biazzo}, {Bignamini}, {Bonavita}, {Borsa}, {del Burgo}, {Chauvin}, {Delorme}, {Gonz{\'a}lez Hern{\'a}ndez}, {Gratton}, {Hagelberg}, {Janson}, {Langlois}, {Lanza}, {Lazzoni}, {Lodieu}, {Maggio}, {Mancini}, {Molinari}, {Molinaro}, {Murgas}, \& {Nardiello}}]{Turrini2023}
{Turrini}, D., {Marzari}, F., {Polychroni}, D., {et~al.} 2023, \aap, 679, A55, \dodoi{10.1051/0004-6361/202244752}

\bibitem[{{Tychoniec} {et~al.}(2020){Tychoniec}, {Manara}, {Rosotti}, {van Dishoeck}, {Cridland}, {Hsieh}, {Murillo}, {Segura-Cox}, {van Terwisga}, \& {Tobin}}]{Tychoniec2020}
{Tychoniec}, {\L}., {Manara}, C.~F., {Rosotti}, G.~P., {et~al.} 2020, \aap, 640, A19, \dodoi{10.1051/0004-6361/202037851}

\bibitem[{{Valdivia-Mena} {et~al.}(2024){Valdivia-Mena}, {Pineda}, {Caselli}, {Segura-Cox}, {Schmiedeke}, {Spezzano}, {Offner}, {Ivlev}, {Kuffmeier}, {Cunningham}, {Neri}, \& {Maureira}}]{Valdivia-Mena2024}
{Valdivia-Mena}, M.~T., {Pineda}, J.~E., {Caselli}, P., {et~al.} 2024, \aap, 687, A71, \dodoi{10.1051/0004-6361/202449395}

\bibitem[{{Van Clepper} {et~al.}(2022){Van Clepper}, {Bergner}, {Bosman}, {Bergin}, \& {Ciesla}}]{vanClepper2022}
{Van Clepper}, E., {Bergner}, J.~B., {Bosman}, A.~D., {Bergin}, E., \& {Ciesla}, F.~J. 2022, \apj, 927, 206, \dodoi{10.3847/1538-4357/ac511b}

\bibitem[{{van Dijk} \& {Miguel}(2025)}]{2025vanDijk}
{van Dijk}, E., \& {Miguel}, Y. 2025, \mnras

\bibitem[{{van 't Hoff} {et~al.}(2020){van 't Hoff}, {Bergin}, {J{\o}rgensen}, \& {Blake}}]{vantHoff2020}
{van 't Hoff}, M. L.~R., {Bergin}, E.~A., {J{\o}rgensen}, J.~K., \& {Blake}, G.~A. 2020, \apjl, 897, L38, \dodoi{10.3847/2041-8213/ab9f97}

\bibitem[{{Vazan} {et~al.}(2015){Vazan}, {Helled}, {Kovetz}, \& {Podolak}}]{Vazan2015}
{Vazan}, A., {Helled}, R., {Kovetz}, A., \& {Podolak}, M. 2015, \apj, 803, 32, \dodoi{10.1088/0004-637X/803/1/32}

\bibitem[{{Vigan} {et~al.}(2021){Vigan}, {Fontanive}, {Meyer}, {Biller}, {Bonavita}, {Feldt}, {Desidera}, {Marleau}, {Emsenhuber}, {Galicher}, {Rice}, {Forgan}, {Mordasini}, {Gratton}, {Le Coroller}, {Maire}, {Cantalloube}, {Chauvin}, {Cheetham}, {Hagelberg}, {Lagrange}, {Langlois}, {Bonnefoy}, {Beuzit}, {Boccaletti}, {D'Orazi}, {Delorme}, {Dominik}, {Henning}, {Janson}, {Lagadec}, {Lazzoni}, {Ligi}, {Menard}, {Mesa}, {Messina}, {Moutou}, {M{\"u}ller}, {Perrot}, {Samland}, {Schmid}, {Schmidt}, {Sissa}, {Turatto}, {Udry}, {Zurlo}, {Abe}, {Antichi}, {Asensio-Torres}, {Baruffolo}, {Baudoz}, {Baudrand}, {Bazzon}, {Blanchard}, {Bohn}, {Brown Sevilla}, {Carbillet}, {Carle}, {Cascone}, {Charton}, {Claudi}, {Costille}, {De Caprio}, {Delboulb{\'e}}, {Dohlen}, {Engler}, {Fantinel}, {Feautrier}, {Fusco}, {Gigan}, {Girard}, {Giro}, {Gisler}, {Gluck}, {Gry}, {Hubin}, {Hugot}, {Jaquet}, {Kasper}, {Le Mignant}, {Llored}, {Madec}, {Magnard}, {Martinez}, {Maurel}, {M{\"o}ller-Nilsson}, {Mouillet}, {Moulin}, {Orign{\'e}},
  {Pavlov}, {Perret}, {Petit}, {Pragt}, {Puget}, {Rabou}, {Ramos}, {Rickman}, {Rigal}, {Rochat}, {Roelfsema}, {Rousset}, {Roux}, {Salasnich}, {Sauvage}, {Sevin}, {Soenke}, {Stadler}, {Suarez}, {Wahhaj}, {Weber}, \& {Wildi}}]{Vigan2021}
{Vigan}, A., {Fontanive}, C., {Meyer}, M., {et~al.} 2021, \aap, 651, A72, \dodoi{10.1051/0004-6361/202038107}

\bibitem[{{Villenave} {et~al.}(2022){Villenave}, {Stapelfeldt}, {Duch{\^e}ne}, {M{\'e}nard}, {Lambrechts}, {Sierra}, {Flores}, {Dent}, {Wolff}, {Ribas}, {Benisty}, {Cuello}, \& {Pinte}}]{Villenave2022}
{Villenave}, M., {Stapelfeldt}, K.~R., {Duch{\^e}ne}, G., {et~al.} 2022, \apj, 930, 11, \dodoi{10.3847/1538-4357/ac5fae}

\bibitem[{{Wahl} {et~al.}(2017){Wahl}, {Hubbard}, {Militzer}, {Guillot}, {Miguel}, {Movshovitz}, {Kaspi}, {Helled}, {Reese}, {Galanti}, {Levin}, {Connerney}, \& {Bolton}}]{2017GeoRL..44.4649W}
{Wahl}, S.~M., {Hubbard}, W.~B., {Militzer}, B., {et~al.} 2017, \grl, 44, 4649, \dodoi{10.1002/2017GL073160}

\bibitem[{{Wakeford} {et~al.}(2018){Wakeford}, {Sing}, {Deming}, {Lewis}, {Goyal}, {Wilson}, {Barstow}, {Kataria}, {Drummond}, {Evans}, {Carter}, {Nikolov}, {Knutson}, {Ballester}, \& {Mandell}}]{wakeford18}
{Wakeford}, H.~R., {Sing}, D.~K., {Deming}, D., {et~al.} 2018, \aj, 155, 29, \dodoi{10.3847/1538-3881/aa9e4e}

\bibitem[{{Wang} {et~al.}(2023){Wang}, {Ormel}, {Huang}, \& {Kuiper}}]{Wang2023}
{Wang}, Y., {Ormel}, C.~W., {Huang}, P., \& {Kuiper}, R. 2023, \mnras, 523, 6186, \dodoi{10.1093/mnras/stad1753}

\bibitem[{{Weidenschilling}(1977)}]{Weidenschilling1977}
{Weidenschilling}, S.~J. 1977, \mnras, 180, 57, \dodoi{10.1093/mnras/180.2.57}

\bibitem[{{Welbanks} \& {Madhusudhan}(2021)}]{Welbanks2021}
{Welbanks}, L., \& {Madhusudhan}, N. 2021, \apj, 913, 114, \dodoi{10.3847/1538-4357/abee94}

\bibitem[{{Welbanks} {et~al.}(2019){Welbanks}, {Madhusudhan}, {Allard}, {Hubeny}, {Spiegelman}, \& {Leininger}}]{welbanks19}
{Welbanks}, L., {Madhusudhan}, N., {Allard}, N.~F., {et~al.} 2019, \apjl, 887, L20, \dodoi{10.3847/2041-8213/ab5a89}

\bibitem[{{Welbanks} {et~al.}(2023){Welbanks}, {McGill}, {Line}, \& {Madhusudhan}}]{welbanks23}
{Welbanks}, L., {McGill}, P., {Line}, M., \& {Madhusudhan}, N. 2023, \aj, 165, 112, \dodoi{10.3847/1538-3881/acab67}

\bibitem[{{Welbanks} {et~al.}(2024){Welbanks}, {Bell}, {Beatty}, {Line}, {Ohno}, {Fortney}, {Schlawin}, {Greene}, {Rauscher}, {McGill}, {Murphy}, {Parmentier}, {Tang}, {Edelman}, {Mukherjee}, {Wiser}, {Lagage}, {Dyrek}, \& {Arnold}}]{welbanks24}
{Welbanks}, L., {Bell}, T.~J., {Beatty}, T.~G., {et~al.} 2024, \nat, 630, 836, \dodoi{10.1038/s41586-024-07514-w}

\bibitem[{{Welbanks} {et~al.}(2025){Welbanks}, {Nixon}, {McGill}, {Tilke}, {Wiser}, {Rotman}, {Mukherjee}, {Feinstein}, {Line}, {Seager}, {Beatty}, {Seligman}, {Parmentier}, \& {Sing}}]{welbanks25}
{Welbanks}, L., {Nixon}, M.~C., {McGill}, P., {et~al.} 2025, arXiv e-prints, arXiv:2504.21788, \dodoi{10.48550/arXiv.2504.21788}

\bibitem[{{Whiteford} {et~al.}(2024){Whiteford}, {Faherty}, {Burningham}, {Petrus}, {Patapis}, {Biller}, {Skemer}, \& {Hinkley}}]{whiteford2024}
{Whiteford}, N., {Faherty}, J., {Burningham}, B., {et~al.} 2024, in AAS/Division for Extreme Solar Systems Abstracts, Vol.~56, AAS/Division for Extreme Solar Systems Abstracts, 626.17

\bibitem[{{Whiteford} {et~al.}(2023){Whiteford}, {Glasse}, {Chubb}, {Kitzmann}, {Ray}, {Phillips}, {Biller}, {Palmer}, {Rice}, {Waldmann}, {Changeat}, {Skaf}, {Wang}, {Edwards}, \& {Al-Refaie}}]{Whiteford2023}
{Whiteford}, N., {Glasse}, A., {Chubb}, K.~L., {et~al.} 2023, \mnras, 525, 1375, \dodoi{10.1093/mnras/stad670}

\bibitem[{{Williams} \& {Best}(2014)}]{Williams2014}
{Williams}, J.~P., \& {Best}, W. M.~J. 2014, \apj, 788, 59, \dodoi{10.1088/0004-637X/788/1/59}

\bibitem[{{Williams} {et~al.}(2019){Williams}, {Cieza}, {Hales}, {Ansdell}, {Ruiz-Rodriguez}, {Casassus}, {Perez}, \& {Zurlo}}]{Williams2019}
{Williams}, J.~P., {Cieza}, L., {Hales}, A., {et~al.} 2019, \apjl, 875, L9, \dodoi{10.3847/2041-8213/ab1338}

\bibitem[{{Williams} \& {Cieza}(2011)}]{Williams2011}
{Williams}, J.~P., \& {Cieza}, L.~A. 2011, \araa, 49, 67, \dodoi{10.1146/annurev-astro-081710-102548}

\bibitem[{{Winter} {et~al.}(2024){Winter}, {Benisty}, {Manara}, \& {Gupta}}]{Winter2024b}
{Winter}, A.~J., {Benisty}, M., {Manara}, C.~F., \& {Gupta}, A. 2024, \aap, 691, A169, \dodoi{10.1051/0004-6361/202452120}

\bibitem[{{Wong} {et~al.}(2022){Wong}, {Chachan}, {Knutson}, {Henry}, {Adams}, {Kataria}, {Benneke}, {Gao}, {Deming}, {L{\'o}pez-Morales}, {Sing}, {Alam}, {Ballester}, {Barstow}, {Buchhave}, {dos Santos}, {Fu}, {Garc{\'\i}a Mu{\~n}oz}, {MacDonald}, {Mikal-Evans}, {Sanz-Forcada}, \& {Wakeford}}]{wong22}
{Wong}, I., {Chachan}, Y., {Knutson}, H.~A., {et~al.} 2022, \aj, 164, 30, \dodoi{10.3847/1538-3881/ac7234}

\bibitem[{{Worthen} {et~al.}(2024){Worthen}, {Chen}, {Law}, {Lu}, {Hoch}, {Chai}, {Sloan}, {Sargent}, {Kammerer}, {Hines}, {Rebollido}, {Balmer}, {Perrin}, {Watson}, {Pueyo}, {Girard}, {Lisse}, \& {Stark}}]{Worthen2024}
{Worthen}, K., {Chen}, C.~H., {Law}, D.~R., {et~al.} 2024, \apj, 964, 168, \dodoi{10.3847/1538-4357/ad2354}

\bibitem[{{Xie} {et~al.}(2023){Xie}, {Pascucci}, {Long}, {Pontoppidan}, {Banzatti}, {Kalyaan}, {Salyk}, {Liu}, {Najita}, {Pinilla}, {Arulanantham}, {Herczeg}, {Carr}, {Bergin}, {Ballering}, {Krijt}, {Blake}, {Zhang}, {{\"O}berg}, {Green}, \& {Jdiscs Collaboration}}]{Xie2023}
{Xie}, C., {Pascucci}, I., {Long}, F., {et~al.} 2023, \apjl, 959, L25, \dodoi{10.3847/2041-8213/ad0ed9}

\bibitem[{{Xue} {et~al.}(2024){Xue}, {Bean}, {Zhang}, {Welbanks}, {Lunine}, \& {August}}]{xue24}
{Xue}, Q., {Bean}, J.~L., {Zhang}, M., {et~al.} 2024, \apjl, 963, L5, \dodoi{10.3847/2041-8213/ad2682}

\bibitem[{{Young} \& {Clarke}(2015)}]{young2015}
{Young}, M.~D., \& {Clarke}, C.~J. 2015, \mnras, 451, 3987, \dodoi{10.1093/mnras/stv1266}

\bibitem[{{Young} \& {Clarke}(2016)}]{young2016}
---. 2016, \mnras, 455, 1438, \dodoi{10.1093/mnras/stv2378}

\bibitem[{{Yu} {et~al.}(2016){Yu}, {Willacy}, {Dodson-Robinson}, {Turner}, \& {Evans}}]{Yu2016}
{Yu}, M., {Willacy}, K., {Dodson-Robinson}, S.~E., {Turner}, N.~J., \& {Evans}, II, N.~J. 2016, \apj, 822, 53, \dodoi{10.3847/0004-637X/822/1/53}

\bibitem[{{Zahnle} {et~al.}(2003){Zahnle}, {Schenk}, {Levison}, \& {Dones}}]{Zahnle2003}
{Zahnle}, K., {Schenk}, P., {Levison}, H., \& {Dones}, L. 2003, \icarus, 163, 263, \dodoi{10.1016/S0019-1035(03)00048-4}

\bibitem[{{Zhang} {et~al.}(2019){Zhang}, {Bergin}, {Schwarz}, {Krijt}, \& {Ciesla}}]{Zhang2019}
{Zhang}, K., {Bergin}, E.~A., {Schwarz}, K., {Krijt}, S., \& {Ciesla}, F. 2019, \apj, 883, 98, \dodoi{10.3847/1538-4357/ab38b9}

\bibitem[{{Zhang} {et~al.}(2021){Zhang}, {Snellen}, {Bohn}, {Molli{\`e}re}, {Ginski}, {Hoeijmakers}, {Kenworthy}, {Mamajek}, {Meshkat}, {Reggiani}, \& {Snik}}]{Zhang2021}
{Zhang}, Y., {Snellen}, I. A.~G., {Bohn}, A.~J., {et~al.} 2021, \nat, 595, 370, \dodoi{10.1038/s41586-021-03616-x}

\bibitem[{Zhang(2024)}]{Zhang2024}
Zhang, Z. 2024, Research Notes of the AAS, 8, 114, \dodoi{10.3847/2515-5172/ad4481}

\bibitem[{{Zinzi} \& {Turrini}(2017)}]{Zinzi2017}
{Zinzi}, A., \& {Turrini}, D. 2017, \aap, 605, L4, \dodoi{10.1051/0004-6361/201731595}

\bibitem[{{Zormpas} {et~al.}(2022){Zormpas}, {Birnstiel}, {Rosotti}, \& {Andrews}}]{Zormpas2022}
{Zormpas}, A., {Birnstiel}, T., {Rosotti}, G.~P., \& {Andrews}, S.~M. 2022, \aap, 661, A66, \dodoi{10.1051/0004-6361/202142046}

\end{thebibliography}
\bibliographystyle{aasjournal}

\end{document}